\documentclass[twocolumn]{aastex631}
\usepackage{amsmath}
\usepackage{fontawesome}

\definecolor{linkcolor}{rgb}{0.1216,0.4667,0.7059}
\definecolor{revcolor}{rgb}{0,0,0}

\definecolor{kwsmcolor}{rgb}{0.5,0,0.5}

\newcommand{\fv}{\boldsymbol{f}}
\newcommand{\Fv}{\boldsymbol{F}}
\newcommand{\Gv}{\boldsymbol{G}}
\newcommand{\Bv}{\boldsymbol{\mathcal{B}}}

\newcommand{\xv}{\boldsymbol{x}}

\newcommand{\uv}{\boldsymbol{u}}
\newcommand{\qv}{\boldsymbol{q}}

\newcommand{\tcv}{\boldsymbol{\phi}}

\newcommand{\nulc}{\hat{\nu}} 
\newcommand{\nulcl}{\nulc_{l}} 

\newcommand{\ntexp}{{\mathsf{n}}} 
\newcommand{\ntexpl}{{\mathsf{n}_l}} 

\newcommand{\bpar}{{\boldsymbol{b}}}
\newcommand{\nbeta}{\acute{\beta}}
\newcommand{\ngamma}{\acute{\gamma}}
\newcommand{\aD}{\acute{\beta}}
\newcommand{\Elow}{E}
\newcommand{\Elowl}{\Elow_l}

\newcommand{\LSD}{lineshape density }

\newcommand{\Bl}{B_\nu}
\newcommand{\dBl}{\dot{B}_\nu}

\newcommand{\Ttyp}{T_\mathrm{wp}}
\newcommand{\ttyp}{t_\mathrm{wp}}
\newcommand{\Tref}{T_\mathrm{ref}}
\newcommand{\tref}{t_\mathrm{ref}}
\newcommand{\Gaa}{\mathcal{G}_{11}}
\newcommand{\Gab}{\mathcal{G}_{12}}
\newcommand{\Gba}{\mathcal{G}_{21}}
\newcommand{\Gbb}{\mathcal{G}_{22}}

\newcommand{\Nlayer}{N_\mathrm{layer}}

\newcommand{\gnu}{\underline{\nu}}
\newcommand{\gbpar}{\underline{\bpar}}
\newcommand{\gElow}{\underline{\Elow}}

\newcommand{\Kzz}{K_{zz}}
\newcommand{\fsed}{f_\mathrm{sed}}
\newcommand{\cond}{{\mathsf{c}}}
\newcommand{\total}{{\mathsf{tot}}}
\newcommand{\vapor}{{\mathsf{v}}}

\usepackage{color}

\definecolor{red}{rgb}{0.7,0.0,0.0}
\definecolor{green}{rgb}{0.0,0.5,0.0}

\usepackage{pifont}
\newcommand{\cmark}{\ding{51}}%
\newcommand{\xmark}{\ding{55}}%
\newcommand{\gcheckmark}{ {\color{green} \cmark }}
\newcommand{\notava}{ {\color{red} \xmark }}

\usepackage{algorithmic}
\received{October 10, 2024}
\revised{March 28th, 2025}
\accepted{April 10th, 2025}
\shorttitle{{\sf ExoJAX}2: Differentiable Spectrum Model}
\shortauthors{Kawahara et al.}
\begin{document}

\title{Differentiable Modeling of Planet and Substellar Atmosphere: High-Resolution Emission, Transmission, and Reflection Spectroscopy with ExoJAX2}

\correspondingauthor{Hajime Kawahara}

\author[0000-0003-3309-9134]{Hajime Kawahara}
\email{kawahara@ir.isas.jaxa.jp}
\affiliation{Department of Space Astronomy and Astrophysics, ISAS/JAXA, 3-1-1, Yoshinodai, Sagamihara, Kanagawa, 252-5210 Japan}
\affiliation{Department of Astronomy, Graduate School of Science, The University of Tokyo, 7-3-1 Hongo, Bunkyo-ku, Tokyo 113-0033, Japan}

\author[0000-0003-3800-7518]{Yui Kawashima}
\affiliation{Department of Astronomy, Graduate School of Science, Kyoto University, Kitashirakawa Oiwake-cho, Sakyo-ku, Kyoto 606-8502, Japan}
\affiliation{Frontier Research Institute for Interdisciplinary Sciences and Department of Geophysics, Tohoku University, 
\\Aramaki aza Aoba 6-3, Aoba-ku, Sendai, 980-8578, Japan}
\affiliation{Department of Space Astronomy and Astrophysics, ISAS/JAXA, 3-1-1, Yoshinodai, Sagamihara, Kanagawa, 252-5210 Japan}
\affiliation{Cluster for Pioneering Research, RIKEN, 2-1 Hirosawa, Wako, Saitama 351-0198, Japan}

\author[0000-0002-4093-8736]{Shotaro Tada}
\affiliation{Astronomical Science Program, The Graduate University for Advanced Studies, SOKENDAI, \\2-21-1 Osawa, Mitaka, Tokyo 181-8588, Japan}

\author[0000-0001-6309-4380]{Hiroyuki Tako Ishikawa}
\affiliation{Department of Physics and Astronomy, The University of Western Ontario, 1151 Richmond St, London, Ontario, N6A~3K7, Canada}


\author[0000-0001-5414-4171]{Ko Hosokawa}
\affiliation{Astronomical Science Program, The Graduate University for Advanced Studies, SOKENDAI, \\2-21-1 Osawa, Mitaka, Tokyo 181-8588, Japan}

\author[0000-0002-8607-358X]{Yui Kasagi}
\affiliation{Institute of Space and Astronomical Science, Japan Aerospace Exploration Agency, \\3-1-1 Yoshinodai, Chuo-ku, Sagamihara, Kanagawa, 252-5210, Japan}

\author[0000-0001-6181-3142]{Takayuki Kotani}
\affiliation{Astrobiology Center, 2-21-1 Osawa, Mitaka, Tokyo 181-8588, Japan}
\affiliation{National Astronomical Observatory of Japan, 2-21-1 Osawa, Mitaka, Tokyo 181-8588, Japan}
\affiliation{Astronomical Science Program, The Graduate University for Advanced Studies, SOKENDAI, \\2-21-1 Osawa, Mitaka, Tokyo 181-8588, Japan}

\author[0000-0003-1298-9699]{Kento Masuda}
\affiliation{Department of Earth and Space Science, Osaka University, Osaka 560-0043, Japan}

\author[0000-0003-4698-6285]{Stevanus K Nugroho}
\affiliation{Astrobiology Center, 2-21-1 Osawa, Mitaka, Tokyo 181-8588, Japan}
\affiliation{National Astronomical Observatory of Japan, 2-21-1 Osawa, Mitaka, Tokyo 181-8588, Japan}

\author[0000-0002-6510-0681]{Motohide Tamura} 
\affiliation{Graduate School of Science, The University of Tokyo,  7-3-1 Hongo, Bunkyo-ku, Tokyo 113-0033, Japan}
\affiliation{Astrobiology Center, 2-21-1 Osawa, Mitaka, Tokyo 181-8588, Japan}
\affiliation{National Astronomical Observatory of Japan, 2-21-1 Osawa, Mitaka, Tokyo 181-8588, Japan}

\author[0000-0001-7692-0581]{Hibiki Yama}
\affiliation{Department of Earth and Space Science, Osaka University, Osaka 560-0043, Japan}

\author[0000-0003-4269-3311]{Daniel Kitzmann}
\affiliation{Space Research and Planetary Sciences, Physics Institute, University of Bern,
 Gesellschaftsstrasse 6, 3012 Bern, Switzerland}

\author[0000-0002-7050-7873]{Nicolas Minesi}
\affiliation{Laboratoire EM2C, CNRS, CentraleSup\'{e}lec, Universit\'{e} Paris-Saclay, 3 rue Joliot Curie, 91190 Gif-sur-Yvette, France}

\author[0000-0003-2528-3409]{Brett M. Morris}
\affiliation{Space Telescope Science Institute, 3700 San Martin Dr, Baltimore, MD 21218}
\nocollaboration{14}

\begin{abstract}
Modeling based on differentiable programming holds great promise for astronomy, enabling advanced techniques such as gradient-based posterior sampling and optimization. This paradigm motivated us to develop {\sf ExoJAX} \citep{2022ApJS..258...31K}, the first auto-differentiable spectrum model of exoplanets and brown dwarfs. {\sf ExoJAX} directly calculates cross-sections as functions of temperature and pressure to minimize interpolation errors in high-dispersion spectra, although initial work focused on narrowband emission spectroscopy. Here, we introduce a fast, memory-efficient opacity algorithm and differentiable radiative transfer for emission, transmission, and reflection spectroscopy.  In the era of data-rich JWST observations, retrieval analyses are often forced to bin high-resolution spectra due to computational bottlenecks. The new algorithm efficiently handles native-resolution data, preserving the full information content and dynamic range. The advances proposed in this paper enable broader applications, demonstrated by retrievals of GL229 B’s high-dispersion emission, WASP-39 b’s JWST mid-resolution transmission at original resolution ($R\sim 2,700$), and Jupiter’s reflection spectrum. We derive a C/O ratio for GL229 B consistent with its host star, constrain WASP-39 b’s radial velocity from molecular line structures, and infer Jupiter’s metallicity in line with previous estimates.
\end{abstract}


\section{Introduction}

Recent advancements in gradient-based techniques for Bayesian inference in astronomy have been widely recognized. Hamiltonian Monte Carlo \citep[HMC;][]{1987PhLB..195..216D,2011arXiv1111.4246H,2017arXiv170102434B} has much higher acceptance rates, lower correlations between samples, and a milder dependence of computational time on the number of parameters than the conventional Metropolis-Hastings algorithm with a random walk proposal distribution.\footnote{See \cite{2017arXiv170102434B} and \cite{2012arXiv1206.1901N} for references on HMC.} HMC has played a crucial role in the precise determination of a mass-radius relation for the TRAPPIST-1 planets using transit timing and photometric data \citep{2020arXiv201001074A}, the modeling of the radial velocity anomaly of a tidally deformed giant star in a binary \citep{2021ApJ...910L..17M}, and an occultation map of Io's surface \citep{2021arXiv210303758B}. However, HMC requires gradient information of the model.

Recent developments in packages for differentiable programming \citep[DP;][]{2024arXiv240314606B}, such as {\sf JAX}, {\sf PyTorch}, and {\sf TensorFlow}, and probabilistic programming languages (PPLs) like {\sf NumPyro}, {\sf PyMC}, and {\sf BlackJAX}, have allowed for more flexible code development for gradient-based modeling and inference. This new technique is rapidly being used in astronomical applications of gradient-based optimization, such as modeling astronomical optical systems \citep{2021ApJ...907...40P} and stellar population synthesis \citep{2023MNRAS.521.1741H}. It has now become possible to efficiently perform high-dimensional Bayesian inference using DP in the field of cosmology \citep{2022ApJ...935...49G,2023OJAp....6E..15C,2023OJAp....6E..20P,2024arXiv240512965P}.

Bayesian inference is an indispensable tool in modern exoplanet atmospheric research, particularly in analyzing low to medium dispersion spectra—a process known as {\it retrieval} \citep[e.g.,][and references therein]{2009ApJ...707...24M,2012MNRAS.420..170L,2013ApJ...775..137L}. To date, about 50 atmospheric retrieval codes have been published \citep[for completeness, see a catalog by][]{2023arXiv230312925M}. Recently, we have applied a DP-based approach to the atmospheric modeling of the high-dispersion spectra of exoplanets and brown dwarfs \citep[][hereafter Paper I]{2022ApJS..258...31K}. Our code, called {\sf ExoJAX} (hereafter {\sf ExoJAX}1), uses the auto-differentiation and accelerated linear algebra package {\sf JAX} \citep{jax2018github}, and therefore allows retrieval using HMC because of the ability to compute gradients of the spectral model. \cite{2024ApJ...961...30Y} presented another differentiable retrieval code, {\sf Diff-$\tau$}, using {\sf TensorFlow} and demonstrated retrieval using variational inference with normalizing flows as an application of the gradient-based retrieval code. \cite{2023arXiv230106575V} also proposed a retrieval code based on normalizing flows with variational inference. In their case, they use neural posterior estimation (NPE) to enable variational inference by learning spectra generated by {\sf petitRADTRANS}. Alternatively, \cite{2022ApJ...941..200G} proposed a novel hybrid data/model-driven approach to analyzing high-resolution spectra using {\sf PyTorch}.

A significant advantage of DP-based codes, particularly from a pragmatic perspective, is their capability to execute gradient-based optimization. Nowadays, numerous rapidly converging gradient optimization techniques have been suggested and implemented \citep[e.g.,][]{kochenderfer2019algorithms}. These include ADAM \citep[Adaptive Moment Estimation;][]{2014arXiv1412.6980K}, which is the current standard for gradient-based optimization in the field of machine learning. 
{Since the best-fit initialization of molecular and atmospheric parameters is not immediately clear from visual inspection alone} in the observational modeling of spectra, efficient model fitting is essential for practical applications.

In this paper, we extend {\sf ExoJAX}1 to be applicable to broadband emission, transmission, and reflection spectroscopy. This is achieved by introducing a new GPU memory-efficient opacity calculator, adding accessibility to atomic databases in addition to molecular ones, and providing a systematic treatment of the radiative transfer scheme. This paper refers to the new version with these extensions as {\sf ExoJAX}2.
{ These extensions enable the atmospheric retrieval of JWST transmission spectra without the need for data binning and also allow for the direct retrieval of high-resolution spectra of brown dwarfs.}

Among the various spectrum models and retrieval codes, the design philosophy of {\sf ExoJAX}2 can be summarized as follows:
\begin{enumerate}
    \item Publicly available\footnote{\cite{hajime_kawahara_2025_15094566} as the frozen version in Zenodo} \href{https://github.com/HajimeKawahara/exojax}{\color{linkcolor}\faGithub} , maintainable, flexible Python package with unit testing. 
    \item Provides end-to-end simulation capabilities—from loading molecular line lists (e.g., ExoMol \citep{2016JMoSp.327...73T}, HITRAN/HITEMP \citep{2010JQSRT.111.2139R}) to generating spectral models comparable to observations -- all within a single package.
    \item Offers an auto-differentiable spectral model compatible with PPLs and gradient-based optimization, providing sufficient wavelength resolution even for high-dispersion data.
\end{enumerate}
In other words, {\sf ExoJAX}2 is not strictly a retrieval code but rather a differentiable spectral model, and from the perspective of retrieval, it has a modular design that allows users to choose their preferred inference method through a PPL.

Although {\sf ExoJAX}2 is essentially a functional library, usability is enhanced by separable modules corresponding to each computational stage. These modules include molecular/atomic/continuum/particulates database IO ({\sf {\it x}\,db}; where {\sf {\it x} = m, a, c, or p}), opacity calculation ({\sf opa}), atmospheric radiative transfer ({\sf art}), and spectral operators such as rotation and instrumental response ({\sf sop}). They correspond to Sections 2, 3, 4, and 6, respectively. {Section 5 discusses considerations specific to differentiable programming and methods for reducing GPU device memory usage. Section 7 applies {\sf ExoJAX2} to three observational datasets: the high-dispersion emission spectrum of Gl 229B, the native-resolution transmission spectrum of WASP-39b observed by JWST using NIRSpec/G395H, and the high-dispersion reflection spectrum of Jupiter, demonstrating Bayesian analysis using HMC-NUTS. Section 8 presents the discussion and summary.}

\section{Line and Continuum Database Interface}

To effectively utilize gradient-based techniques for atmospheric modeling, having efficient and flexible access to comprehensive opacity data is crucial. {\sf ExoJAX}2 addresses this requirement by enhancing its database interfaces, which form the backbone of high-precision spectral simulations. In this section, we introduce these improvements, highlighting how they enable more extensive and flexible handling of line and continuum opacities. As summarized in Table \ref{tab:availability_line_continuum}, these enhancements significantly broaden the range of available opacity data within {\sf ExoJAX}2.

\begin{table}[]
  \caption{Line and continuum opacities available in {\sf ExoJAX}1 and 2}
  \begin{center}
  \begin{tabular}{lcc} 
  \hline\hline
database/opacity &   {\sf ExoJAX}1 & {\sf ExoJAX}2 \\
\hline
\multicolumn{3}{c}{\it molecular line} \\
ExoMol &   \gcheckmark & \gcheckmark \\
HITEMP &   \gcheckmark & \gcheckmark \\
HITRAN  air broadening & \gcheckmark & \gcheckmark \\
HITRAN  non-air broadening$\,^\star$ & \notava & \gcheckmark \\
\hline
\multicolumn{3}{c}{\it metal line} \\
Kurucz  &  \notava  & \gcheckmark \\
VALD3   &  \notava  & \gcheckmark \\
\hline
\multicolumn{3}{c}{\it continuum} \\
CIA & \gcheckmark$\,^\dagger$ & \gcheckmark$\,^\dagger$\\
H- &  \notava & \gcheckmark$\,^\ast$ \\
Rayleigh scattering &  \notava & \gcheckmark \\
Cloud (Mie) scattering &  \notava &\gcheckmark$^\ddagger$ \\
\hline
\hline
  \end{tabular}
  \end{center}
  \tablecomments{$\star$: H/He/CO2/H2O \citep{2022ApJS..262...40T}.
$\dagger$:\cite{2012JQSRT.113.1276R,2019Icar..328..160K}, $\ast$:\cite{1988A&A...193..189J}, $\ddagger$: based on \cite{ackerman2001precipitating} cloud model, but, the Mie scattering computation from {\sf PyMieScatt} \citep{2018JQSRT.205..127S}.}
\label{tab:availability_line_continuum}
\end{table}

\subsection{Molecular Lines}

In Paper I, we introduced a built-in I/O system that sourced data from molecular databases such as ExoMol, HITRAN, and HITEMP \citep{2016JMoSp.327...73T, 2009JQSRT.110..533R}. In {\sf ExoJAX}2, we adopted an approach that accesses molecular and atomic databases via an external API, {\sf radis.api}, which is currently integrated with the fast opacity calculator {\sf Radis} \citep{pannier2019radis} and is being co-developed by both the {\sf Radis} and {\sf ExoJAX} teams. Within {\sf Radis}, the opacity calculator uses DIT \citep[Discrete Integral Transform; ][]{2021JQSRT.26107476V}-- the precursor of the new algorithm in this paper -- as its core engine. The API leverages {\sf Vaex} \citep{2018A&A...618A..13B} to load molecular datasets using lazy out-of-core data frames and accommodates non-air pressure-induced broadenings caused by hydrogen, helium, carbon dioxide, and water \citep{2022ApJS..262...40T}. {\sf ExoJAX}2 can also factor in the pressure shift caused by a non-air background atmosphere. The associated velocity shift may be inconsequential for the transmission spectrum but can be substantial for the emission spectrum because the pressure of the atmospheric layer that affects the spectrum is lower in the former, while it is higher in the latter. For instance, in a hydrogen-rich atmosphere, the velocity shift for CO K-band absorption can be up to $\Delta V \lesssim 0.3 (P/1 \mathrm{bar})$ km/s.

One significant benefit of end-to-end simulations is their ability to maximize the information present in molecular databases. {\sf ExoJAX 2} can sift through molecular lines based on optional data from the databases, such as vibrational quantum numbers or electronic states. This functionality proves especially valuable for tasks like identifying non-LTE, as demonstrated in a recent high-dispersion spectral analysis of an ultra-hot Jupiter \citep{2023arXiv230511071W}.

\subsection{Metal Lines} 

Metal lines frequently appear in the high-dispersion spectra of ultrahot planets and cool dwarfs. Thus, {\sf ExoJAX }2 supports the utilization of atomic line lists, incorporating those from Kurucz \citep{2005MSAIS...8...14K} and VALD3 \citep{2015PhyS...90e4005R}. Table \ref{tab:gamma_parameters} lists the five pressure broadening methods supported by {\sf ExoJAX }2.

\begin{table*}[tb!]
\centering
\caption{Description of gamma parameters for Kurucz and VALD databases in {\sf ExoJAX}.}
\label{tab:gamma_parameters}
\begin{tabular}{ll}
\hline\hline
Method & Description \\
\hline
gamma\_vald3 & Uses van der Waals gamma from line lists (VALD or Kurucz) or estimates from \cite{1955psmb.book.....U}. \\
gamma\_uns & Approximated gamma based on \cite{1955psmb.book.....U}. \\
gamma\_KA3 & Derived from the third equation in \cite{1981SAOSR.391.....K} (page 4). \\
gamma\_KA4 & From the fourth equation in \cite{1981SAOSR.391.....K} (page 4). \\
gamma\_KA3s & As gamma\_KA3, but without differentiating iron group elements. \\
\hline
\end{tabular}
\end{table*}

 \subsection{Continuous Opacity}\label{ss:cia}
{\sf ExoJAX }1 includes only collision-induced absorption (CIA) as its continuous opacity source. {\sf In ExoJAX }2, in addition to CIA, we support Rayleigh scattering, Mie scattering due to clouds, and the H- continuum, as summarized in Table \ref{tab:availability_line_continuum}. These newly added sources of continuum opacity can be utilized in visible light transmission spectroscopy, and are applicable to cloudy planets and brown dwarfs.

The scattering cross-section of the Rayleigh scattering for gas is expressed as a function of the polarizability of the gas, denoted by $\alpha$, as
\begin{align}
\label{eq:rayleigh_alpha}
    \sigma_\mathrm{ray}(\nu) = \frac{ 128 \pi^5 }{3} \nu^4 \alpha^2 F_k
\end{align}
where $F_k$ is the King factor \cite[e.g.][]{liou2002introduction}, which explains the effect of the depolarization. 

In the context of Mie scattering, the opacity is determined by assuming a log-normal distribution for the size distribution of the condensates. 
\begin{eqnarray}
    n_0 (r) d r = \frac{N_0}{r \sqrt{2 \pi} \log \sigma_g} e^{- \frac{(\log{r} - \log{r_g})^2}{2 \log^2 \sigma_g}} dr,
\end{eqnarray}
where $N_0$ is the reference number density of the condensates ($\mathrm{cm^{-3}}$). This involves using the size radius $r_g$ and log-variance $\sigma_g$ as parameters to calculate grids for the volume extinction coefficient $\beta^\mathrm{ext}_0$ and volume scattering coefficient, 
\begin{align}
    \beta^\mathrm{ext}_0 &= \int_0^\infty \pi r^2 Q_\mathrm{ext} n_0 (r) dr \\
    \beta^\mathrm{scat}_0 &= \int_0^\infty \pi r^2 Q_\mathrm{scat} n_0 (r) dr
\end{align}
and the asymmetry factor $g$ using {\sf PyMieScatt} \citep{2018JQSRT.205..127S}, where $Q_\mathrm{ext}$ and $Q_\mathrm{scat}$ are Mie efficiency for extinction and scattering\footnote{
We use the functions for homogeneous spheres, in particular, assuming a log-normal distribution, {\sf Mie\_Lognormal} in {\sf PyMieScatt}. We regard the bulk asymmetric parameter, i.e., the scattering coefficient weighted asymmetric factor as $g$, which is one of the outputs of {\sf Mie\_Lognormal}. 
{The functions in PyMieScatt are not differentiable as they are, so a grid of $\sigma_g$ and $r_g$ is created, and the values are interpolated using {\sf jax.numpy.interp} to implement them in a differentiable form.}
}. 
We note that $\beta^\mathrm{ext}_0$ and $\beta^\mathrm{ext}_0$ are those values for the reference condensates number density $N_0$ (we adopt $N_0 = 1 \mathrm{cm^{-3}}$ as default).

The values of the extinction coefficient, single scattering albedo, and asymmetry factor for a given $r_g$, $\sigma_g$ for the reference condensate number density can be obtained by performing bilinear interpolation on these grids. Then, we can compute the layer cloud optical depth by
\begin{align}
    d \tau_\mathrm{cld} = - \beta_\mathrm{ext} dr =  - \beta^\mathrm{ext}_0 \frac{\xi_c n}{N_0} dr =  \frac{\xi_c \sigma^\mathrm{ext}_\mathrm{cld}}{\mu m_H g} dP, 
\end{align}
where $\beta_\mathrm{ext}$ is the extinction coefficient for the condensate number density of $N_c = \xi_c n$, $\xi_c$ is the condensate volume mixing ratio, $n$ is the gas number density, $\mu$ is the mean molecular weight, $m_H$ is the atomic mass unit, $g$ is the gravity, $dP$ is the layer pressure difference, and $\sigma^\mathrm{ext}_\mathrm{cld} \equiv \beta^\mathrm{ext}_0/N_0$ is the cloud extinction cross section in the unit of $\mathrm{cm}^2$. We also define the cloud scattering cross-section 
by $\sigma_\mathrm{cld}^\mathrm{scat} \equiv \beta^\mathrm{scat}_0/N_0$.

Recent high-dispersion spectral characterizations may probe the atmospheres of extremely hot planets like WASP-33b \citep[e.g.][]{2017AJ....154..221N,2021ApJ...910L...9N}.
{\sf ExoJAX }2 incorporates support for the H${^-}$ continuum  \citep{1988A&A...193..189J}, particularly relevant for ultrahot Jupiters.

\section{Opacity Calculator for Huge Line Lists}\label{sec:opacity}

One of the key features of {\sf ExoJAX} is its ability to perform end-to-end analysis of observed spectra using molecular and atomic databases as input. To achieve this, in {\sf ExoJAX}1, at each step of the MCMC or optimization process, the opacity (or cross-section) is recalculated directly from the line information in the databases, as shown in the left panel of Figure \ref{fig:diffdit}. {\sf ExoJAX} accelerates Bayesian inference by performing core calculations on a GPU device. However, this implementation requires device memory usage proportional to the number of spectral lines and atmospheric layers. This high memory requirement makes broadband spectrum computation difficult because GPU device memory is usually much smaller than the host's RAM.\footnote{In this paper, `device' refers to the GPU and its global memory, while `host' denotes the CPU and RAM \citep[e.g.,][]{sanders2010cuda}. For instance, an RTX A100 GPU has 80GB of CUDA memory.}

To overcome the device memory issue, we adopt a strategy of pre-computing the line strength density -- a condensed representation of opacity contributions from spectral lines -- before loading it into device memory, as shown in the right panel of Figure \ref{fig:diffdit}.

Table \ref{tab:memory} presents the estimated device memory usage for the methods used in {\sf ExoJAX}1 (LPF Direct and {Modified version of Discrete Integral Transform;} MODIT) and the new method proposed in this paper ({Precomputation of line density with MODIT}; PreMODIT), along with the actual usage demonstrated during the high-resolution spectroscopic retrieval of the H-band for the brown dwarf Gl 229B in Section 6.1. Compared to the previous methods, the new PreMODIT method effectively overcomes previous memory limitations by fitting within the device memory limits of current GPUs, enabling efficient broadband spectrum computations.

 \begin{table*}[]
  \caption{Comparison of required device memory for a single spectrum computation \label{tab:memory}}
  \begin{center}
  \begin{tabular}{lcccccc} 
  \hline\hline
  method & target lines& & dominant memory usage & Gl229B H-band & forward diff& reverse diff\\
  \hline
LPF Direct & metals & &$2J N_\mathrm{free} N_\mathrm{layer} N_\nu \sum_i  N_\mathrm{line}^{(i)}$ & $3 \times 10^8$~GB & \gcheckmark & \gcheckmark\\
MODIT & & &$K  N_\mathrm{free} N_\nu \sum_i N_\mathrm{line}^{(i)} N_\mathrm{par}$ &  $3 \times 10^7$~GB  & \gcheckmark & \gcheckmark \\
PreMODIT & molecules  & (LBD)  &$2 J \mathrm{max}_j (N_\nu^{(j)} N_\mathrm{brd}^{(j)} N_\mathrm{elower}^{(j)} )$ & 0.1~GB  & \gcheckmark & \gcheckmark \\
- & & (opacity) &  $K N_\mathrm{free} N_\mathrm{layer} \mathrm{max}_j (N_\nu^{(j)} N_\mathrm{brd}^{(j)} )$ & 50~GB  & \gcheckmark & \gcheckmark \\
\hline
PreMODIT (opart) & molecule  & (opacity) &  $K N_\mathrm{free} \mathrm{max}_j (N_\nu^{(j)} N_\mathrm{brd}^{(j)} )$ & 0.3~GB  & \gcheckmark & \notava \\
\hline
  \end{tabular}
  \end{center}
  \tablecomments{$N_\mathrm{info} \ge 5$: the number of the line information entries. $J = 8$ byte. $K=4 \times 8 $ byte (FFT for complex128 plus float64 with the double buffer size). $N_\mathrm{par} = 10$: the number of the parameters of the line property. $N_\mathrm{free}$: the number of the free parameters for HMC or the optimization.
  The parameters for the calculation of Gl229B H-band is as follows: $N_\mathrm{free} = 10$, $N_\mathrm{layer} = 200$, $N_\nu = 8 \times 10^4$.
  For $\mathrm{H_2O}$ (ExoMol), $\mathrm{CH_4}$ (HITEMP), and $\mathrm{NH_3}$ (ExoMol), $N_\mathrm{line} = 6 \times 10^5, 9 \times 10^5, 1 \times 10^8$ and $N_\mathrm{brd} = 6, 10, 4$, $N_\mathrm{elower} = 10, 8, 7$. {The columns of forward diff and reverese diff indicate the capability of forward-mode and reverse-mode differentiations. }}
  \label{tab:device_memory_usage}
\end{table*}

\subsection{Formulation}\label{ss:formulation}

\begin{figure}
    \centering
    \includegraphics[width=\linewidth]{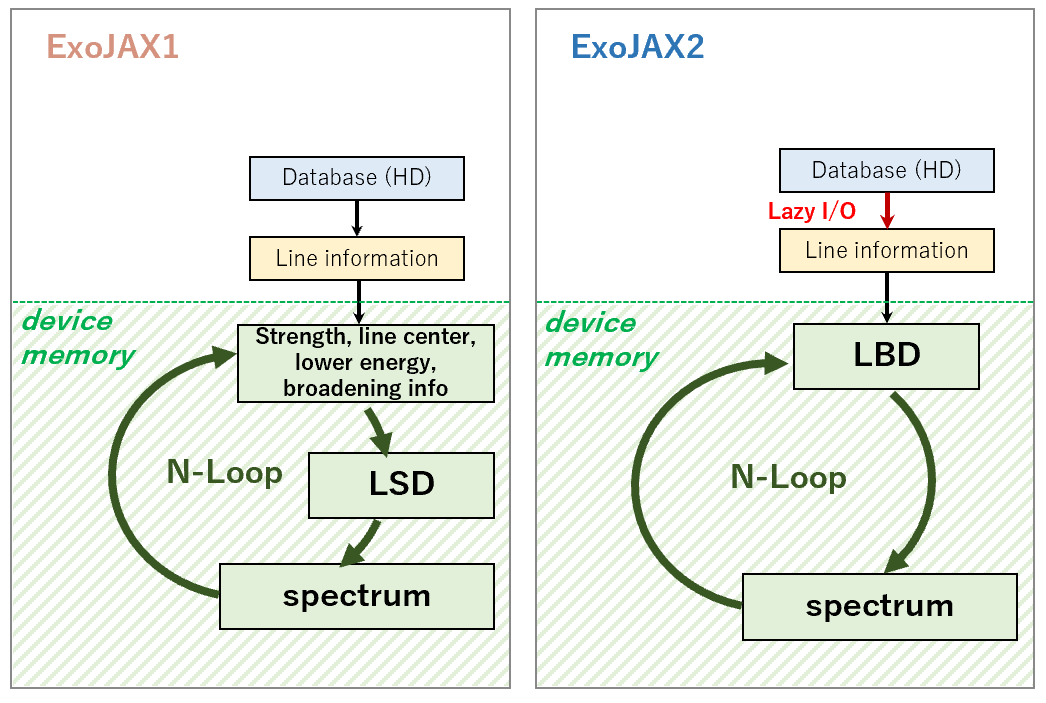}
    \caption{Difference of the major drivers of the opacity calculators between {\sf ExoJAX}1 ({Modified version of Discrete Integral Transform;} MODIT) and {\sf ExoJAX}2 ({Precomputation of line density with MODIT;} PreMODIT). The quantities in the shaded regions are stored in the GPU device memory. While the \LSD (LSD in the panel) is recalculated each time the physical conditions (i.e. $T$, abundance, etc) are changed in MODIT, we only need to calculate the line basis density (LBD) the first time in {\sf ExoJAX}2. This reduces the computational cost and device memory usage during iterative calculations at HMC-NUTS, optimization, and so on.}
    \label{fig:diffdit}
\end{figure}

{\sf ExoJAX }1 had two auto-differentiable opacity calculators. One method calculated the Voigt profile for each line, called direct-LPF (line profile), and the other method, MODIT densified the line information and then calculated the cross-section\footnote{A modified method based on Discrete Integral Transform  \cite[DIT;][]{van2021discrete}}. As shown in Paper I, MODIT is advantageous for numbers of lines larger than a few hundred.  This paper further improves this method to enable a wider bandwidth for the opacity calculation.

Here, we modify MODIT to reduce the use of the device memory and to accelerate the computational time for a very large number of the lines $(\gtrsim 10^5)$. The line strength of the $l$-th line at temperature $T$ is expressed as
\begin{eqnarray}
\label{eq:linestT}
S_l (T; \Elow_l, \nulcl)  &=&  \frac{g(\nulc_l,T)}{q_r(T)} f(\Elowl,T) S_{\mathrm{ref},l}
 \end{eqnarray}
 where 
 \begin{align}
  q_r(T) &\equiv Q(T)/Q(T_\mathrm{ref})
 \end{align}
is the relative partition function to the partition function $Q(T_\mathrm{ref})$ at temperature $T$, $T_\mathrm{ref}$ is  the reference temperature, and $S_{\mathrm{ref},l} = S_l (T_\mathrm{ref}; \Elow_l, \nulcl)$ is the line strength at $T=T_\mathrm{ref}$. For HITRAN/HITEMP, the original format selects $T_\mathrm{ref}$ = 296K; however, as will be discussed later, we redefine the optimal $T_\mathrm{ref}$ to match the temperature range under consideration and recalculate $q(T)$ and $S_{\mathrm{ref},l}$ accordingly. In equations (\ref{eq:linestT}), we also defined temperature-dependent functions by
 \begin{eqnarray}
 \label{eq:fE}
f(\Elow,T) &\equiv& e^{ - c_2 \Elow  (T^{-1} - T_\mathrm{ref}^{-1}) } \\ 
g(\nulc,T) &\equiv& \frac{1- e^{- c_2 \nulc/ T}}{1-e^{- c_2 \nulc/T_\mathrm{ref}}},
\end{eqnarray}
where $c_2 = h c/k_B = 1.4387773 \, \mathrm{cm \, K}$, $k_B$ is the Boltzmann constant, and $c$ is the speed of light.
The total cross-section over the lines is expressed as
\begin{eqnarray}
\label{eq:dit}
\sigma (\nu) &=& \sum_l S_l (T; \Elow_l, \nulcl) V(\nu - \nulcl, \bpar_l) \nonumber\\
&=&\int d \bpar  \int d \nulc \int d \Elow \sum_l \delta_D(\nulc - \nulcl, \bpar - \bpar_l) \nonumber \\ &\times&S_l  (T; \Elow_l, \nulcl) V(\nu - \nulcl, \bpar_l),
\end{eqnarray}
where $\bpar_l$ is the broadening parameter sets, including the thermal Doppler broadening 
\begin{eqnarray}
\label{eq:thermal}
\beta_l = \sqrt{\frac{k_B T}{M m_u}} \frac{\nulcl}{c},
\end{eqnarray}
and van der Waals broadening $\gamma_l$. In Equation (\ref{eq:thermal}), $M$ is the molecular/atomic weight and $m_u$ indicates the atomic mass constant.

In the framework of DIT/MODIT, the distribution of the line strength is approximated by the density of the line strength called the lineshape density \citep{van2021discrete}.
\begin{eqnarray}
\sum_l S_{l} (T; \Elow_l, \nulcl)  \delta_D(\nulc - \nulcl, \bpar - \bpar_l)  \to \mathfrak{S}(\nulc,\bpar,T)
\end{eqnarray}
As clear in the above definition, the \LSD is a function of the temperature. Therefore, we need to recalculate the \LSD every time we change a T-P profile, as shown in the left panel of Figure \ref{fig:diffdit}. This fact limits the computationablity with large numbers of molecular lines, as the computation time is proportional to the number of lines at $N_\mathrm{line} \gtrsim 10^5$ (Figure 4 in Paper I). Another drawback of this specification is that all information on each molecular line must be kept in a GPU device memory. For instance, assuming that the number of the atmospheric layer is 100 for $N_\mathrm{line} = 10^8$, corresponding to $N_\mathrm{line}$ of methane in $\Delta \lambda = 100$ \AA \, around 1.64$\mu$m of ExoMol/YT34to10, we need 4 (or 8) $\times 100 \times 10^8$ = 40 (or 80) GB of the line strength information for single (or double) precision floating number. In MODIT, we need to store them in a GPU device memory, corresponding to the current  CUDA memory size (40/80 GB for NVIDIA RTX A100).

\begin{figure*}
    \centering
    \includegraphics[width=\linewidth]{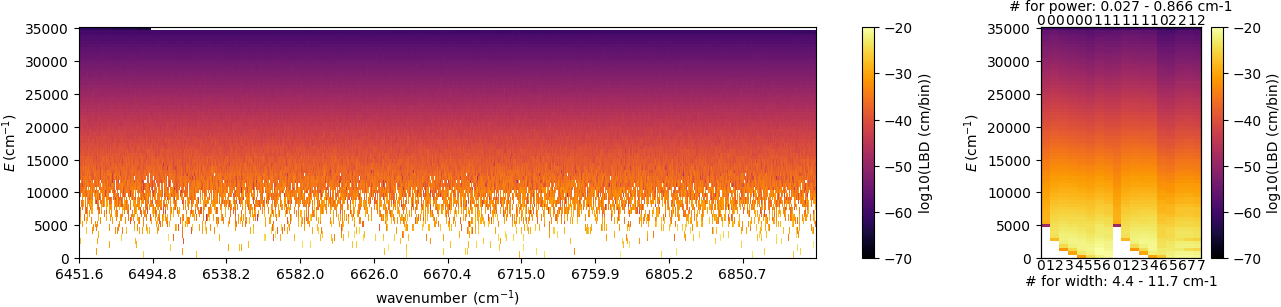}
    \caption{ { An example of the Line Basis Density (LBD; ${\mathfrak{L}} (\nulc,\Elow,\bpar)$) for water using ExoMol/POKAZATEL.  Left panel: The LBD integrated along the broadening parameter ($\bpar$) axis. The horizontal axis represents the wavenumber $\nulc$, and the vertical axis represents $\Elow$. White regions indicate the absence of spectral lines.  Right panel: The LBD integrated along the $\Elow$ axis. The horizontal axis represents the index in the broadening parameter direction. Since the broadening direction has two axes, i.e., reference width and power law index, these indices are shown at the bottom (width) and top (power law index) of the panel. }}
    \label{fig:premodit_lbd_h2o}
\end{figure*}

\begin{figure}
    \centering
    \includegraphics[width=\linewidth]{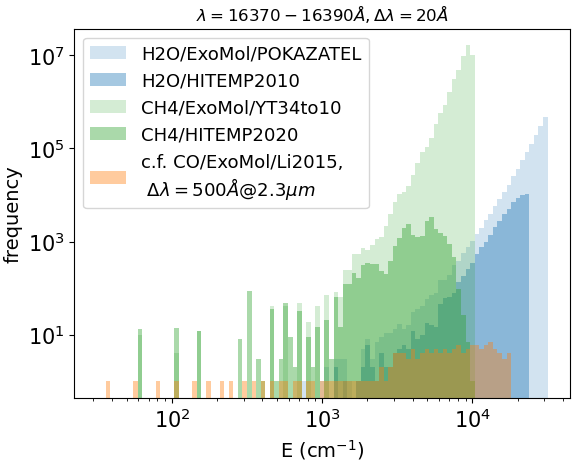}
    \caption{The distribution of the lower state energy ($\Elow$) of methane (green) and water (blue) in 16370--16390 \AA ($\Delta \lambda$ = 20 \AA) taken from HITEMP and ExoMol. We also show the $\Elow$ distribution of carbon monoxide (Li2015) around 2.3 micron ($\Delta \lambda$ = 500 \AA).}
    \label{fig:elower_distribution}
\end{figure}

In our modification of the MODIT algorithm, we precompute the \LSD for the reference temperature as
\begin{eqnarray}
\label{eq:premodit_ap}
\sum_l S_{\mathrm{ref},l} \delta_D(\nulc - \nulcl, \Elow - \Elow_l, \bpar - \bpar_l)  \to {\mathfrak{L}}(\nulc,\Elow,\bpar).
\end{eqnarray}
We apply the temperature correction to ${\mathfrak{L}}(\nulc,\Elow,\bpar)$, instead of recomputing the \LSD, every time we change the temperature. We do not need to transfer the line datasets such as $S_{\mathrm{ref},l}$, $\Elowl$ into the device memory. We call ${\mathfrak{L}}(\nulc,\Elow,\bpar)$ the line basis density (LBD) because it generates \LSD.

Substituting Equation (\ref{eq:linestT}) to Equation (\ref{eq:dit}), the cross-section can be expressed as  
\begin{align}
&\,\sigma (\nu)
= \int d \bpar \int d \nulc  \, \sum_l V(\nu - \nulcl, \bpar_l)  \frac{ g(\nulcl, T)}{q_r(T)}   \nonumber \\ 
&\times \int d \Elow f(\Elowl, T) S_{\mathrm{ref},l} \delta_D(\nulc - \nulcl,  \Elow - \Elowl, \bpar - \bpar_l) \\
&\approx \int d \bpar  \int d \nulc \, \mathfrak{S} (\nulc,\bpar, T)  V(\nu - \nulc, \bpar) \\
\label{eq:premodit_FT}
&= \int d \bpar \, \mathrm{FT}_\nu^{-1} \left[ \mathrm{FT}_\nu (\mathfrak{S})  \mathrm{FT}_\nu (V) \right]
\end{align}
where the \LSD is expressed using the LBD as  
\begin{eqnarray}
\label{eq:sslog}
\mathfrak{S} (\nulc,\bpar, T) &=& \frac{g(\nulc,T)}{q_r(T)}  \left[ \int d \Elow f(\Elow,T) {\mathfrak{L}} (\nulc,\Elow,\bpar) \right].
\end{eqnarray}


 MODIT uses the logarithm of the wavenumber $q= R_0 \log{\nu}$ as the wavenumber coordinate, where $R_0$ is the spectral resolution of the wavenumber grid. Then, the discretization of $q$, that is, an evenly spaced grid on the logarithm scale of the wavenumber (ESLOG)\footnote{Another advantage is that ESLOG has the same velocity width for the wavenumber bin, which can be computed by $\Delta v = c \log{(R^{-1} + 1)}$, where $R$ is the spectral resolution of the ESLOG bin $R = \nu/\Delta \nu$. This makes the spin rotation broadening to be a simple convolution. }, enables us to use the common Doppler width for a given temperature and isotope because the Voigt profile is expressed as  
\begin{eqnarray}
 V(\nu,\gamma_l,\beta_l(T)) d \nu = V(q,\ngamma_l,\nbeta(T)) d q,
\end{eqnarray}
where 
\begin{eqnarray}
\aD &\equiv& \frac{R_0 \beta}{\nulcl} = \sqrt{\frac{k_B T}{M m_u}} \frac{R_0}{c} \\
\ngamma_l &\equiv& \frac{R_0 \gamma_l}{\nulcl},
\end{eqnarray}
are the normalized gamma parameters and $\aD$ is common for all of the lines in an isotope. Therefore, $V(q,\ngamma,\nbeta)$ depends on the line properties only through $\ngamma$ (Paper I). Then, the broadening parameters $\bpar$ are for the Lorentzian. 

For ExoMol, the normalized gamma parameter of the $l$-th line for the pressure broadening 
\begin{eqnarray}
\label{eq:broadening_exomol}
\ngamma_{p,l} (\bpar; T, P) = R_0 \acute{\alpha}_{\mathrm{ref},l} \left( \frac{T}{T_\mathrm{ref, HIT}} \right)^{-\ntexpl} \left( \frac{P}{1 \,\mathrm{bar}} \right),
\end{eqnarray}
can be derived using the broadening parameters, $\ntexp$ and $\acute{\alpha}_\mathrm{ref}$, where $\acute{\alpha}_{\mathrm{ref},l} \equiv \alpha_\mathrm{ref,l}/\nu_l$ and  $\ntexpl$ is the temperature exponent, $\alpha_\mathrm{ref}$ is the HWHM of the pressure broadening at the reference temperature ($T_\mathrm{ref, HIT} = $296 K) and pressure (1 bar). For HITRAN/HITEMP, $\ngamma_{p,l} (\bpar; T, P)$ is calculated in the same manner as for ExoMol. However, note that by default, the values of $\alpha_\mathrm{ref}$ and $\ntexpl$ used are those for Earth air (i.e. $\mathrm{N}_2$ and $\mathrm{O}_2$ atmosphere). Note that the different background atmosphere from Earth air may cause systematics in the pressure broadening \citep{2019ApJ...872...27G}. For some molecules in HITRAN, broadening coefficients for hydrogen or helium background atmosphere are provided and can also be used.

{
Figure \ref{fig:premodit_lbd_h2o} illustrates an example of the LBD for water within a specific wavenumber range. In this range, the ExoMol/POKAZATEL database contains approximately $1.4 \times 10^8$ spectral lines, resulting in a densely populated LBD. This is an overwhelmingly larger number compared to the typical number of metal lines, highlighting the advantage of density-based opacity calculation methods over line-by-line approaches for molecular lines. To calculate an LBD such as the one shown in Figure  \ref{fig:premodit_lbd_h2o}, the LBD must first be discretized.}

\subsection{Discretization of the Line Density}\label{ss:discretization}

Next, we consider implementing Equation (\ref{eq:sslog}).  To avoid misunderstandings, underscores are used for the grid variables. Specifically, $\gnu_j, \gbpar_k, \gElow_h$ represent the discretized grid values of $\nu, \bpar, \Elow$ respectively, when the LBD is discretized. The discretized version of Equation (\ref{eq:premodit_FT}) is expressed as
\begin{eqnarray}
\sigma (\nu_j) = \sum_k \, \mathrm{DFT}_\nu^{-1} \left[ \mathrm{DFT}_j (\mathfrak{S}(\gnu_j,\gbpar_k, T))  \mathrm{DFT}_j (V_k) \right] \nonumber \\
\end{eqnarray}
where $\mathrm{DFT}_j$ is the discrete Fourier transform along the $j$ axis (i.e. wavenumber). The Fourier transform of the Voigt profile {for the $k$-th broadening parameter, $\mathrm{DFT}_j (V_k)$} is described in Paper I. Also, we defined the discretized lineshape density as
\begin{align}
\label{eq:discretized_lsd}
\mathfrak{S}_{jk} &\equiv \mathfrak{S}(\gnu_j,\gbpar_k, T) = \frac{g(\gnu_j,T)}{q_r(T)} 
\sum_{h=1}^{h_\mathrm{max}} \, f(\gElow_h,T) {\mathfrak{L}}_{jkh}, 
\end{align}
where ${\mathfrak{L}}_{jkh} \equiv \mathfrak{L} (\gnu_j,\gbpar_k,\gElow_h)$ is 
the discretized LBD, which is precomputed before sending to the GPU device\footnote{Exactly speaking, we send the logarithms of the coefficients of ${\mathfrak{L}}_{jkh}$ to the device memory.}, and $h_\mathrm{max}$ is the index of the maximum lower energy
\footnote{PreMODIT uses the three-dimensional array of $\nu$, ${\bf b}$, and $\Elow$. Hence, if the $\Elow$ to be used can be made smaller, this will save the device memory. In practice, the maximum of $\Elow$ in the water databases, POKAZATEL and HITEMP are excessively large ($\sim 3 \times 10^4 \mathrm{cm}^{-1}$ in Fig. \ref{fig:elower_distribution}).  At a given temperature and pressure, the maximum $\Elow$ required can be determined by {\sf ExoJAX}2 from the condition that the error in line intensity is within 1\%.}
. 
Because we calculate the cross-section from ${\mathfrak{L}}_{jkh}$ in the device, instead of the individual line information, we do not need to store the individual line information in the device memory, as shown in the right panel in Figure \ref{fig:diffdit}. 

In the framework of DIT, the line strength of each line is distributed linearly across two adjacent density grids of wavenumber and logarithms of Gaussian and Lorentzian width \citep{van2021discrete},
$\Delta S_1^{l} = w_{1,l} S_l$ and 
$\Delta S_2^{l} = w_{2,l} S_l$, 
where $w_{1,l}$ and $w_{2,l}$ are the weights of the line $l$ to the neighbouring grid points 1 and 2. Therefore, those should obey $w_{1,l} + w_{2,l} = 1$. The linear weights are defined as 
\begin{eqnarray}
w_{1,l} &=& \frac{ x_2 - x_l }{x_2 - x_1} \\
w_{2,l} &=&  \frac{ x_l - x_1 }{x_2 - x_1}, 
\end{eqnarray}
where $x_1$ and $x_2$ are arbitrary quantities at the neighboring grid points 1 and 2, and $x_l$ represents the quantity for line $l$. In the wavenumber grid, $x_l$ corresponds to the line center $\hat{\nu}_l$, and $x_1$ and $x_2$ are the neighboring wavenumber grid points to the line center, for instance. In our case, we used this definition for the wavenumber ($x = \nu$) and broadening parameters ($x = $ the component of $\bpar$), consistent with the original DIT framework.

For the LBD, we also need to discretize $\Elow$ and store the line information into the LBD along the $\Elow$ grid\footnote{ One might consider discretizing $T$ in equation (\ref{eq:sslog}) directly instead of $\Elow$. However, this method requires more memory usage (See Appendix \ref{ap:tgrid}). It should be noted that using a table of cross-section $\sigma(\nu)$ discretized directly with $T$ requires a further subdivision of the temperature grid to achieve the necessary accuracy.}. To make the density along the $\Elow$ grid is not straightforward because the contribution to the line strength is not linear of $\Elow$, but has the form as Equation (\ref{eq:fE}). 

We define the temperature-dependent weights of the line $l$ to the grid points 1 and 2 given by
\begin{eqnarray}
\label{eq:wg1}
w_{1,l}(T) &=& \frac{ x_2(T) - x_l(T) }{x_2(T) - x_1(T)}  \\
\label{eq:wg2}
w_{2,l}(T) &=& \frac{ x_l(T) - x_1(T) }{x_2(T) - x_1(T)} 
\end{eqnarray}
where $x_l(T) \equiv f(\Elow_l,T) , x_1(T) \equiv f(\gElow_{1,l},T),  x_2(T) \equiv f(\gElow_{2,l},T)$, where $\gElow_{1,l}$ and $\gElow_{2,l}$ are $\Elow$ at the neighbouring different grid points of $E_l$, and $\gElow_{1,l} \le E_l \le \gElow_{2,l}$.

Consider the reconstruction of the \LSD at temperature $T$ from the LBD. For simplicity, we ignore the wavenumber and broadening parameter dependencies on the LBD ${\mathfrak{L}}_{\Ttyp} (\gElow_h)$. The LBD is constructed using the weights at temperature $T=\Ttyp$ as follows:
\begin{eqnarray}
\label{eq:lbd_s}
&\,&{\mathfrak{L}}_{\Ttyp} (\gElow_h) = \nonumber \\ 
&\,& \sum_l w_{1,l} (\Ttyp) S_l \delta(\gElow_h, \gElow_{1,l}) + w_{2,l} (\Ttyp) S_l \delta(\gElow_h, \gElow_{2,l}) \nonumber \\
\end{eqnarray}
where 
\begin{eqnarray}
\delta(x,y) =
\left\{
\begin{array}{ll}
1 & \, \mbox{for } x = y  \\
0 & \, \mbox{for } x \ne y
\end{array}
\right.
\end{eqnarray}

\subsubsection{Discretization Error of $\Elow$}

The error due to the discretization of the LBD in the $\Elow$-direction can be evaluated by assessing the integral part of Equation (\ref{eq:sslog}). This part densifies the following direct summation of the line strength,
\begin{align}
s(T) = \sum_l f(\Elowl,T) S_{l,0}.
\end{align}
Therefore, the error can be evaluated by comparing the corresponding term 
\begin{align}
\label{eq:linestrength_from_LBD}
  \tilde{s} (T) = \sum_{h=1}^{h_\mathrm{max}} f(\gElow_h, T) {\mathfrak{L}}_{\Ttyp} (\gElow_h),
\end{align}
 in Equation (\ref{eq:discretized_lsd}), which is the discretized form of the integral part, with $s(T)$.

The line strength summation evaluated using Equation (\ref{eq:linestrength_from_LBD}) at the temperature $T=\Ttyp$, which was used for weight calculation, and the reference temperature $T=\Tref$ for the input line strength, will match the direct sum of the individual line strengths $s(T)$, as shown in the following calculations. 
 
First, the line strength sum from the discretized LBD evaluated at $T = \Ttyp$ is expressed as
\begin{align}
\label{eq:directsum}
&\tilde{s} (\Ttyp) = \sum_h f(\gElow_h, \Ttyp) {\mathfrak{L}}_{\Ttyp} (\gElow_h) \\
&= \sum_l [ f(\gElow_{1,l}, \Ttyp) w_{1,l} (\Ttyp) + f(\gElow_{2,l}, \Ttyp) w_{2,l} (\Ttyp) ] S_l \nonumber \\
&= \sum_l f(E_l, \Ttyp) S_l = s(\Ttyp)
\end{align}
Second, the line strength evaluated at the reference temperature $T = \Tref$ can be transformed as follows.
\begin{eqnarray}
&\,& \tilde{s} (\Tref) = \nonumber \\
&\,& \sum_l [ f(E_{1,l}, \Tref) w_{1,l} (\Tref) + f(E_{2,l}, \Tref) w_{2,l} (\Tref) ] S_l \nonumber \\
&\,& = \sum_l S_l = s(\Tref),
\end{eqnarray}
where we used $f(E,\Tref) = 1$ and $w_{1,l} (T) + w_{2,l} (T) = 1$.

\begin{figure}
    \centering
    \includegraphics[width=\linewidth]{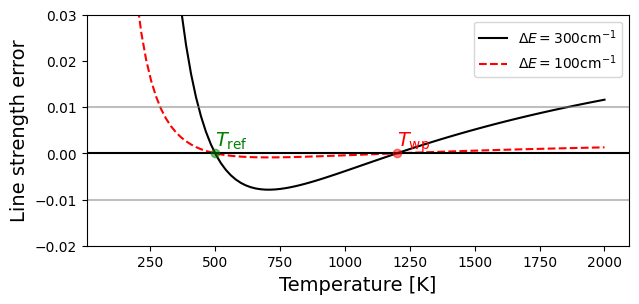}
    \caption{The error of the line strength summation derived from the LBD for different $\Delta E$. The solid and dashed lines correspond to $\Delta E=300$ and $100 \mathrm{cm}^{-1}$. We adopt $\Tref$ and $\Ttyp$ to 500 and 1200 K. The gray lines indicate $\pm 1$ \% error. In the case of $\Delta E=300 \mathrm{cm}^{-1}$, the robust temperature range that keeps the error within 1\% is 430 K - 1850 K. }
    \label{fig:fig_elower_grid_error}
\end{figure}

From the above discussion, a small error is expected for the temperature range between $\Ttyp$ and $\Tref$. The error in $s(T)$ from Equation (\ref{eq:linestrength_from_LBD}) for an arbitrary temperature $T$, $\Delta_S (T; \Delta E) = \tilde{s}(T)/s(T) -1$, depends on the interval  of the $\Elow$ grid, $\Delta E$, and can be analytically evaluated (Equation (\ref{eq:error_line_s}) in Appendix \ref{ap:err_line_intensity}). Figure \ref{fig:fig_elower_grid_error} shows an example of $\Delta_S (T; \Delta E)$ for $\Tref=500$K and $\Ttyp=1200$K. As expected, the smaller the $\Delta E$, the smaller the error, and the error is small between $\Tref$ and $\Ttyp$ compared to the outside of the range. This analytical evaluation formula has been verified by comparison with the line strengths directly calculated using actual molecular lines. 

Considering the case of $\Delta E=300 \mathrm{cm}^{-1}$ in Figure \ref{fig:fig_elower_grid_error}, when the maximum value of E is 10,000 $\mathrm{cm}^{-1}$, approximately 33 grids are needed for 430-1850K to maintain an error within 1 \%. On the other hand, for temperature-based grids, approximately 180 grids are required in the same temperature range (See Appendix \ref{ap:tgrid}).

To further reduce memory usage and computational cost, it is desirable to set the maximum $\Delta E$ such that the error remains within an acceptable range. To achieve this, first define the temperature range, and then determine $\Tref$, $\Ttyp$, and $\Delta E$ using the evaluation formula $\Delta_S(T; \Delta E)$ in Equation (\ref{eq:error_line_s}), ensuring that the line strength remains within the specified maximum error across this temperature range. While the error remains small between $\Tref$ and $\Ttyp$, it generally increases outside this range, so the optimal values of $\Tref$ and $\Ttyp$ should be close to the boundary of the temperature range.

\subsubsection{Discretization Error of Broadening Parameters}

To see the discretization error of the normalized gamma parameter, we rewrite the equation (\ref{eq:broadening_exomol}) as 
\begin{eqnarray}
\label{eq:broadening_exomol_re}
\ngamma_{p,l} (\bpar; T, P) &=& R_0 \acute{\alpha}_{\mathrm{ref, br},l} \left( \frac{T}{T_\mathrm{ref, br}} \right)^{-\ntexpl} \left( \frac{P}{1 \,\mathrm{bar}} \right), \\
\acute{\alpha}_{\mathrm{ref, br},l} &=& \acute{\alpha}_{\mathrm{ref},l} \left( \frac{T_\mathrm{ref, br}}{T_\mathrm{ref, HIT}} \right)^{-\ntexpl}
\end{eqnarray}
Then the error propagates as
\begin{eqnarray}
\Delta \ngamma_{p,l} = \Delta (\log \acute{\alpha}_{\mathrm{ref, br}}) - (\log{T} - \log{T_\mathrm{ref, br}}) \Delta \ntexp.
\end{eqnarray}
In this expression, the error is proportional to the factor of $(\log{T} - \log{T_\mathrm{ref, br}})$. Therefore, we take the midpoint of logarithms of the maximum and minimum temperatures as $\log{T_\mathrm{ref, br}} = (\log{T_\mathrm{max}} + \log{T_\mathrm{min}})/2$. This decreases the error from $\Delta \ntexp$.

In the original DIT \citep{van2021discrete}, the grid interval of the logarithm of the line width is set to $C_\gamma=0.2$ as a fiducial value. In our case, $C_\gamma$ depends on both the intervals of $\ntexp$ and $\acute{\alpha}_{\mathrm{ref, br}}$. Therefore, we set up those intervals so that the mean interval of the width would be $\overline{\Delta \ngamma_{p,l}} = C_\gamma$.  We found that $(\log{T_\mathrm{max}} - \log{T_\mathrm{ref, br}})\Delta \ntexp = 3 C_\gamma/4$ and $\Delta \acute{\alpha}_{\mathrm{ref, br}} =  3 C_\gamma/4$  satisfies this condition, assuming that these parameters are uniformly distributed around the grids. In {\sf ExoJAX}2, we adopt 0.2 to $C_\gamma$ as a default value same as {\sf radis}. However, this default value might be oversampling for the analysis of real high-dispersion data. For instance, we found that the spectrum assuming $C_\gamma = 1.0$ agrees with that with $C_\gamma=0.1$ within a 1 \% accuracy in the case of GL229B \citep{2024arXiv241011561K}. Because the number of the broadening parameter grids is proportional to the device memory use, we recommend the users change $C_\gamma$ to the optimal value within their required accuracy.

\section{Atmospheric Radiative Transfer}

In {\sf ExoJAX}1, only the emission spectrum with pure absorption was supported. In {\sf ExoJAX}2, various types of radiative transfer were implemented with automatic differentiability. Table \ref{tab:rt} summarizes the radiative transfer available in {\sf ExoJAX}2.

\begin{table}[]
  \caption{Radiative transfer available in {\sf ExoJAX}2}
  \label{tab:rt}
  \begin{center}
  \begin{tabular}{llll} 
  \hline\hline
type & {\sf ExoJAX}1 & {\sf ExoJAX}2 & Section\\
& & \multicolumn{2}{c}{(this paper)} \\
\hline
\multicolumn{4}{c}{{\bf Emission with pure absorption}} \\
 {\it flux-based} & \gcheckmark  & \gcheckmark  & Paper I\\
 {\it intensity-based} &  \notava  & \gcheckmark & \S \ref{ss:pure_intensity}\\
 \hline
 \multicolumn{4}{c}{{\bf Emission including scattering}  } \\
 {\it flux-adding} &  \notava  & \gcheckmark & \S \ref{ss:emission_scattering}\\
 {\it effective-transmission} &  \notava  & \gcheckmark & \S \ref{ap:ett_lart}\\
 \hline
 \multicolumn{4}{c}{\bf Reflection} \\
  {\it flux-adding}  &  \notava  & \gcheckmark & \S \ref{ss:emission_scattering}\\
\hline
 \multicolumn{4}{c}{\bf Transmission with pure absorption} \\
 &  \notava  & \gcheckmark & \S \ref{ss:transmission} \\
\hline
  \end{tabular}
  \end{center}

\end{table}

{In this section, we present the re-derivation of the $\mathsf{N}$-stream radiative transfer scheme with pure absorption, two-stream radiative transfer formalisms, and transmission spectrum calculations. Whereas Section 3 introduced newly proposed methods such as pre-MODIT and LBD -- each of which necessarily involves certain approximations to manage computational efficiency -- here we follow the original theoretical foundations without imposing additional simplifying assumptions. In other words, the methodology in this section can be considered `exact' in the same sense as the original works, since the mathematical formulation remains consistent with their derivations. By clearly distinguishing these re-derivations from the approximate methods, we aim to avoid any misconceptions that compromises are being made where none exist.}

\subsection{Setup of atmospheric layers and preconditioning for spectral calculations}

\begin{figure*}
    \centering
    \includegraphics[width=0.45\linewidth]{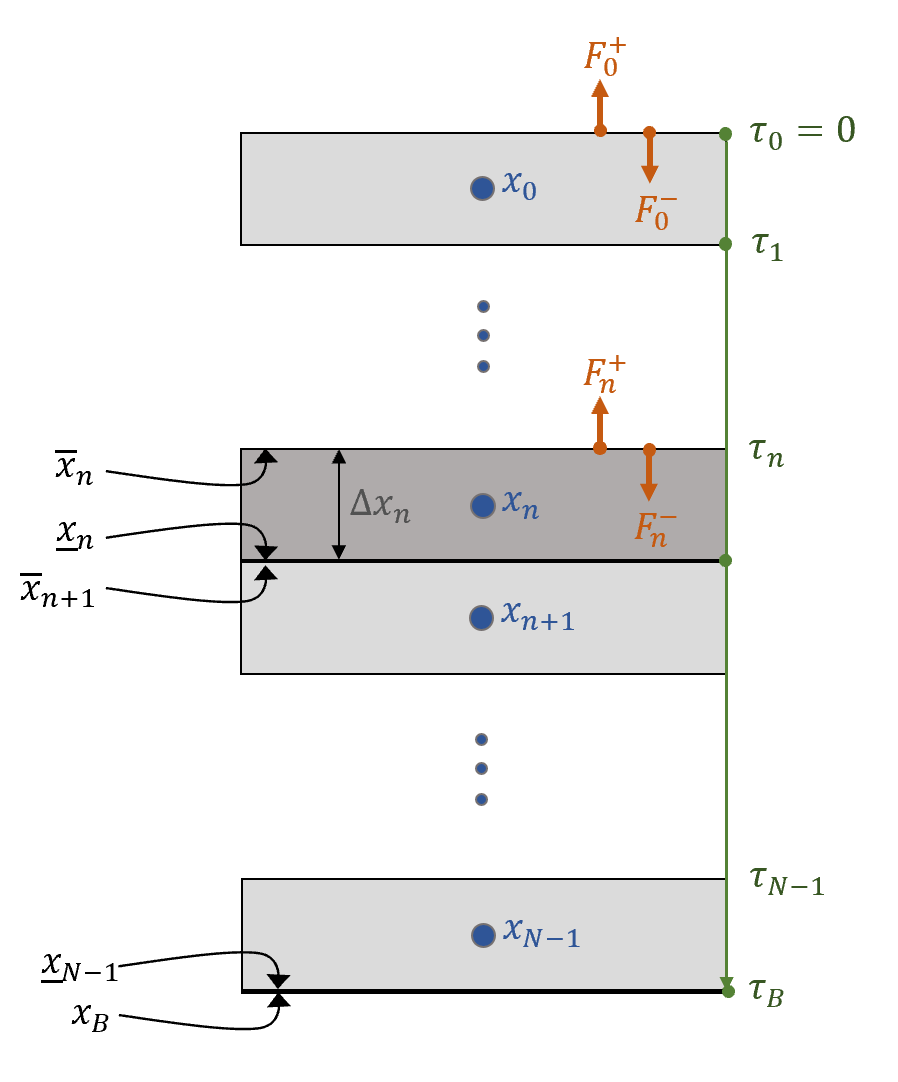}
     \includegraphics[width=0.54\linewidth]{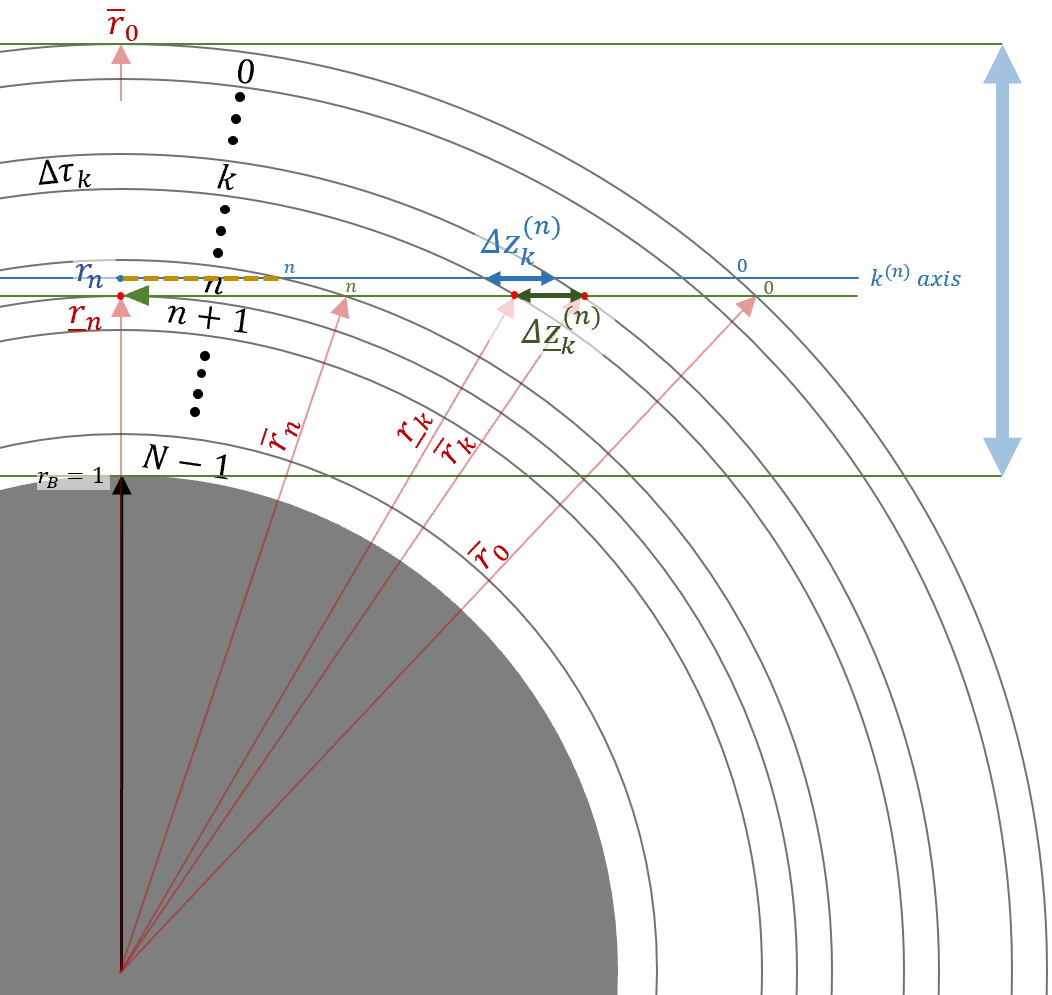}
    \caption{{\sf ExoJAX}2 layer coordinate system. Left: the notation rule in the layer and its boundaries. Right: coordinate system for transmission spectroscopy. The contribution of the atmospheric layers to the transit radius, $\Delta R_{p,\nu}$ is evaluated along the green vertical arrow.}
    \label{fig:transmission_coord}
\end{figure*}

We assume an atmospheric layer model with logarithmic pressure evenly spaced along the vertical axis. Figure \ref{fig:transmission_coord} (left) illustrates an example of the symbols used in the layer model of {\sf ExoJAX}2. The representative value for the physical quantity $x$ in the $n$-th layer is defined as $x_n$. If there is a gradient within the layer, $x_n$ is taken as the central value in the logarithmic pressure coordinate. On the other hand, the physical quantities at the upper and lower boundaries of each layer are defined as $\overline{x}_n$ and $\underline{x}_n$, respectively. For instance, in the isothermal layer model, $T_n = \overline{T}_n = \underline{T}_n$. However, when allowing for a gradient within the layer and imposing smooth connection conditions, we take $\overline{T}_{n+1} = \underline{T}_n$. When imposing smooth connection conditions and also representing the source function in the form $B_\nu(\tau_\nu) = a \tau_\nu + b$ ($a, b \in \mathbb{R}$ ) within a layer, it is referred to as {the linsap (linear-source approximation) layer model} in this paper. The value of the lower boundary of the $(N-1)$-th layer is denoted by $x_B$. If the planet has a surface, $x_B$ is interpreted as the surface value of $x$. 

For pressure, our default setting assumes that the layers are evenly divided along the logarithmic pressure. Then, we obtain 
\begin{align}
\label{eq:pressure_overline}
    \overline{P}_{n} &= 10^{\log_{10} P_n - r \Delta q} = k^r P_{n}\\
    \label{eq:pressure_underline}
    \underline{P}_{n} &= 10^{\log_{10} P_n + (1-r) \Delta q} = k^{r-1} P_{n} \\
    \label{eq:pressure_delta}
    \Delta P_n &= (k^{r-1} - k^{r}) P_n,
\end{align}
where $\Delta q = \Delta \log_{10} (P/\mathrm{bar})$ is a constant interval of the log pressure, $k \equiv 10^{- \Delta q}$ is the (constant) pressure decrease rate (i.e. $P_{n} = k P_{n+1}$), and $r$ is the reference point of the layer representative pressure (the midpoint $r=1/2$ as a default value). See Appendix \ref{ap:s:logp} for a more detailed description.

There are exceptions to the above notation rule. For the flux, intensity, and optical depth, to avoid complexity, the value at the upper boundary of each layer is denoted without the overline, as $F^+_n$, $I^+_n$, and $\tau_n$.  In practice, we compute the optical depth of the $n$-th layer $\Delta \tau_n = \tau_{n+1} - \tau_{n}$ from $n=0$ to $n=N-1$ and then we define the optical depth as $\tau_n \equiv \sum_{i=0}^{n-1} \Delta \tau_i$ for $1 \le n \le N$, $\tau_0 \equiv 0$, and $\tau_B = \tau_N$.

For transmission spectroscopy, we compute the vertical height of the layer, $\Delta h_n$, as described in Appendix \ref{ap:layer_height}. Because this process is iterative, we use {\sf jax.lax.scan} to make the function differentiable. 

Before computing the spectra, {\sf ExoJAX}2 calculates the optical depth profile along the vertical direction, $\Delta \tau_n$ for $n=0,1, \cdots, N-1$.
The optical depth for molecules/atoms is expressed as
\begin{eqnarray}
    \Delta \tau_n = \frac{ \eta_n \sigma_n  }{m g_n} \Delta P_n,
\end{eqnarray}
where $\sigma_n, \Delta P_n, \eta_n,$ and $g_n$ are the cross-section, pressure difference, mass mixing ratio, and gravity of the $n$-th layer, and $m$ is the molecular mass. 

\subsection{Differentiable emission model with pure absorption} \label{ss:pure_intensity}
Emission spectrum calculations in the case of pure absorption can be broadly classified into two methods: one that transfers intensity ($I$) between layers and computes the outgoing flux $F$ by integrating in the angular direction at the top of atmosphere (ToA) -- referred to as the ``intensity-based'' method \citep[e.g.][]{2008JQSRT.109.1136I,2019A&A...627A..67M, 2020ApJ...890..174K} and another that transfers the flux itself, referred as the ``flux-based'' method \citep[e.g.][]{2017AJ....154...91L}. While {\sf ExoJAX}1 adopted the latter\footnote{The flux-based method is characterized by its continuity with radiative spectrum calculations that incorporate scattering (\S \ref{ss:emission_scattering}).}, {\sf ExoJAX}2 has made calculations using the former method possible as well. The intensity-based approach facilitates calculations not just for two-stream but also easily handles four-stream or higher. The flux-based calculation method is described in Paper I, so here we will describe the intensity-based radiative transfer, which is the default in {\sf ExoJAX}2.

\subsubsection{Isothermal layer} \label{sss:iso}
The intensity-based transfer is obtained by discretizing the formal integral solution of the Schwarzschild equation,
\begin{align}
    \mu \frac{d I_\nu}{d \tau} = I_\nu - \mathcal{J}_\nu.    
\end{align}

In the pure absorption case, $ \mathcal{J}_\nu $ is replaced by the Planck function $ B(\tau) $, and by setting $ \tau^\prime = \tau/\mu $ and integrating with respect to $ \tau^\prime $ from 0 to $ \tau_B^\prime = \tau_B/\mu$ ($\tau_B$ is the optical depth at the bottom of the layer model), the following formal solution is obtained as
\begin{align}
\label{eq:intensity_transfer}
         I_0 (\mu) &= I (\tau_B^\prime, \mu) e^{-\tau_B^\prime} +\int^{\tau_B^\prime}_{0} B (\tau) e^{-\tau^\prime} d \tau^\prime \\
    \label{eq:I_discrete_nemesis}
    &\approx B(T_B) e^{- \tau_{B}/\mu} 
    + \sum_{n=0}^{N-1} B(T_n) (e^{-\tau_{n}/\mu} - e^{-\tau_{n+1}/\mu}). \nonumber \\
    &\,\mbox{(intensity-based, isothermal layer)}
\end{align}
We note that the last approximation assumes the isothermal distribution in a layer. 

For the $\mathsf{N}$-stream, the outgoing flux is obtained as
\begin{align}
    F_\mathrm{out} &= 2 \pi \int_0^1 \mu I_0(\mu)  d \mu \\
    \label{eq:f0_gq}
    &\approx 2 \pi \sum_{i=1}^{\mathsf{N}^\prime} w_i \mu_i I_0 (\mu_i), 
\end{align}
where $\mathsf{N}^\prime = \mathsf{N}/2$. In evaluating the final integral of the above equation for $ \mathsf{N} \geq 4 $, it is assessed using the Gauss-Legendre quadrature. That is, $ \mu_i $ and $ w_i $ are the nodes and weights of the Gauss-Legendre quadrature, respectively. For $ \mathsf{N} = 2 $, the node of the Gauss-Legendre quadrature becomes $ x_1 = 0.5 $, but we adopt $ \mu_1 = 2/3 $ and $ w_1 = 3/4 $, which is closer to the analytical solution.

\subsubsection{Linear Source Approximation} \label{sss:linsap}
 The non-isothermal layer model requires to sum up each intensity contribution of the layers: 
\begin{align}
\label{eq:intensity_transfer_OK}
         I^+_n (\mu) &= \hat{T}_n I^+_{n+1} (\mu) + \Delta I^+_n 
\end{align}
where
\begin{align}
          \hat{T}_n &\equiv e^{- \Delta \tau_n ^\prime} 
\end{align}
is the transmission of the $n$-th layer. In the case of the linear source approximation (linsap), the source function profile within the $n$-th layer can be written as
\begin{align}
    B_n (\tau) = \frac{\underline{B}_n-\overline{B}_n}{\Delta \tau_n} (\tau - \tau_n) + \overline{B}_n.
\end{align}
Then, we obtain
\begin{align}
\label{eq:deltaIplus}
    \Delta I^+_n &= \int_{\tau^\prime_n}^{\tau^\prime_n+\Delta \tau^\prime_n} B_n (\tau) e^{-\tau^\prime} d \tau^\prime \\
      &= \beta_n \overline{B}_n + \gamma_n \underline{B}_n = \beta_n \overline{B}_n + \gamma_n \overline{B}_{n+1} \\
          \label{eq:recursiveOK}
      &= \beta_n B(\overline{T}_n) + \gamma_n B(\overline{T}_{n+1}) 
\end{align}
where
\begin{align}
    \beta_n &\equiv 1  - \frac{1 - \hat{T}_n  }{\Delta \tau^\prime_n} \\
    \gamma_n &\equiv - \hat{T}_n  + \frac{1 - \hat{T}_n  }{\Delta \tau^\prime_n}  \\
     &\,\mbox{(intensity-based, linsap)} \nonumber
\end{align}
The boundary condition for the thermal surface is given by 
\begin{align}
\label{eq:intensity_transfer_boundary}
         I^+_{N-1} (\mu) &= \hat{T}_{N-1} B(T_B) + \Delta I^+_{N-1}.
\end{align}
The above formulation is the linear version of the short characteristic method \citep{1987JQSRT..38..325O}, and used in {\sf Helios-r2} \citep{2020ApJ...890..174K}. As can be seen from these expressions, the intensity-based transfer with the linsap requires the specification of both $\overline{T}_n$ for Equation (\ref{eq:recursiveOK}) and $T_n$ for opacity computation. 

Defining $\Delta I^+_N$ by the upward surface intensity $B(T_B)$\footnote{In practice, we can use Equation (\ref{eq:recursiveOK}) with $T_N = T_B$, $\beta_N=1$ and $\gamma_N=0$.}, $I_0^+(\mu)$ can be derived from equations (\ref{eq:intensity_transfer_OK}, \ref{eq:recursiveOK},
\ref{eq:intensity_transfer_boundary}). 
As described in Paper I (Section 3), defining cumulative transmission as
$\tcv = (1, \hat{T}_0,  \hat{T}_0 \hat{T}_1,\cdots,\hat{T}_0 \hat{T}_1 \cdots \hat{T}_{N-2}, \hat{T}_0 \hat{T}_1 \cdots \hat{T}_{N-1})^\top$ and the source vector $\qv \equiv (\Delta I^+_0, \Delta I^+_1, \cdots , \Delta I^+_N)^\top$,  we can compute $I_0^+ (\mu)$ using the cumulative product operator ({\sf jax.numpy.cumprod}) and the inner product
\begin{eqnarray}
\label{eq:rerr_tq}
I_0^+ (\mu) = \tcv \cdot \qv = \sum_{i=0}^{N} w_i, 
\end{eqnarray}
where $w_i = \phi_i q_i$ can be interpreted as the contribution function from the $i$-th layer for $i \le N-1$ and the surface ($i=N$) to the planet spectrum for the stream with $\mu$. The final outgoing flux can be computed by the Gaussian quadrature in Equation (\ref{eq:f0_gq}).

\subsubsection{Comparison}
\begin{figure*}
    \centering
    \includegraphics[width=\linewidth]{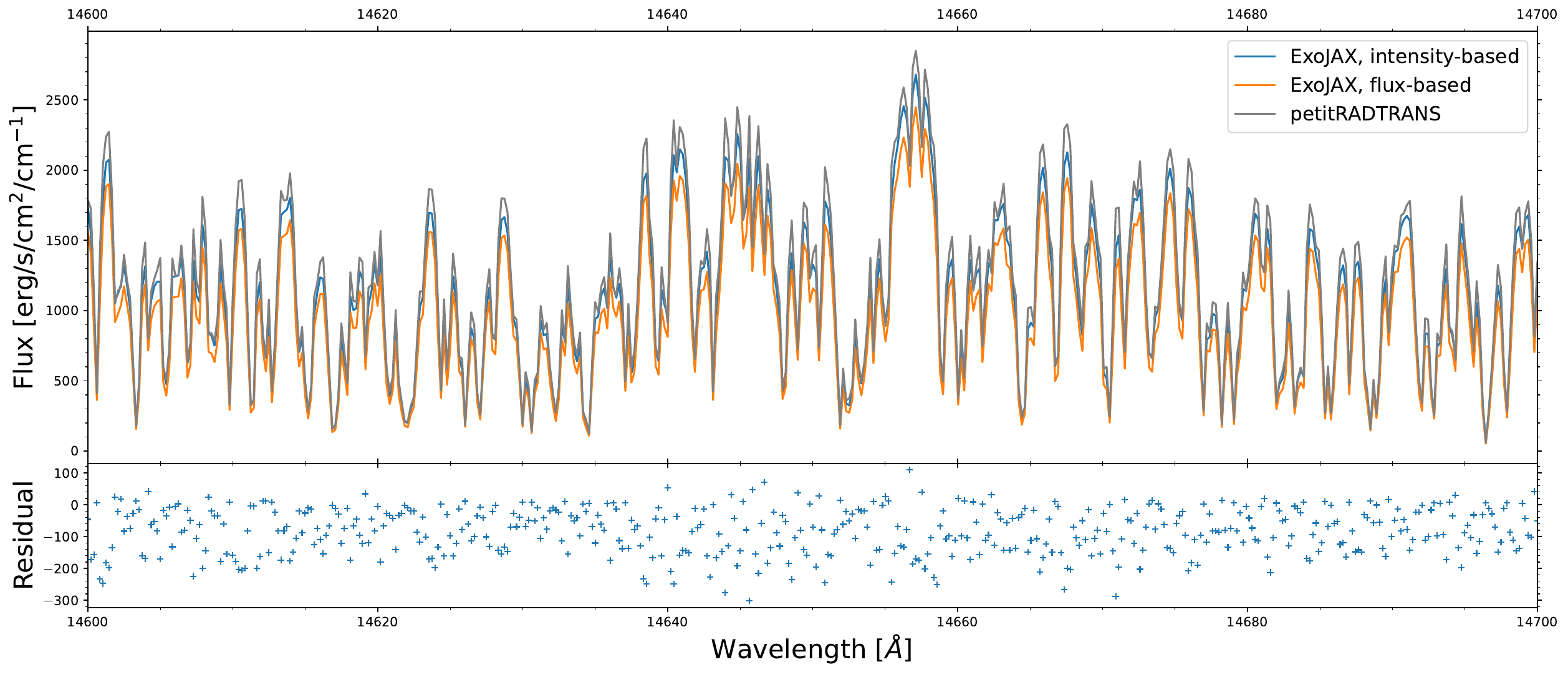}
    \caption{Comparison of the spectrum models calculated with the intensity-based ($\mathsf{N}=6$ stream) and flux-based methods of {\sf ExoJAX}2 and that with {\sf petitRADTRANS}.
    The bottom panel shows the residual between ``{\sf ExoJAX, intensity-based}" and ``{\sf petitRADTRANS}" models.}
    \label{fig:RT_comp}
\end{figure*}

Figure~\ref{fig:RT_comp} shows the comparison of the spectrum models calculated with the intensity-based ($\mathsf{N}=6$ stream) and flux-based methods of {\sf ExoJAX}2 and that with {\sf petitRADTRANS} \citep{2019A&A...627A..67M} for reference.

In this calculation, we consider the line absorption of $\mathrm{H_2O}$ using POKAZATEL line list \citep{2018MNRAS.480.2597P} from ExoMol \citep{2016JMoSp.327...73T} in addition to the Collision-Induced Absorption (CIA) of $\mathrm{H_2}$--$\mathrm{H_2}$ and $\mathrm{H_2}$--$\mathrm{He}$, taking those data from HITRAN \citep{2019Icar..328..160K}.
We adopted the parameters related to the thermal profile as $T_0 = 1000$~K, $\alpha = 0.1$ (see Eq.~(63) of Paper I for the definition of these two parameters).
We set the gravity as $\log{g} = 5.00$ in the cgs unit, and the volume mixing ratio of $\mathrm{H_2O}$ throughout the atmosphere as $x_\mathrm{H_20} = 10^{-3}$, assuming that the remaining mass is composed of $\mathrm{H_2}$ and $\mathrm{He}$ with their relative volume mixing ratio of 6:1.
Instrumental resolution is set as $R \sim 90,000$.

The comparison between the two codes shows general agreement, with the spectra differing by less than 10\%. The intensity-based method in {\sf ExoJAX}2 produces spectra that closely match those from {\sf petitRADTRANS}, which also uses an intensity-based approach, indicating consistent results between the two codes.

\subsection{Reflection and Scattering using a Two-stream Approximation}\label{ss:emission_scattering}


In {\sf ExoJAX}2, not only emission with pure absorption is provided, but also radiative transfer that includes scattering and reflection. In this subsection, we start from the general equation of the two-stream approximation that includes scattering by \cite{1989JGR....9416287T}, and derive the recursion formula equivalent to the flux-adding treatment by \cite{2018JQSRT.211...78R}. The flux-adding treatment represents the equation of the two-stream approximation that includes scattering as the {\it effective} reflection and source sum for each layer. 

The characteristics of the scattering are described by the two quantities: One is the single scattering albedo given by 
\begin{align}
    \omega = \frac{\Delta \tau_\mathrm{cld}^\mathrm{scat} + \Delta \tau_\mathrm{ray}}{\Delta \tau_\mathrm{cld} + \Delta \tau_\mathrm{ray} + \sum_x {\Delta \tau_x}},
\end{align}
where $\tau_\mathrm{cld}^\mathrm{scat}$ and $\Delta \tau_\mathrm{cld}$ are the scattering and extinction layer opacity of clouds,  $\Delta \tau_\mathrm{ray}$ is the Rayleigh scattering layer opacity, $\Delta \tau_x$ is the layer opacity of the other absorption, such as by molecular lines, the CIA, $\mathrm{H}^-$ continuum, and so on. The other is the asymmetric factor $g$.

\subsubsection{Formulation of the two-stream scattering}\label{sss:twoscat}

The solutions of upward and downward streams, for the Toon89-type two-stream approximation can be expressed as
\begin{eqnarray}
\label{eq:2stream_1}
F^+ (\tau) &=& c_1 \zeta^+ e^{\lambda \tau} + c_2 \zeta^- e^{-\lambda \tau} + \pi \mathcal{B}^+ (\tau)  \\
\label{eq:2stream_2}
F^- (\tau) &=& c_1 \zeta^- e^{\lambda \tau} + c_2 \zeta^+ e^{-\lambda \tau} + \pi \mathcal{B}^- (\tau)
\end{eqnarray}
where $\zeta^\pm$ are called the coupling coefficients \citep{heng2017exoplanetary}. Notably, the 2-stream approximation of the spherical harmonics (SH) method also takes the same form \citep[e.g. Equation (16) in ][]{2024ApJ...960..131R}. However, {\sf ExoJAX}2 does not yet support the SH method.

In the above equations, $\zeta^{\pm}$ and $\lambda$ are related to $\gamma_1$ and $\gamma_2$, the coefficients of the differential equation for $F^{\pm}$, as follows. 
\begin{align}
    \zeta^\pm &= \frac{1}{2} \left( 1 \pm \sqrt{\frac{\gamma_1 - \gamma_2}{\gamma_1 + \gamma_2}} \right)\\
    \lambda &= \sqrt{\gamma_1^2 - \gamma_2^2},
\end{align}
The coefficients $\gamma_1$ and $\gamma_2$ are determined by the single scattering albedo $\omega$ and the asymmetric parameter $g$. This relation depends on the method of moment closure \citep{1989JGR....9416287T}. When using the hemispheric mean, the relation 
$\gamma_1 = 2 - \omega (1 + g)$ and $\gamma_2 = \omega (1 - g) $
holds.  Also, the reduced source function is expressed as
\begin{align}
\mathcal{B}^\pm (\tau) = 
    \frac{ 2 (1-\omega)}{\gamma_1 - \gamma_2} \left( \Bl(\tau) \pm \frac{1}{\gamma_1 + \gamma_2} \dBl (\tau) \right).
\end{align}
In the above equation, the second term is neglected in the case of an isothermal layer.

Equations (\ref{eq:2stream_1}, \ref{eq:2stream_2}) can be expressed as
\begin{eqnarray}
    \Fv(\tau) = Q(\tau) \xv + \pi \Bv (\tau)
\end{eqnarray}
where $\Fv (\tau) = (F^+ (\tau), F^-(\tau))^\top $, $\Bv (\tau) = (\mathcal{B}^+ (\tau), \mathcal{B}^- (\tau))^\top$, $\xv \equiv (c_1, c_2)^\top$, and 
\begin{eqnarray}
    Q(\tau) = \left(
\begin{array}{cc}
\zeta^+ e^{\lambda \tau} & \zeta^- e^{-\lambda \tau}  \\
\zeta^- e^{\lambda \tau} & \zeta^+ e^{-\lambda \tau}  \\
\end{array}
\right)
\end{eqnarray}

Then, we consider the $N$-layers model with the optical width difference $\Delta \tau_n$ and define the optical depth \underline{at the top of the $n$-th layer} by $\tau = \tau_n = \sum_{i=0}^{n-1} \Delta \tau_i$ for $n \ge 1$ and $\tau_0 = 0$. 

Let us consider the $n$-th layer. 
\begin{eqnarray}
\label{eq:Fv1}
    \Fv (\tau_n) &=& Q_n(\tau_n) \xv_n + \pi \Bv_n (\tau_n) = \Fv_n\\
\label{eq:Fv2}
    \Fv (\tau_n+\Delta \tau_n) &=& Q_n(\tau_n + \Delta \tau_n) \xv_n + \pi \Bv_n (\tau_n + \Delta \tau_n) \nonumber \\
    &=& \Fv_{n+1}
\end{eqnarray}
where $Q_n$ is defined as $Q(\tau)$ of the $n$-th layer. Therefore, we set the internal parameters of the $n$-th layer $\zeta^\pm_n$ and $\lambda_n$,
\begin{eqnarray}
    Q_n(\tau) = \left(
\begin{array}{cc}
\zeta^+_n e^{\lambda_n \tau} & \zeta^-_n e^{-\lambda_n \tau}  \\
\zeta^-_n e^{\lambda_n \tau} & \zeta^+_n e^{-\lambda_n \tau}  \\
\end{array}
\right).
\end{eqnarray}

From Equations (\ref{eq:Fv1}) and (\ref{eq:Fv2}), we obtain the linear form of the recurrence relation,
\begin{eqnarray}
    \label{eq:rtstandard}
    \Fv_{n+1} &=& \mathcal{G} (\Delta \tau_n) \Fv_n + \pi \Gv_n,
\end{eqnarray}
where 
\begin{eqnarray}
   \Gv_n \equiv \Bv_n (\tau_n + \Delta \tau_n) - \mathcal{G} (\Delta \tau_n) \Bv_n (\tau_n)  
\end{eqnarray}
 and $\Gv_n = (G_n^+, G_n^-)^\top$.
We also defined the transfer function 
\begin{align}
\label{eq:transfer_matrix}
\mathcal{G} (\Delta \tau_n) &\equiv Q(\tau_n + \Delta \tau_n) Q^{-1}(\tau_n) \\
&=  Q_n(\tau_n)
 \left(
\begin{array}{cc}
e^{\lambda_n \Delta \tau_n}  & 0  \\
0 & e^{-\lambda_n \Delta \tau_n}  \\
\end{array}
\right) Q_n^{-1}(\tau_n) \nonumber \\
\end{align}
This is the eigendecomposition of $ \mathcal{G} (\Delta \tau_n) \qv_i = \lambda^\prime_i \qv_i$, where $\qv_i$ is the $i$-th column vector of $Q_n(\tau_n)$. Therefore we can normalize $\qv_i$, in particular, $\qv_1$ by $e^{\lambda_n \tau_n}$ and  $\qv_2$ by $e^{-\lambda_n \tau_n}$.  By defining
\begin{eqnarray}
    Z_n \equiv Q_n(0) = \left(
\begin{array}{cc}
\zeta^+_n  & \zeta^-_n  \\
\zeta^-_n & \zeta^+_n  \\
\end{array}
\right),
\end{eqnarray}
we rewrite the above equation 
as
\begin{eqnarray}
\label{eq:transfer_matrix2}
&\,&\mathcal{G} (\Delta \tau_n) = 
 Z_n
 \left(
\begin{array}{cc}
e^{\lambda_n \Delta \tau_n}  & 0  \\
0 & e^{-\lambda_n \Delta \tau_n}  \\
\end{array}
\right) Z_n^{-1} \\
\label{eq:gzero}
&=& \frac{1}{{\zeta^+_n}^2 - {\zeta^-_n}^2}\left(
\begin{array}{cc}
  {\zeta^+_n}^2 e^{t_n} - {\zeta^-_n}^2 e^{-t_n} & \,  - \zeta^+_n \zeta^-_n (e^{t_n} - e^{-t_n}) \\
\zeta^+_n \zeta^-_n (e^{t_n} - e^{-t_n})& \, {\zeta^+_n}^2 e^{-t_n} - {\zeta^-_n}^2 e^{t_n} \\
\end{array}
\right), \nonumber \\
\end{eqnarray}
where
\begin{eqnarray}
t_n \equiv \lambda_n \Delta \tau_n
\end{eqnarray}
is a function of $\Delta \tau_n$, but not of $\tau_n$.

For the isothermal layer, $\Bv_n (\tau_n) = \Bv_n (\tau_n + \Delta \tau) \equiv \mathcal{B}_n \uv$, $\mathcal{B}_n = \mathcal{B}^+ (\tau_n) = \mathcal{B}^- (\tau_n)$, where $\uv$ is one vector $\uv \equiv (1,1)^\top$. The source matrix $\Gv_n$ can be reduced as  
\begin{align}
\Gv_n &= \mathcal{B}_n (I - \mathcal{G} (\Delta \tau_n) ) \uv \\
&= \frac{\mathcal{B}_n}{\zeta_n^+ + \zeta_n^-} \left(
\begin{array}{c}
     \zeta_n^+ (1 - e^{t_n}) + \zeta_n^- (1 - e^{-t_n}) \\
    \zeta_n^+ (1 - e^{-t_n}) + \zeta_n^- (1 - e^{t_n})
\end{array}
\right)
\end{align}
where $I$ is the identity matrix. 

\subsubsection*{Transfer in a single layer}

The linear form (\ref{eq:rtstandard}) is mathematically concise, however, its physical meaning is unclear. We transform into a form representing the transfer of light in a single layer. 

Equation (\ref{eq:rtstandard}) yields 
\begin{align}
    F^+_n = \Gaa^{-1} F^+_{n+1} - \Gaa^{-1} \Gab F^-_n -\Gaa^{-1} \pi G_n^+,
\end{align}
where $\mathcal{G}_{ij}$ is the $(i-j)$ component of $\mathcal{G}(\Delta \tau_n)$ by dropping $n$ from the symbols. Substituting this equation to 
$F^-_{n+1} = \Gba F^+_{n} + \Gbb F^-_n + \pi G_n^-$, we obtain 
\begin{align}
F^-_{n+1} &= \Gba \Gaa^{-1} F^+_{n+1} + (\Gbb - \Gba \Gaa^{-1} \Gab) F^-_n \nonumber \\
&+ \pi G_n^- - \Gba \Gaa^{-1} \pi G_n^+ .
\end{align}

In the two-stream case, Equation (\ref{eq:gzero}) leads to $\Gba = -\Gab$ and $\Gaa^{-1} = \Gbb - \Gba \Gaa^{-1} \Gab \equiv \mathcal{T}_n $ and $\Gba \Gaa^{-1} = - \Gaa^{-1} \Gab \equiv \mathcal{S}_n $. Then, we can express the radiative transfer within the $n$-th layer as  
\begin{eqnarray}
\label{eq:twosq1}
 F^+_n &=& \mathcal{T}_n F^+_{n+1} + \mathcal{S}_n F^-_n - \mathcal{T}_n \pi G_n^+ \\
 \label{eq:twosq2}
 F^-_{n+1} &=& \mathcal{T}_n F^-_{n} + \mathcal{S}_n F^+_{n+1} + \pi G_n^- - \mathcal{S}_n \pi G_n^+,
\end{eqnarray}
where
\begin{align}
\label{eq:transmission_onelayer}
 &\mathcal{T}_n \equiv 
 \frac{{{\zeta^+_n}}^2 -{{\zeta^-_n}}^2 }{{\zeta^+_n}^2  - (\zeta^-_n\mathsf{T}_n)^2 } \mathsf{T}_n \\
 \label{eq:scattering_onelayer}
&\mathcal{S}_n  \equiv 
\frac{\zeta^+_n \zeta^-_n }{{\zeta^+_n}^2  - (\zeta^-_n\mathsf{T}_n)^2 } (1-\mathsf{T}_n^2)
 \end{align}
 can be regarded as the transmission between the layer bottom and top and the scattering from the opposite direction of the flux\footnote{Defining the opaque limit ($\mathsf{T}_n =0$) of the scattering in Equation (\ref{eq:scattering_onelayer}), as $S_\infty \equiv \zeta_-/\zeta_+$, one can rewrite Equation (\ref{eq:scattering_onelayer_}) to a similar form used in the flux-adding treatment in \cite{2023PSJ.....4...10R} as
\begin{align}
\label{eq:scattering_onelayer_}
 &\mathcal{T}_n = \frac{S_\infty ( 1 - e^{-2 \lambda_n \Delta \tau_n})}{1 - S_\infty^2  e^{-2 \lambda_n \Delta \tau_n} }. 
 \end{align}
 For the hemispheric mean, we obtain
 \begin{align}
     S_\infty &= \frac{\sqrt{1-\omega g}-\sqrt{1-\omega}}{\sqrt{1-\omega g}+\sqrt{1-\omega}} \\
     \lambda_n &= 2 \sqrt{(1-\omega g)(1-\omega)}.
 \end{align}
 } . In the above expressions, we defined the transmission function \citep[][]{heng2017exoplanetary},
 \begin{eqnarray}
 \label{eq:opacity_transfer}
 \mathsf{T}_n &\equiv& e^{-\lambda_n \Delta \tau_n} 
\end{eqnarray} 
We note that Equations (\ref{eq:twosq1}, \ref{eq:twosq2}) are essentially the same as the analytic expressions of the two-stream approximation derived by \cite{heng2017exoplanetary}\footnote{Equation (3.58) in \cite{heng2017exoplanetary}. $\lambda_n$ in our form  corresponds to  $\mathcal{D}$ in \cite{heng2017exoplanetary}.}.

For the isothermal layer, we can reduce Equations (\ref{eq:twosq1}) and (\ref{eq:twosq2}) to
\begin{eqnarray}
\label{eq:twosq1iso}
 F^+_n &=& \mathcal{T}_n F^+_{n+1} + \mathcal{S}_n F^-_n + \pi \hat{\mathcal{B}}_n\\
 \label{eq:twosq2iso}
 F^-_{n+1} &=& \mathcal{T}_n F^-_{n} + \mathcal{S}_n F^+_{n+1} + \pi \hat{\mathcal{B}}_n
\end{eqnarray}
where 
\begin{eqnarray}
    \hat{\mathcal{B}}_n \equiv (1 - \mathcal{T}_n - \mathcal{S}_n) \mathcal{B}_n.
\end{eqnarray}

\subsubsection{Flux-Adding Treatment}\label{ss:flux-adding}

For the solution approach of the two-stream approximation including scattering, the flux-adding treatment \citep{2018JQSRT.211...78R,2023PSJ.....4...10R}, which utilizes the form of reflection, has been proposed. This is derived analogously from the classical adding technique. The flux-adding treatment assumes that the upward flux in a given layer can be expressed as the sum of the downward flux's reflection and the source from the layer.
\begin{align}
    \label{eq:fa1}
    F_n^+ &= \hat{R}_n^+ F_n^- + \hat{S}_n^+ \\
    \label{eq:fa2}
    F_n^- &= \hat{R}_n^- F_n^+ + \hat{S}_n^-.
\end{align}

Replacing $F_n^+$ in Equation (\ref{eq:twosq1iso}) by Equation (\ref{eq:fa1}) and multiplied by $\mathcal{T}_n$, the following expression is derived.
\begin{align}
    \mathcal{T}_n^2 F_{n+1}^+ = (\hat{R}_n^+ - \mathcal{S}_n) \mathcal{T}_n F_n^- + \mathcal{T}_n (\hat{S}_n^+ - \pi \hat{\mathcal{B}}_n).
\end{align}
By using Equation (\ref{eq:twosq2iso}) to eliminate $F_n^-$ from the preceding expression, the following result is derived:
\begin{align}
    \label{eq:recursive_fa}
    F_{n+1}^+ &= \frac{\hat{R}_n^+ - \mathcal{S}_n}{  \mathcal{T}_n^2 -\mathcal{S}_n^2 +  \mathcal{S}_n \hat{R}_n^+} F_{n+1}^- \nonumber \\
    &+ \frac{\hat{\mathcal{B}}_n(\mathcal{S}_n - \mathcal{T}_n- \hat{R}_n^+ ) + \mathcal{T}_n \hat{S}_n^+}{  \mathcal{T}_n^2 -\mathcal{S}_n^2 +  \mathcal{S}_n \hat{R}_n^+}.
\end{align}
Let the coefficient of the first term on the right-hand side of the preceding expression be denoted as $\hat{R}_{n+1}^+$ and the second term as $\hat{S}_{n+1}^+$. This leads to the following recursive relation.
\begin{align}
    \label{eq:fa_Rplus}
    \hat{R}_n^+ &= \mathcal{S}_n + \frac{\mathcal{T}_n^2 \hat{R}_{n+1}^+}{1-\mathcal{S}_n \hat{R}_{n+1}^+} \\
    \label{eq:fa_Splus}
    \hat{S}_n^+ &= \hat{\mathcal{B}}_n + \frac{\mathcal{T}_n (\hat{S}_{n+1} + \hat{\mathcal{B}}_n \hat{R}_{n+1}^+)}{1 - \mathcal{S}_n \hat{R}_{n+1}^+}.
\end{align}
To derive (\ref{eq:fa_Splus}), we used Equation (\ref{eq:fa_Rplus}) to eliminate $R_n^+$ in the second term of Equation (\ref{eq:recursive_fa}).

Therefore, once we assume the reflectivity (i.e. albedo) of the bottom boundary of the bottom layer ($n=N-1$), we can compute $\hat{R}^+_0$ and $\hat{S}^+_0$ and can obtain the outgoing flux at the ToA as 
\begin{align}
    F_0^+ = \hat{R}^+_0 F_\star + \hat{S}^+_0,
\end{align}
where $F_\star$ is the incoming stellar flux. 

Although we do not use $\hat{R}^-_n$ and $\hat{S}^-_n$ for the out-going flux computation, we can also derive them. In Equation (\ref{eq:twosq2iso}), replace $n$ with $n-1$ and then eliminate $F_n^-$ using Equation (\ref{eq:fa2}). Further, by substituting $n$ with $n-1$ in Equation (\ref{eq:twosq1iso}) and extracting $\mathcal{T}_{n-1} F_n^+$, and then inserting this into the above expression, the following equation is derived.
\begin{align}
    \label{eq:recursive_fa2}
    F_{n-1}^- &= \frac{\hat{R}_n^- - \mathcal{S}_{n-1}}{  \mathcal{T}_{n-1}^2 -\mathcal{S}_{n-1}^2 +  \mathcal{S}_{n-1} \hat{R}_n^-} F_{n-1}^+ \nonumber \\
    &+ \frac{\hat{\mathcal{B}}_{n-1}(\mathcal{S}_{n-1} - \mathcal{T}_{n-1}- \hat{R}_n^- ) + \mathcal{T}_{n-1} \hat{S}_n^-}{  \mathcal{T}_{n-1}^2 -\mathcal{S}_{n-1}^2 +  \mathcal{S}_{n-1} \hat{R}_n^-}
\end{align}
By denoting the coefficient of the first term on the right-hand side as $\hat{R}_{n-1}^-$ and the second term as $\hat{S}_{n-1}^-$, the following recursive equations are obtained.
\begin{align}
    \label{eq:fa_Rminus}
    \hat{R}_n^- &= \mathcal{S}_{n-1} + \frac{\mathcal{T}_{n-1}^2 \hat{R}_{n-1}^-}{1 - \mathcal{S}_{n-1} \hat{R}_{n-1}^-} \\
    \label{eq:fa_Sminus}
    \hat{S}_n^- &= \hat{\mathcal{B}}_{n-1} + \frac{\mathcal{T}_{n-1} (\hat{S}_{n-1} + \hat{\mathcal{B}}_{n-1} \hat{R}_{n-1}^-)}{1 - \mathcal{S}_{n-1} \hat{R}_{n-1}^-}
\end{align}

Equations (\ref{eq:fa_Rplus}), (\ref{eq:fa_Splus}), (\ref{eq:fa_Rminus}), and (\ref{eq:fa_Sminus}) are essentially the same as Equations (7), (8), (4) and (5) in \cite{2018JQSRT.211...78R}.

As we have seen, in the flux-adding treatment, the recurrence relations can be formulated and solved by including scattering as an effective reflection and source sum for each layer. Similarly, the other recurrence relation can also be formulated by including scattering as an effective transmission and source sum for each layer. Although we have not yet identified practical benefits in this method, it is a natural extension of flux-based pure absorption and is theoretically interesting, so we will describe it in Appendix \ref{ap:ett_lart}. We note that we confirmed that this method and the flux-adding treatment provide consistent results for radiation involving scattering.

\subsection{Transmission Spectrum} \label{ss:transmission}

Figure \ref{fig:transmission_coord} shows the geometry and coordinate system for transmission spectroscopy. The optical depth of the $n$-th layer along the vertical direction can be converted to that along the chord direction by multiplying the geometric weight, $\Delta \underline{z}^{(n)}_k/\Delta h_k$, where $\Delta \underline{z}^{(n)}_k$ is the horizontal width of the $k$-th layer from the observer to the $n$-th layer. Therefore, the chord optical depth at the lower boundary of the $n$-th layer, $\underline{t}_n = t(\underline{r}_n)$, is expressed as 
\begin{align}
\label{eq:chord_optical_depth}
    \underline{t}_n &= 2 \sum_{k=0}^{n} \kappa_k \rho_k \Delta \underline{z}^{(n)}_k = \sum_{k=0}^{n} \underline{C}_{nk}  \Delta \tau_k \\ 
    &= \overline{t}_{n+1}\,\,\,
 \mbox{for $n=0,\cdots,N-1$} 
\end{align}
where $\underline{t}_{N-1} = \overline{t}_N$ is the chord optical depth at $R=R_0$ and we assume $\overline{t}_0 = 0$. We used $\Delta \tau_k = \kappa_k \rho_k \Delta h_k$ and defined a lower triangle matrix, whose element is expressed as
\begin{eqnarray}
    \underline{C}_{nk} = 
    \left\{
    \begin{array}{ll}
    \displaystyle{
    \frac{2 \Delta \underline{z}^{(n)}_k}{\Delta h_k}
     = 2 \frac{\sqrt{\overline{r}_{k}^2 - \underline{r}_{n}^2} - \sqrt{\underline{r}_{k}^2 - \underline{r}_{n}^2}}{\Delta h_k}} & \mbox{\,, $k \le n$} \nonumber \\
     0  & \mbox{\,, $k > n$}
     \end{array}
     \right.
\end{eqnarray}
Then, the matrix $\underline{C}$, which we call the (upper) chord geometry matrix, can convert the vector of $\Delta \tau_i$ to that of the chord optical depth as $\underline{C} \boldsymbol{\Delta \tau}$ except for at the top of the atmosphere $\overline{r}_0$, where $\boldsymbol{\Delta \tau} = (\tau_0, \tau_1, \cdots, \tau_{N})^\top$. We assume that the chord of the top of the atmosphere is zero. The chord optical depth vector at the upper boundary is defined by $\overline{\boldsymbol{t}} = (\overline{t}_0, \overline{t}_1, \cdots, \overline{t}_{N})^\top$.   In the above description, we defined the upper boundary of the $n$-th layer by $\overline{r}_n$, i.e., $\Delta h_n = \overline{r}_{n} - \overline{r}_{n+1}$. 

While the chord optical depth vector at the layer boundary is sufficient for the trapezoidal integration of the total occultation, the chord optical depth at the midpoint of the layer is required when using Simpson's rule for the integration. We define
\begin{eqnarray}
    C_{nk} = 
    \left\{
    \begin{array}{ll}
    \displaystyle{
   2 \frac{\sqrt{\overline{r}_{n}^2 - {r}_{n}^2} }{\Delta h_k}}  & \mbox{\,, $k = n$}  \nonumber \\
    \displaystyle{
    \frac{2 \Delta z^{(n)}_k}{\Delta h_k}
     = 2 \frac{\sqrt{\overline{r}_{k}^2 - {r}_{n}^2} - \sqrt{\underline{r}_{k}^2 - {r}_{n}^2}}{\Delta h_k}} & \mbox{\,, $k < n$} \\
     0  & \mbox{\,, $k > n$}
     \end{array}
     \right.
\end{eqnarray}
and
$
    \boldsymbol{t} = C \boldsymbol{\Delta \tau},
$
 where we note that the optical length at the deepest element ($k=n$), as indicated by the yellow dashed line in Figure \ref{fig:transmission_coord}, is a different form from the case of $k < n$. 

The transit radius at the wavenumber of $\nu$ is given by
\begin{eqnarray}
    R_{p,\nu} = \sqrt{ R_{0}^2 + \Delta R_{p,\nu}^2 },
\end{eqnarray}
where $R_0=\underline{r}_{N-1}$ is the lower boundary of the bottom layer and is assumed to be completely opaque.
The contribution of the atmospheric layers to the transit radius is expressed as
\begin{eqnarray}
   \Delta R_{p,\nu}^2 &\equiv& 2 \int_{R_{0}}^\infty [ 1 - \mathcal{T}_\nu(r)] r d r \\ 
   \label{eq:deltaR}
   &\approx& 2 \int_{\underline{r}_{N-1}}^{\overline{r}_0} [ 1 - \mathcal{T}_\nu(r)] r d r,
\end{eqnarray}
where 
\begin{equation}
\label{eq:Trans_tau_chord}
  \mathcal{T}_\nu(r) = e^{-t(r)}
\end{equation}
is the transmission at $r$ and $t(r)$ is the chord optical depth at $r$ and $\underline{r}_0$ is the lower boundary of the ToA. We evaluate Equation (\ref{eq:deltaR}) by the trapezoid integral or the Simpson integral. 

{
\section{Special Considerations for Differentiable Implementation in ExoJAX2 and Further Reduction of Memory Usage}
This section discusses programming considerations specific to differentiable programming. Additionally, it examines further reductions in GPU device memory usage.

\begin{figure}
    \centering
    \includegraphics[width=\linewidth]{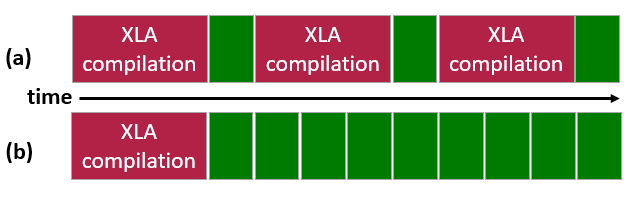}
    \caption{  
(a) XLA compilation occurs on every function execution.  
(b) XLA compilation occurs only before the first function call.\label{fig:jaxconsider}}
\end{figure}

\subsection{Scan, JIT, XLA compilation}
The key considerations for implementing ExoJAX in differentiable programming using JAX are as follows.
In JAX, to enable automatic differentiation through loop computations, it is generally necessary to use the `scan' function ({\sf jax.lax.scan}), which originates from functional programming \citep[Section 5 in ][]{2024arXiv240314606B}. 

To maximize computational efficiency, it is necessary to enable XLA (Accelerated Linear Algebra), a framework designed for executing linear algebra computations efficiently on CPUs, GPUs, and TPUs. This is achieved using JIT (Just-in-Time) compilation. When using XLA, the function is compiled upon its first invocation. However, if the compilation time exceeds the function’s computation time or data transfer time, the following points should be considered:
Depending on the implementation, XLA compilation may occur every time the function is called as shown in Panel (a) of Figure \ref{fig:jaxconsider}.  In cases such as retrieval or optimization, where the same function is repeatedly invoked, XLA compilation time can become the bottleneck in execution. To prevent this issue, it is essential to use a memory profiler like {\sf Perfetto} to verify that XLA compilation does not occur again after the first function call as shown in Panel (b) of Figure \ref{fig:jaxconsider}.

\subsection{Forward and Reverse-Mode Differentiation}

\begin{figure*}
    \centering
    \includegraphics[width=\linewidth]{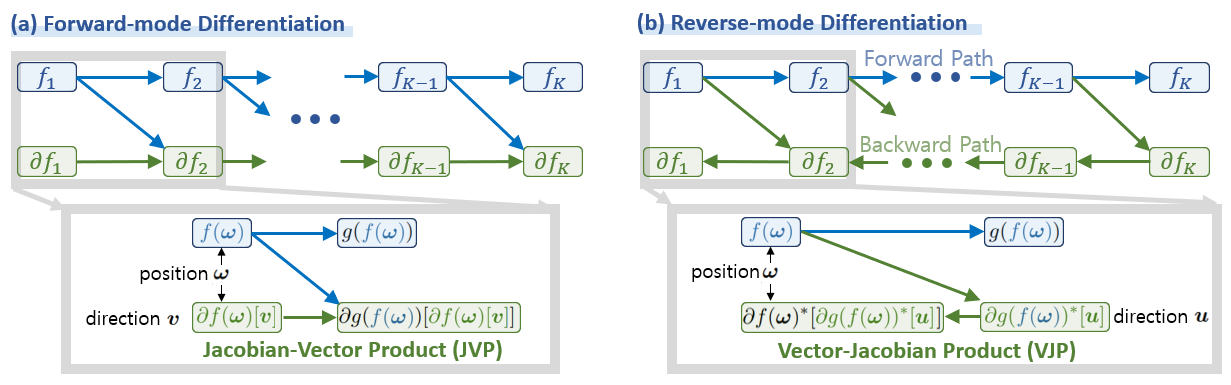}
    \caption{ Computational graph of (a) Forward-mode and (b) Reverse-mode differentiation. Panel (a): In forward-mode differentiation, computation progresses from left to right, and at each step, differentiation is performed using the Jacobian Vector Product (JVP). Since JVP only requires the function value ($f_i$ and $\partial f_i$ to compute $f_{i+1}$ and $\partial f_{i+1}$) and its derivative from the previous step, these are the only values needed for step updates. Panel (b): In reverse-mode differentiation, the Vector-Jacobian Product (VJP) is used. Because the function values after each update are required for backward step updates, the computed function values ($f_1$ to $f_{K-1}$) must first be stored while composing the function from left to right. Then, during the backward pass (right to left), these stored intermediate results are used to propagate the derivatives through the backward path.
\label{fig:diff}}
\end{figure*}

As shown in Figure \ref{fig:diff}, there are two types of computation methods for automatic differentiation: forward-mode and reverse-mode.
Forward-mode differentiation works by propagating derivatives from the inputs of a function toward its outputs. Reverse-mode differentiation (or backpropagation) does the opposite: it propagates derivatives from the outputs backward to the inputs.
Machine learning frameworks like {\sf TensorFlow} and {\sf PyTorch} primarily support reverse-mode because it is computationally more efficient when the number of input parameters is large compared to the number of output parameters (e.g., many inputs for neural nets and a single output for a cost function).

However, in reverse-mode differentiation, the computation first evaluates intermediate function values in the forward path. Then, during the backward path, derivatives are computed from the output side to the input side using these intermediate function values. The reason for this is that reverse-mode differentiation propagates function derivatives using the Vector-Jacobian Product (VJP), which requires the function values at the propagation points (corresponding to $f({\bf \omega})$ in Panel (b) of Figure \ref{fig:diff}). Therefore, reverse-mode differentiation requires storing all intermediate results in the computation graph, making it less efficient in terms of (device) memory usage. In contrast, forward-mode differentiation can discard results at each step, making it more memory-efficient because function derivatives are propagated using the Jacobian Vector Product (JVP), which represents directional derivatives. In JVP, only the function value at the propagation source ($f({\bf \omega})$ in Panel (a) of Figure \ref{fig:diff}) is required for the propagation computation \citep[see also Section 8 in ][]{2024arXiv240314606B}. Since {\sf JAX} supports both reverse-mode and forward-mode differentiation, {\sf ExoJAX2} can utilize both differentiation modes as well. Unlike neural networks, the number of input parameters in atmospheric retrievals is usually not very large, meaning that in many cases, there is little difference in computation time between the two modes. 
In particular, as explained in \S \ref{ss:opart}, when solving opacity and radiative transfer layer by layer, forward-mode differentiation is well suited because it allows device memory to be freed layer by layer.

\begin{table}[]
  \caption{{GPU device memory reduction techniques}}
  \label{tab:reduce_memory}
  \begin{center}
  \begin{tabular}{lcc} 
  \hline\hline
   type & Opart& $\nu$--stitching\\
   \hline
   Emission &\gcheckmark & \gcheckmark\\
   Reflection &\gcheckmark & \gcheckmark\\
   Transmission &  \notava  & \gcheckmark \\
\hline
  \end{tabular}
  \end{center}
\end{table}

\subsection{Layer-by-layer opacity calculation and atmospheric radiative transfer (Opart)}\label{ss:opart}
As shown in Table \ref{tab:device_memory_usage}, when opacity calculation and radiative transfer are computed separately, device memory usage is constrained by the cross-section matrix for all layers, resulting in memory usage proportional to the number of layers $N_\mathrm{layer}$. This can lead to device memory overflow, particularly when GPUs with limited device memory, such as high-end GPUs like NVIDIA A100 (80GB) or H100 (96GB), are unavailable, or when increasing the number of layers to several hundred or more. 

One solution to this issue is to calculate the opacity and radiative transfer for each layer sequentially while computing the final flux. This approach eliminates the need to store the opacity of a given layer for use in subsequent layers, thereby making the device memory usage independent of the number of layers. Fortunately, 
 emission for pure absorption cases, such an algorithm can be structured as demonstrated below. 

Combining Equation (\ref{eq:I_discrete_nemesis}) into Equation (\ref{eq:f0_gq}), we can write the outgoing flux by the layer-wise summation form as 
\begin{align}
    F_\mathrm{out} 
    &= \sum_{n=0}^{N-1} 2 \pi B(T_n) \, \xi (\tau_n, \Delta \tau_{n}) + F_B \\
     F_B &= 2 \pi \sum_{i=1}^{\mathsf{N}^\prime} w_i \mu_i B(T_B) e^{-\tau_B/\mu_i},
\end{align}
where
\begin{align}
 \xi (\tau_n, \Delta \tau_{n}) &\equiv  \sum_{i=1}^{\mathsf{N}^\prime} w_i \mu_i (e^{-\tau_n/\mu_i} - e^{-(\tau_{n} + \Delta \tau_n)/\mu_i}).
\end{align}
As shown in the equation above, in the \(n\)-th layer, it suffices to compute \(\xi(\tau_n, \Delta \tau_n)\). This requires only the opacity calculation for the \(n\)-th layer (\(\Delta \tau_n\)) and the cumulative opacity up to that layer (\(\tau_n\)) to be stored. The opacity values for individual layers ( 0 to \(n-1\)) do not need to be retained. This implementation can be achieved using {\sf jax.lax.scan}.
We refer Opart to this approach, which combines {\sf opa} and {\sf art}, as {\sf opart}.

In the two-stream approximation for reflection or emission with scattering, as explained in \S 4.3, the layer-by-layer strategy for updating opacity and radiative transfer can also be applied. Specifically, in the layer-wise update equations (\ref{eq:fa_Rplus}) and (\ref{eq:fa_Splus}), opacity only requires the $\Delta \tau_n$ of the layer being updated, as determined by Equation (\ref{eq:opacity_transfer}). Similarly, the single scattering albedo and the asymmetry parameter only need the values of the layer being updated. As a result, only the opacity, single scattering albedo, and asymmetry parameters of a single layer need to be stored in device memory at any given time. This makes it possible to implement the {\sf Opart} method even for the two-stream approximation.

However, in the case of transmission, this strategy cannot be applied. This is because, as given by Equation (\ref{eq:Trans_tau_chord}), the contribution of each layer to the squared radius is nonlinear with respect to the chord opacity $t$. Additionally, since the chord opacity $t$ is given by a weighted sum of the opacity of each layer $\Delta \tau_n$, it is not possible to store the contributions layer by layer.

To save the device memory usage, {\sf Opart} needs to use forward-mode differentiation instead of reverse-mode differentiation. For gradient optimization, {\sf jax.jacfwd} or {\sf jax.jvp} is used to compute gradients in forward-mode, which are then utilized for optimization. In the case of HMC-NUTS, the forward differentiation mode implemented in {\sf NumPyro} can be employed.

}

\subsection{$\nu$--Stitching}
{
The other approach to reducing device memory usage is to partition the wavenumber grid into multiple segments and later recombine them through wavenumber grid stitching (\(\nu\)--stitching). To achieve memory reduction with this method, forward-mode differentiation is required. In \(\nu\)--stitching, aliasing parts from adjacent segments must be merged. Therefore, the cross-sectional area of each segment is computed based on the available "open cross section" of the aliasing part, and the segments are subsequently combined using the Overlap--and--Add (OLA) method (Appendix \ref{ap:ola}). {\sf ExoJAX}2 implements and utilizes OLA in a form that is differentiable. An advantage of \(\nu\) stitching is that it can be applied to the calculation of transmission spectra, where opart cannot be used.
Table \ref{tab:reduce_memory} summarizes the applicability of different device memory reduction strategies for each radiative transfer method. 
}

\section{Other improvements from {\sf ExoJAX }1}
This section describes several improvements beyond those noted in the previous section. 

{\sf ExoJAX}2 is now compatible with the reverse-mode of differentiation. This improvement enabled us to use a gradient-based optimizer, such as {\sf JaxOpt} \citep{jaxopt_implicit_diff}. While it is not necessary for the retrieval itself, it is highly valuable in actual spectral analysis. Specifically, it is often challenging to pinpoint a plausible model for the spectrum obtained from observations, and this requires a great deal of trial and error. By using gradient optimization, it becomes possible to quickly determine whether a particular model is likely to explain the observed spectrum before conducting MCMC. 

In {\sf ExoJAX}1, the rotational and instrumental responses were computed using a transformation matrix between the input and output wavenumber grids. This algorithm requires 4 (or 8) $\times N_\mathrm{\nu} \times N_\mathrm{\nu^\prime}$ byte of the GPU memory for FP32 (or FP64), where $N_\mathrm{\nu}$ and $N_\mathrm{\nu^\prime}$ are the numbers of the input and output wavenumber grids. In broadband opacity computation, the GPU memory overflows. We rewrite the rotational and instrumental responses by FFT-based convolutions, which require much less GPU memory, $\sim \mathcal{O}(N_\mathrm{\nu})$. In broadband spectra, the lengths of rotation kernels or IP kernels are typically much shorter compared to the length of the spectrum. In such cases, using FFT-based convolution is inefficient in terms of computation and memory usage. In these situations, the OLA, which divides the spectrum into appropriate lengths and convolves each before summing them, proves effective. 

Since {\sf ExoJAX}1, all source code is open and maintained on {\sf github}. This makes it easy for users to add necessary functions on their own. The development of {\sf ExoJAX}2 incorporates Test Driven Development \citep[TDD;][]{beck2003test}. 
This approach enhances code reliability and maintainability, ensuring that new features and updates integrate smoothly without introducing regressions.
To enable continuous software development, three types of tests were prepared at {\sf ExoJAX}2 \citep[e.g.][]{percival2020architecture}; unit tests, integration tests, and end-to-end tests. The unit tests are provided to make the software more maintainable \citep{kohrikov2020unit}. Using {\sf github actions}, all unit test codes are automatically performed when the code update is submitted to the develop and master branches. 

In addition, the molecular database I/O ({\sf mdb}), opacity calculation ({\sf opa}), radiative transfer ({\sf art}), spectral operators ({\sf sop}), and atmospheric microphysics ({\sf amp}) classes are designed separately.

\section{Demonstration and Applicability}

The primary aim of this paper is to present the extended algorithm in {\sf ExoJAX}2. However, in this section, we wish to illustrate some practical examples that highlight the advantages brought about by the aforementioned extensions. It should be noted, however, that the scientific discussions specific to each analysis will be reserved for separate papers.

\subsection{High-resolution Spectroscopy of a Clear-Sky T-dwarf Emission}

First and foremost, in {\sf ExoJAX}1, due to the device memory usage constraints, Bayesian analysis of high-resolution spectroscopy could typically only be conducted over a narrow bandwidth of about $\mathcal{O}(10^1) \mathrm{nm}$. With {\sf ExoJAX}2, these constraints have been significantly relaxed as discussed in Section \ref{sec:opacity}, enabling full Bayesian analyses over broader bandwidths using HMC-NUTS.

\begin{figure*}
    \centering
    \includegraphics[width=\linewidth]{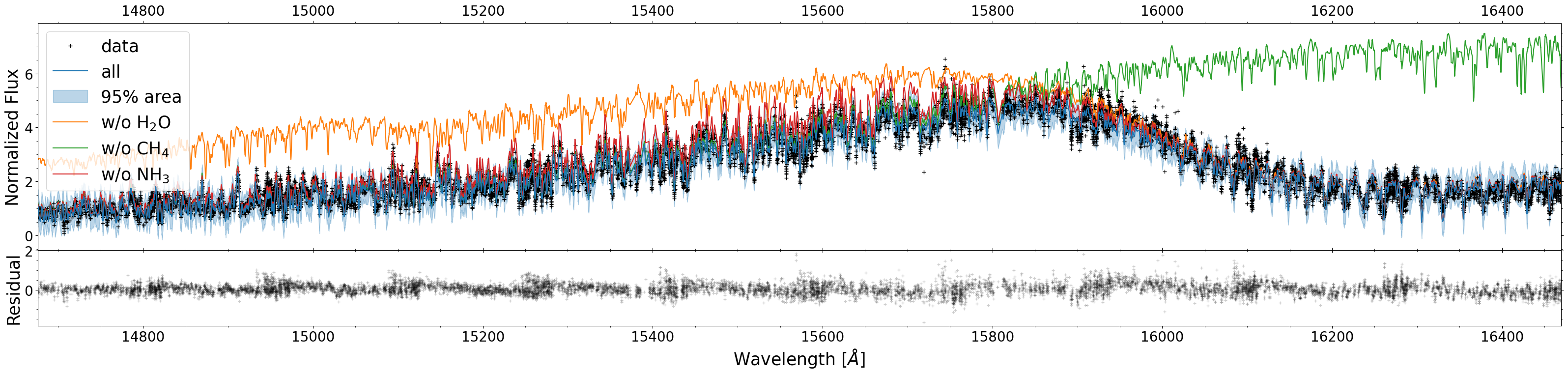}
    \caption{The high-resolution spectrum of the T-dwarf Gl229 B analyzed using {\sf ExoJAX}2 \citep{2024arXiv241011561K} with $\sim 10^8$ total molecular absorption lines considered. 
    The blue solid line shows the median value, while the blue shaded regions indicate the 95\% credible interval.
    Black points represent the observed Subaru/IRD data. The periodic increases in the residuals observed correspond to the gaps between orders.
    }
    \label{fig:gl229B}
\end{figure*}

Figure \ref{fig:gl229B} displays an example of the high-resolution spectroscopy of the H-band for the T-dwarf Gl229B, alongside the credible interval obtained using HMC-NUTS with {\sf ExoJAX}2 \citep{2024arXiv241011561K}. In this example, molecular absorptions include methane \citep[from HITEMP;][]{2020ApJS..247...55H}, water \citep[from ExoMol/POKAZATEL;][]{2018MNRAS.480.2597P}, and ammonia \citep[from ExoMol/CoYuTe;][]{2015JQSRT.161..117A, 2019MNRAS.490.4638C}, with a total of $\sim$0.1 billion lines. For this example, the single-precision computation was sufficient, and with the use of the NVIDIA A100 (PCIe), 500 warm-up iterations and 1,000 sampling iterations were completed in 68 hours. The total number of iterations, i.e., the total number of spectra and their gradients calculated, was approximately 0.125 million. This corresponds to a computational cost of approximately 2 seconds per iteration.

The full Bayesian analysis of the high-resolution spectrum of Gl229B led to the following conclusions: (1) The C/O ratio of Gl229B ($\mathrm{C/O} = 0.76 \pm 0.1$) is consistent with that of the primary star, Gl229A,  (2) mass estimation is possible using only the spectrum, and (3) the combination of the spectral analysis and astrometry supports the binary hypothesis of Gl229B\footnote{During the review process of this paper, the binarity of Gl229B was demonstrated through both imaging and spectroscopic separation\citep{xuan2024cool, whitebook2024discovery}.}. For more details, we refer to \cite{2024arXiv241011561K}.

In addition to the ability of full Bayesian analysis, {\sf ExoJAX}2 allows optimization based on the gradient method through compatibility of {\sf JAXopt} \citep{jaxopt_implicit_diff}. We realized that this was useful to search for the initial parameters of the HMC-NUTS.

\subsection{Medium-resolution Transmission Spectroscopy of a Hot Saturn}

\begin{table}[tb!]
\centering
\caption{Model parameters and their prior distribution in the analysis of WASP-39 b's transmission spectrum using HMC-NUTS. }
\begin{tabular}{lll}
\hline\hline
Symbol & Description & Prior\\
\hline
$R_\mathrm{s}$ & host star's radius ($R_{\odot}$) & $\mathcal{N}(0.939, 0.022)$ \\
$M_\mathrm{p}$ & planetary mass ($M_{\mathrm{J}}$) & $\mathcal{N}(0.281, 0.032)$ \\
$\mathrm{RV}$ & radial velocity ($\mathrm{km\ s^{-1}}$) & $\mathcal{U}(-200.0, 0.0)$ \\
$T$ & temperature ($\mathrm{K}$) & $\mathcal{U}(500, 2000)$ \\
$R_\mathrm{p}$ at 10 bar & height at 10 bar ($R_{\mathrm{J}}$) & $\mathcal{U}(1.0, 1.5)$ \\
$\log_{10} P_\mathrm{cloud}$ & cloud top pressure (bar) & $\mathcal{U}(-11.0, 1.0)$\\
$\log_{10}\mathrm{H_{2}O}$ & VMR of $\mathrm{H_{2}O}$ & $\mathcal{U}(-15.0, 0.0)$\\
$\log_{10}\mathrm{CO}$ & VMR of $\mathrm{CO}$ & $\mathcal{U}(-15.0, 0.0)$\\
$\log_{10}\mathrm{CO_{2}}$ & VMR of $\mathrm{CO_{2}}$ & $\mathcal{U}(-15.0, 0.0)$ \\
$\log_{10}\mathrm{SO_{2}}$ & VMR of $\mathrm{SO_{2}}$ & $\mathcal{U}(-15.0, 0.0)$\\
$\log_{10}\mathrm{H_{2}S}$ & VMR of $\mathrm{H_{2}S}$ & $\mathcal{U}(-15.0, 0.0)$\\
$\log_{10}\mathrm{CH_{4}}$ & VMR of $\mathrm{CH_{4}}$ & $\mathcal{U}(-15.0, 0.0)$\\
$\log_{10}\mathrm{NH_{3}}$ & VMR of $\mathrm{NH_{3}}$ & $\mathcal{U}(-15.0, 0.0)$ \\
$\log_{10}\mathrm{HCN}$ & VMR of $\mathrm{HCN}$ & $\mathcal{U}(-15.0, 0.0)$\\
$\log_{10}\mathrm{C_{2}H_{2}}$ & VMR of $\mathrm{C_{2}H_{2}}$ & $\mathcal{U}(-15.0, 0.0)$\\
\hline
\end{tabular}
\tablecomments{$\mathcal{N}(a, b)$: the normal distribution with a mean of $a$ and a standard deviation of $b$, truncated to ensure that the values are non-negative. $\mathcal{U}(a,b)$: the uniform distribution from $a$ to $b$. $\mathrm{VMR}$: Volume Mixing Ratio.}
\label{tab:prior_wasp39b}
\end{table}

\begin{figure*}
    \centering
    \includegraphics[width=\linewidth]{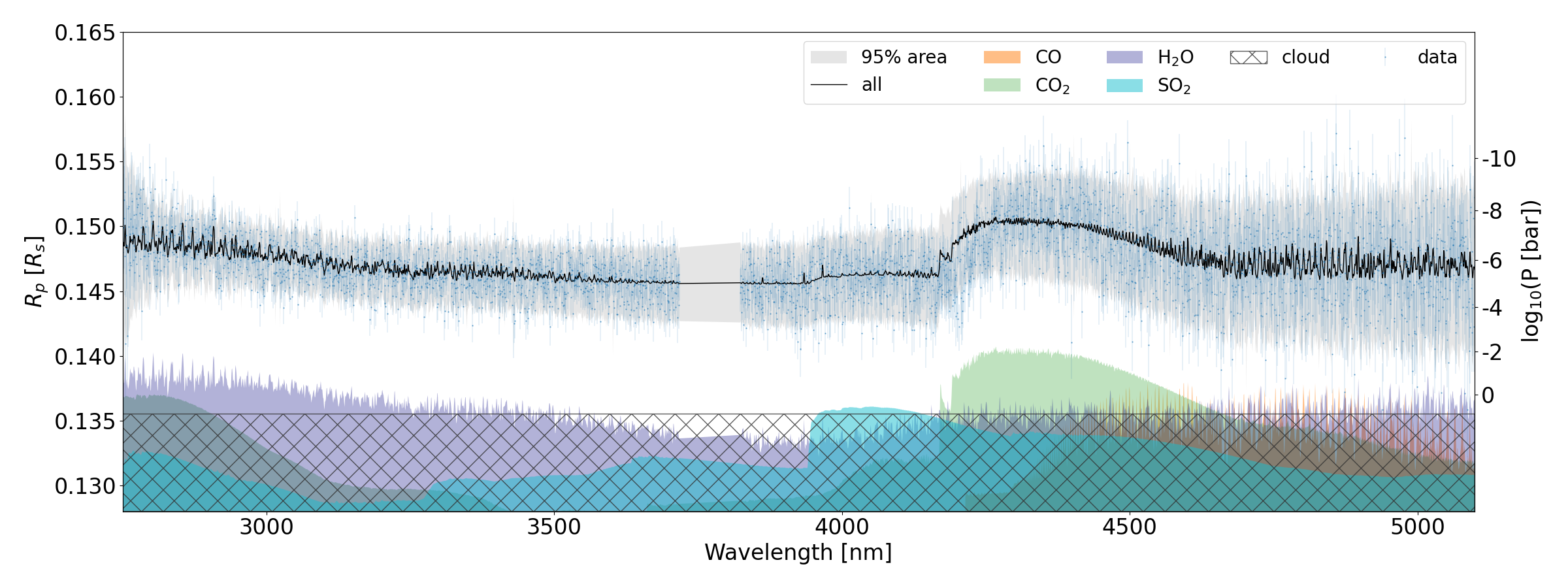}
    \caption{
    Medium-resolution transmission spectrum of WASP-39 b analyzed using {\sf ExoJAX}2 and the {\sf HMC-NUTS} method. The black solid line represents the model derived from the median values of each parameter obtained through MCMC sampling, with the 95\% credible interval shaded in gray. The blue points show the observed planetary radii from JWST NIRSpec/G395H observations. The vertical axis on the left shows the planetary radius, while the vertical axis on the right displays the corresponding pressure. The contributions of $\mathrm{CO}$, $\mathrm{CO_2}$, $\mathrm{H_2O}$, $\mathrm{SO_2}$, gray clouds, and collision-induced absorption (CIA) are represented by colored shading with an offset of -0.01 $R_s$.}
    \label{fig:WASP39b}
\end{figure*}

\begin{figure}
    \centering
    \includegraphics[width=\linewidth]{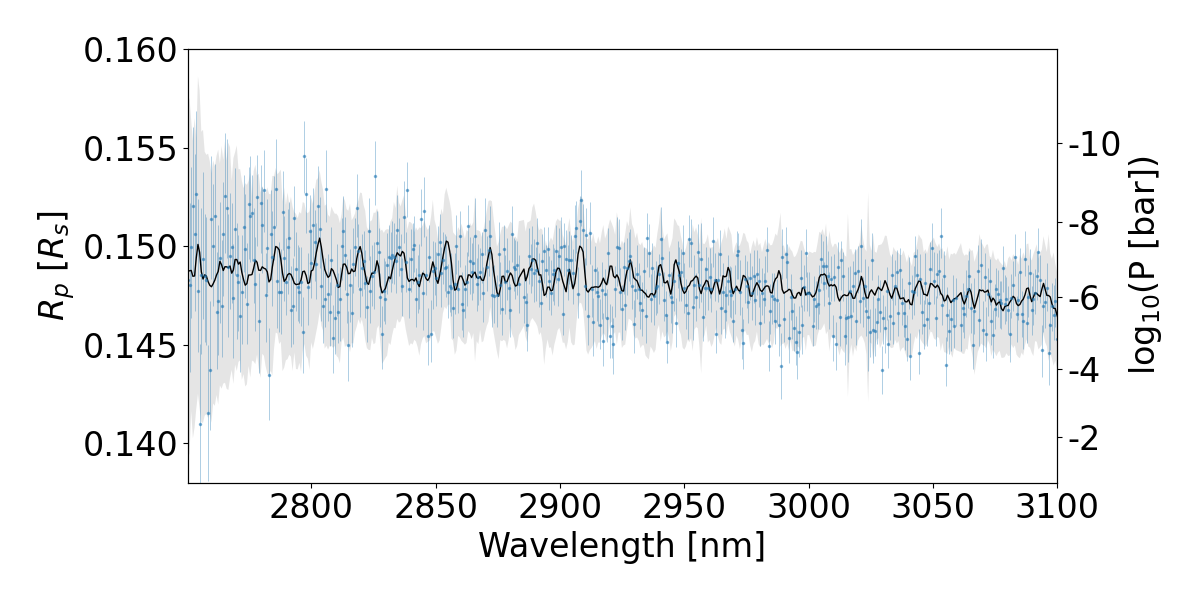}
    \caption{Enlarged view of the 2750-3100 nm region in Figure \ref{fig:WASP39b}. The black solid line and the gray shaded area are the same as in Figure \ref{fig:WASP39b}.
    \label{fig:WASP39b_H2O}}
\end{figure}

\begin{figure*}
    \centering
    \includegraphics[width=\linewidth]{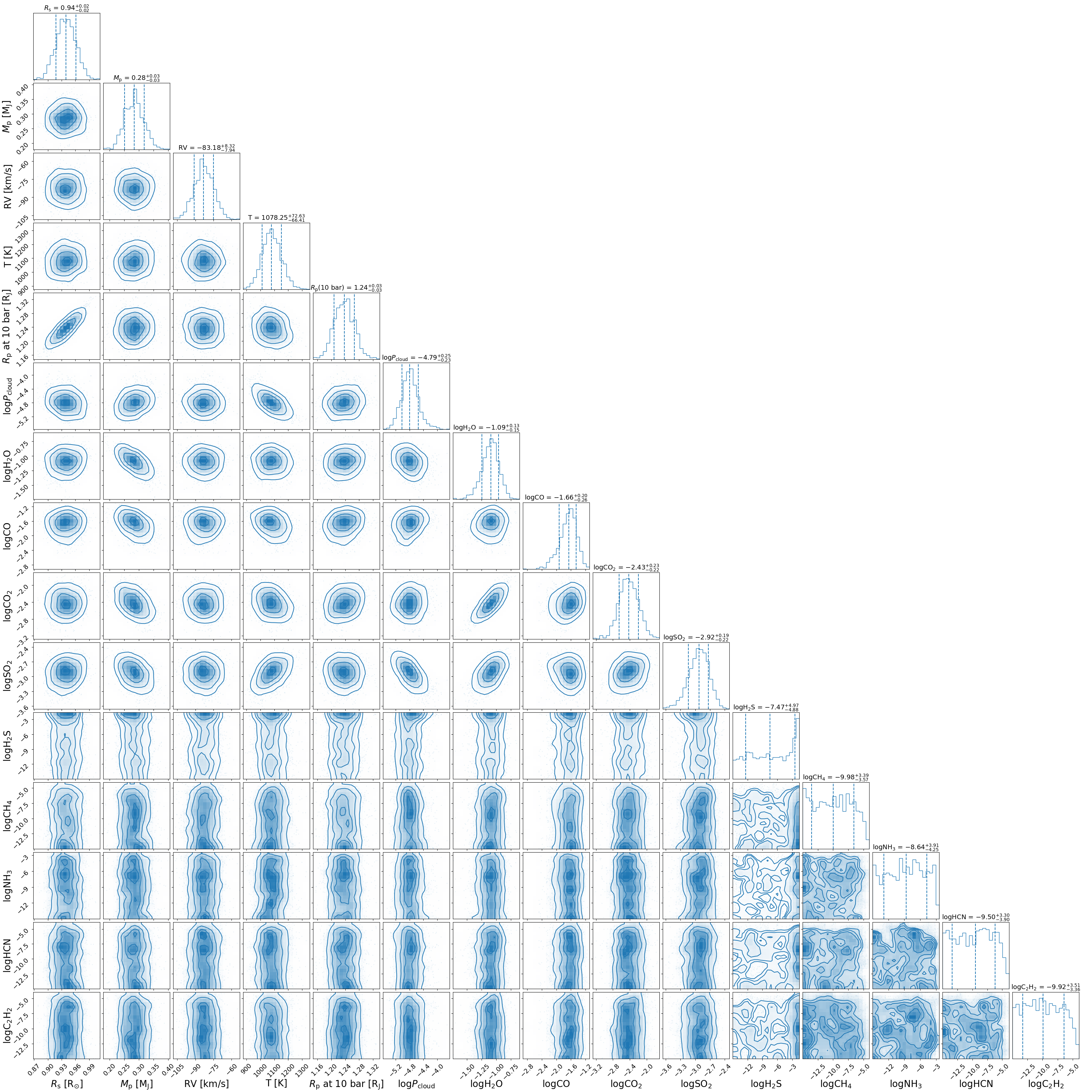}
    \caption{Posterior distributions for the atmospheric parameters obtained from the transmission spectrum of WASP-39 b, observed using the JWST NIRSpec/G395H. The posterior distributions were derived using a Markov Chain Monte Carlo (MCMC) method with the No-U-Turn Sampler (HMC-NUTS) algorithm. The diagonal panels show the one-dimensional marginalized distributions for each parameter, while the off-diagonal panels illustrate the two-dimensional correlations between parameter pairs.
    \label{fig:WASP39b_corner}}
\end{figure*}

As an example of the transmission spectrum model in \S \ref{ss:transmission}, we present a Bayesian analysis of JWST data for the hot Saturn WASP-39b. These data were obtained through the JWST Transiting Exoplanet Community Director’s Discretionary Early Release Science (JTEC ERS) Program \citep[][ERS-1366]{2016PASP..128i4401S, 2018PASP..130k4402B}. 
We analyzed the data obtained using the NIRSpec/G395H \citep{2023Natur.614..664A}. 
The details of the light curve reduction process from the raw data are the same as in the separate paper by \cite{2025arXiv250308988T} -- which discusses atmospheric inhomogeneities inferred from the color dependence of timing variations.

In this demonstration, we explore atmospheric retrieval at the original spectral resolution of NIRSPEC/G395H \( R \sim 3000 \) by using the light curves without binning.
Figure \ref{fig:WASP39b} presents an HMC-NUTS fit of the {\sf ExoJAX} model to the transmission spectrum.
A spectral resolution of NIRSpec/G395H \citep{2022A&A...661A..80J, 2023PASP..135c8001B} allows for the resolution of fine structures of the molecular lines. The line-by-line modeling by {\sf ExoJAX}2 captures these structures, such as those formed by water, as shown in Figure \ref{fig:WASP39b_H2O}.

Our model assumed a gray cloud that is opaque below a certain pressure, independent of wavelength, and the CIA (\S \ref{ss:cia}) of the hydrogen and helium atmosphere \citep[$\mathrm{H_2}$--$\mathrm{H_2}$ and $\mathrm{H_2}$--$\mathrm{He}$,][]{2012JQSRT.113.1276R, 2019Icar..328..160K}. We also assumed a constant temperature profile characterized by $T$. This model included nine molecular species: $\mathrm{H_2O}$, $\mathrm{CO}$, $\mathrm{CO_2}$, and $\mathrm{CH_4}$ \citep[HITEMP,][]{2010JQSRT.111.2139R, 2020ApJS..247...55H}, $\mathrm{SO_2}$ \citep[ExoMol/ExoAmes,][]{2016MNRAS.459.3890U, 2018JQSRT.208..152T, 2013JQSRT.130....4R}, $\mathrm{H_2S}$ \citep[ExoMol/AYT2,][]{2016MNRAS.460.4063A, 2018JQSRT.218..178C}, $\mathrm{NH_3}$ \citep[ExoMol/CoYuTe,][]{2015JQSRT.161..117A, 2019MNRAS.490.4638C}, $\mathrm{HCN}$ \citep[ExoMol/Harris,][]{2006MNRAS.367..400H, 2014MNRAS.437.1828B, 1985CP.....93..115M, 1973JChPh..58..442C, 1980JChPh..73.1494C} and $\mathrm{C_2H_2}$ \citep[ExoMol/aCeTY,][]{2020MNRAS.493.1531C, 2016JQSRT.168..193W, 2017JQSRT.203....3G}. The free parameters in this model were $T$, the radius of the host star ($R_\mathrm{s}$), the planetary mass ($M_\mathrm{p}$), the radial velocity (RV), the planetary radius $R_p$ at an altitude of $10$ bar, the logarithm of the cloud top pressure $\log P_{\mathrm{cloud}}$, and the logarithms of the volume mixing ratio (VMR) of the nine species. The prior distributions are given in Table \ref{tab:prior_wasp39b}. The prior distributions for $R_\mathrm{s}$ and $M_\mathrm{p}$ were based on \citet{2018A&A...613A..41M}.

The atmosphere was divided logarithmically into 120 layers, ranging from $10\ \mathrm{bar}$ to $10^{-11}\ \mathrm{bar}$. The wavelength range was equally divided into 5 regions in wave number space. The model spectra were calculated at a spectral resolution 10 times higher than the instrumental spectral resolution (ranging from $\sim2000$ to $\sim3500$) for each region and then broadened to the instrumental spectral resolution using Gaussian functions.

The opacity of each atmospheric layer for the gray cloud was approximated by a sigmoid function to make it differentiable, as follows: 
\begin{equation} 
\Delta \tau_{k} = \frac{\tau_{\mathrm{cloud}}}{1+(\tau_{\mathrm{cloud}}-1)\exp  \left(\frac{\log_{10} P_{k} - \log_{10} P_{\mathrm{cloud}}}{w}\right)}, 
\end{equation}
which equals 1 at $P_{} = P_{\mathrm{cloud}}$, where $P_{k}$ is the pressure of the $k$-th layer, $\tau_{\mathrm{cloud}}$ is the cloud opacity, and $w$ is the width of the sigmoid. We used $\tau_{\mathrm{cloud}} = 50$ and $w = 1/25$ for this example.

The number of warmup iterations used for the HMC-NUTS calculation was 1,000, and the number of samples generated from the Markov chain was 1,500. The total number of steps, i.e., the total number of spectra and their gradients calculated, was approximately 2.1 million, with a computation time of about 80 hours using an NVIDIA/A100 80GB. This corresponds to a computational cost of 0.14 seconds per step. 

Posterior sampling from HMC-NUTS (Figure \ref{fig:WASP39b_corner}) provides the credible intervals for these parameters: $T = 1087^{+73}_{-66}\ \mathrm{K}$, the logarithms of VMRs of $\mathrm{H_2O}$, $\mathrm{CO}$, $\mathrm{CO_2}$, and $\mathrm{SO_2}$ were $-1.09^{+0.13}_{-0.15}, -1.66^{+0.20}_{-0.26}, -2.43^{+0.23}_{-0.22},  -2.92^{+0.19}_{-0.22}$, respectively (68\% credible interval). The VMRs of $\mathrm{H_2S}$, $\mathrm{CH_4}$, $\mathrm{NH_3}$, $\mathrm{HCN}$ and $\mathrm{C_2H_2}$ were unconstrained. By capturing the fine structures caused by water seen in Figure \ref{fig:WASP39b_H2O}, the radial velocity was constrained to $-83 \pm 8 \ \mathrm{km/s}$ (68\% credible interval). This value is consistent with the theoretical prediction of $-87.3 \mathrm{km/s}$ \citep{2023ApJ...955L..19E}.

The inferred elemental abundances and ratios are $\log(\mathrm{O/H})=-1.23^{+0.10}_{-0.17}$, $\log(\mathrm{C/H})=-1.87^{+0.15}_{-0.30}$, $\log(\mathrm{S/H})=-3.11^{+0.15}_{-0.53}$, $\mathrm{C/O}=0.232^{+0.083}_{-0.087}$, and $\mathrm{S/O}=0.0129^{+0.0055}_{-0.0084}$. Therefore, $\mathrm{O/H}$, $\mathrm{C/H}$, and $\mathrm{S/H}$ correspond to $121^{+33}_{-39}$, $46^{+19}_{-23}$, and $58^{+25}_{-41} \times$ solar, respectively, while $\mathrm{C/O}$ and $\mathrm{S/O}$ correspond to $0.39^{+0.14}_{-0.15}$, and $0.48^{+20}_{-31} \times$ solar \citep{2021A&A...653A.141A}. 
In summary, our results indicate that the atmosphere of WASP-39 b exhibits a supersolar metallicity in the range of 20-150 $Z_\mathrm{\odot}$, and is oxygen-rich.

\begin{figure}
    \centering
    \includegraphics[width=\linewidth]{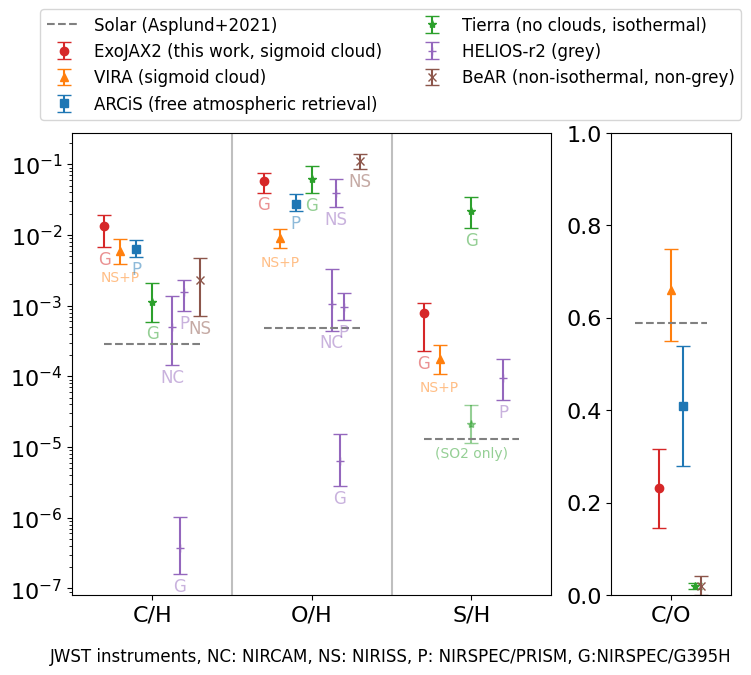}
    \caption{Comparison of C, O, S abundances relative to Hydrogen (left) and C/O with the other retrieval results of WASP-39 b: {\sf VIRA} \citep{2024MNRAS.530.3252C} (sigmoid clouds), {\sf ARCiS} \citep[free atmospheric retrieval][]{2024A&A...685A..64K}, {\sf Tierra} \citep[][isothermal]{2023ApJ...950L..17N}, {\sf Helios-r2} \citep[grey][]{2024A&A...687A.110L}, and {\sf BeAR} \citep{fisher24}. The abbreviations below the error bars indicate the types of JWST instruments used as follows: NC = NIRCam, NS = NIRISS, P = NIRSPEC/PRISM, G = NIRSPEC/G395H. The light green error bars for Tierra represent the S/H value when considering only $\mathrm{SO_2}$, excluding the contribution from $\mathrm{H_2S}$. The solar values are indicated by the dashed lines \citep{2021A&A...653A.141A}. 
    \label{fig:comparison}}
\end{figure}

The observational data of WASP-39 b from JWST's ERS program has been used for atmospheric retrieval in several papers to date. Figure \ref{fig:comparison} compares C/H, O/H, S/H, and C/O values, limited to cases where reference values from previous retrieval studies are provided or can be converted. Recent studies on atmospheric retrieval suggest that the posterior distributions can vary significantly depending on the wavelength range \citep[e.g.][]{fisher24}, instruments \citep[e.g.][]{2023ApJ...950L..17N}, or differences in opacity \citep[e.g.][]{2022ExA....53..447B} used. The purpose of this comparison is not to make claims about WASP-39 b but rather to illustrate how {\sf ExoJAX}2 provides estimation results in relation to previous retrieval studies using the JWST/ERS data for the same target.

First, despite the fact that the results from the retrieval code {\sf VIRA} \citep{2024MNRAS.530.3252C} were obtained using different instruments and settings (NIRISS and NIRSpec/PRISM), which differ in wavelength range and resolution, they generally align with the abundance estimates from {\sf ExoJAX}2. However, only O/H shows a value that is an order of magnitude larger in {\sf ExoJAX}2. This discrepancy is suspected to be due to the resolution of the fine structure of water lines by NIRSpec/G395H's higher spectral resolution (Figure \ref{fig:WASP39b_H2O}), but further analysis is required to confirm this. Next, the results from {\sf ARCiS}'s free atmospheric retrieval \citep[C/O and metalicity; Fig. 3 of their paper][]{2024A&A...685A..64K}, when converted to C/H and O/H, as well as C/O, are consistent with the findings of {\sf ExoJAX}2. However, S/H could not be converted from the literature, and thus it has not been included.
We also compare the results from analyzing only JWST/NIRISS data using the {\sf BeAR} retrieval code \citep{fisher24}. These results show a lower C/H and higher O/H compared to {\sf ExoJAX}2, but the differences are not significant enough to be considered inconsistent.

Regarding {\sf Tierra}, the S/H value differs significantly, and it is important to note that the Tierra model does not account for clouds. The large S/H is attributed to the high abundance of $\mathrm{H_2 S}$. Due to the substantial differences between the atmospheric models (clouds/no clouds) used in this paper and {\sf Tierra}'s model, the values are shown for reference only. Finally, in \cite{2024A&A...687A.110L}, using {\sf Helios-r2}, retrievals were performed for each instrument separately, and the inferred abundances differ by up to four orders of magnitude between instruments. In particular, the posterior from G395H shows significantly lower abundance. Although the data analyzed in this paper differs in binning, such a large discrepancy is unlikely to be explained solely by this difference. The impact of differences between instruments, data processing (binning), and models on retrieval results represents a challenge for future research.

\subsection{High-resolution Reflection Spectroscopy of Jupiter}

\begin{figure}
    \centering
    \includegraphics[width=\linewidth]{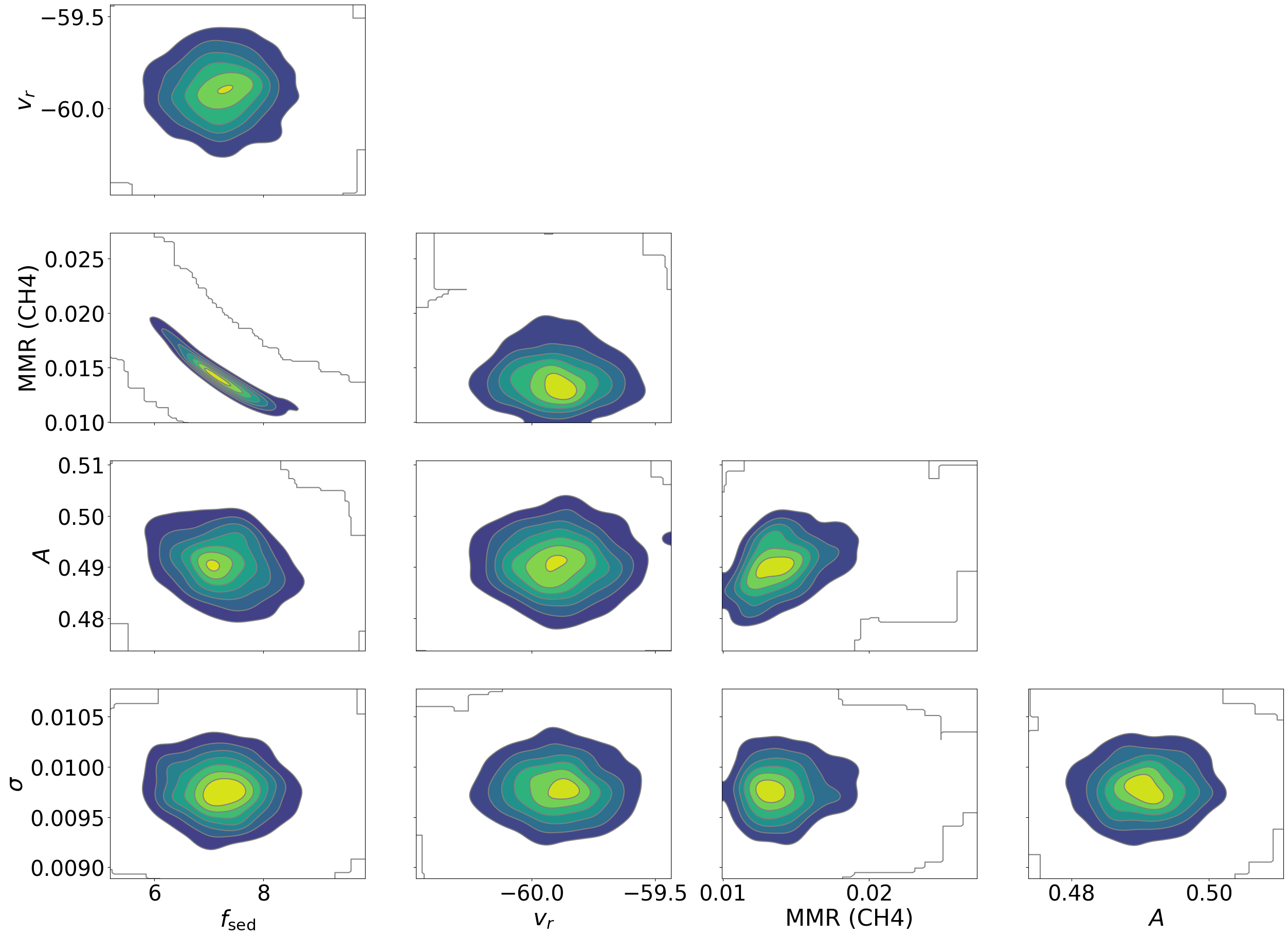}
    \includegraphics[width=\linewidth]{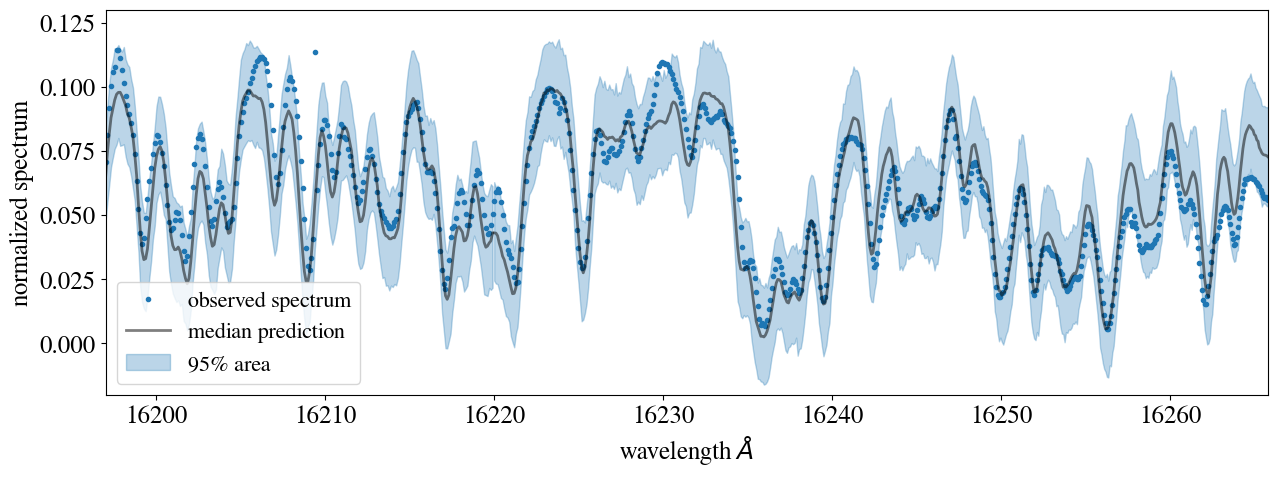}    
    \caption{Top panel: Posterior distribution from reflection spectrum modeling using HMC-NUTS. We have five parameters, $\fsed$, $v_r$ (relative radial velocity), the mass mixing ratio of methane (MMR), the normalization of the spectrum $A$, and the standard deviation of the Gaussian noise $\sigma$. Bottom panel: Jupiter's reflection spectrum (dots) and predictions using HMC-NUTS. The solid line represents the median of the samples, and the shaded region indicates the 95\% credible area.\label{fig:reflection}}
\end{figure}

As an example of analyzing reflection spectra with {\sf ExoJAX}, we present the near-infrared spectrum of Jupiter. This data was obtained by one of the authors (T. Kotani) with petitIRD, InfraRed Doppler \citep{2018SPIE10702E..11K} before its commissioning at the Subaru Telescope, connected to a 20 cm telescope before Subaru.
The data processing was conducted using {\sf PyIRD} (Kasagi et al. in prep\footnote{https://github.com/prvjapan/pyird}). In this dataset, wavelength calibration could not be achieved with high precision due to the insufficient number of Thorium-Argon lines. Therefore, we recalibrated the wavelength using a reflection spectrum model generated by {\sf ExoJAX}2. The residuals of wavelength were modeled using a third-order polynomial, and the coefficients were optimized using {\sf ADAM} based on the residuals from the reflection model. Despite the above corrections, the wavelength calibration was insufficient at the edges of the order. Therefore, in this demonstration, we will analyze the reflected light from Jupiter using a narrow wavelength range (16170–16262 \AA) in a single order. 

The incident solar spectrum was obtained using the disk-integrated high-resolution spectrum (resolution $<$ 0.01 nm)\footnote{http://bdap.ipsl.fr/voscat\_en/solarspectra.html} from \cite{2023RemS...15.3560M}.
The gravity and planetary radius values used were those of Jupiter, with the mean molecular weight assumed to be 2.22. 
The clouds were assumed to be ammonia clouds, and an Ackerman-Marley-like (AM) cloud model from Appendix \ref{ap:clouds} was employed. In this cloud model, the parameters $\Kzz$ (eddy diffusion coefficient), $\fsed$ (the ratio of the sedimentation and
the eddy velocities), and $\sigma_g$ are the model parameters, however, for the reflection spectra in this wavelength range, the dependences of $\Kzz$ and $\sigma_g$ on the spectrum are not significant. We fixed the eddy diffusion coefficient to the observed value of $\Kzz = 10^5 \mathrm{cm^2/s}$  \citep{2005JGRE..110.8001M}.  The $\sigma_g$ was fixed to 2. In other words, under these conditions, $\fsed$ is the only parameter related to the microphysics of the AM cloud model.

The atmospheric model used a layer model that was divided logarithmically into 200 layers, ranging from $10^{-3}$ bar to 30 bar.  However, since the pressure at the base of the ammonia clouds is around 1 bar, the regions at higher pressures do not affect the spectrum \citep{2019ApJ...887..166H}. The temperature-pressure distribution was based on actual measurements from the Cassini and Galileo probes \citep{1998JGR...10322857S}.
Methane was assumed as the only atmospheric absorber, using the CH4 database from the latest one from the ExoMol database \citep[i.e. MM linelist][]{2024MNRAS.528.3719Y}. The velocity components considered were the relative velocity between the observer and Jupiter. Broadening due to Jupiter's rotation and the instrument's resolution was assumed to be Gaussian, with a resolution of R = 25,000.

The prior distributions are as follows: $\log_{10} \fsed \sim \mathcal{U}(0,2)$, the relative radial velocity $v_r \sim \mathcal{U} (-40, -70)$ km/s, $\log_{10}$ MMR (CH4) $\sim \mathcal{U} (-3,-1)$, the spectrum normalization $A \sim \mathcal{U} (0,1)$, the standard deviation of the Gaussian noise $\sigma \sim \mathrm{Exp} (1.0)$.

Figure \ref{fig:reflection} shows the posterior distribution from HMC-NUTS using {\sf ExoJAX} and the predictions for the spectrum. The 5–95\% credible interval for the methane mass mixing ratio is 0.011–0.017, which corresponds to $Z = 2.7–4.1 Z_\odot$ when using the solar abundance from \cite{1989GeCoA..53..197A}. This estimate is consistent with the observed value of approximately $3 Z_\odot$ \citep{2003P&SS...51..105A}. The inverse correlation observed in the posterior distributions of $\fsed$ and the methane mixing ratio can be interpreted as follows: when $\fsed$ increases, the cloud top pressure rises, which increases the optical path length, thereby requiring a lower methane mixing ratio.

\section{Discussion and Summary}

In this paper, we have focused on demonstrating {\sf ExoJAX}2's application through the spectral analysis of actual planets and brown dwarfs. However, {\sf ExoJAX}2 is differentiable from the cross-section level, enabling several additional applications. One such application is the modeling of telluric lines in high-resolution spectroscopy. Moreover, since it allows the calculation of Voigt profiles at user-specified resolutions, it can also be applied to the analysis of molecular line experimental data. These applications will be discussed in future papers. We have also demonstrated only posterior inference using HMC-NUTS. However, we have confirmed that other inference methods included in JAX-compatible PPLs and packages, such as Stochastic Variational Inference (SVI) in {\sf NumPyro} {, normalizing-flow enhanced Markov chain Monte Carlo \citep{Wong:2022xvh}, and nested sampling using  {\sf JAXNS} \citep{2020arXiv201215286A} }, can be directly applied to {\sf ExoJAX}2's spectral model. How {\sf ExoJAX}2 can be further utilized in DP-based Bayesian inference methods like HMC-NUTS and SVI remains a topic for future research.

{Although {\sf ExoJAX 2} incorporates a number of innovations -- such as GPU-accelerated computations, pre-caching of molecular line lists (pre-MODIT), autodifferentiable implementations of radiative transfer, code porting to JAX, and advanced sampling algorithms (e.g., HMC-NUTS) -- we acknowledge that the framework, in its present form, does not yet provide fully self-consistent atmospheric modeling; a model that performs radiative transfer calculation over a bolometric wavelength range and incorporates consistent temperature -- pressure profiles, coherent cloud formation and dissipation models, and either chemical equilibrium or non-equilibrium processes. Nonetheless, these current developments represent an important step toward bridging the traditional gap between physically robust forward-modeling (historically tied to computationally expensive grids) and rapid retrieval algorithms. Our ongoing work aims to refine the physical modules within {\sf ExoJAX} and integrate them seamlessly, with the ultimate goal of delivering an end-to-end retrieval framework that achieves both computational efficiency and the highest level of physical self-consistency possible.}

In this paper, we have significantly enhanced the functionality of the differentiable spectral model for exoplanets and brown dwarfs, {\sf ExoJAX}. Specifically, we introduced a new opacity calculation algorithm, PreMODIT, which drastically reduces GPU device memory usage. Additionally, we extended the radiative transfer scheme, which was previously limited to pure absorption spectra, to support transmission spectroscopy and a two-stream approximation for scattering and reflected light. We used {\sf ExoJAX}2 to model the high-resolution emission spectrum of the benchmark T-type brown dwarf Gl 229B, the medium-resolution transmission spectrum of the hot Saturn WASP-39b observed by JWST, and Jupiter's reflection spectrum using HMC-NUTS, demonstrating retrieval through differentiable programming.

We thank Michael Gully-Santiago, Masayuki Kuzuhara, Ken Osato, Erwan Pannier, Benjamin Pope, Ji Wang, Jason Wang, Shota Miyazaki, and Dirk van den Bekerom for the fruitful discussions.
 This study was supported by JSPS KAKENHI grant nos. 21H04998, 23H00133, 23H01224 (H.K.), 21K13984, 22H05150, 23H01224 (Y.K.), 23K25920 (T.K.), 24H00242 (M.T.), 22K14092 (S.K.N), The Mitsubishi Foundation (202310018), and the PROJECT Research from the Astrobiology Center (AB0505).

\software{RADIS \citep{pannier2019radis}, Vaex \citep{2018A&A...618A..13B}, ExoJAX \citep{2022ApJS..258...31K}, PDR \citep{m_st_clair_2024_12115247}, matplotlib, numpy, JAX \citep{jax2018github}, NumPyro \citep{phan2019composable}, JaxOpt \citep{jaxopt_implicit_diff}}

\appendix

\section{Opacity Calculation Errors}
\subsection{Interpolation of Tabulated Cross-Section}\label{ap:taberror}

\begin{figure*}
    \centering
    \includegraphics[width=0.49\linewidth]{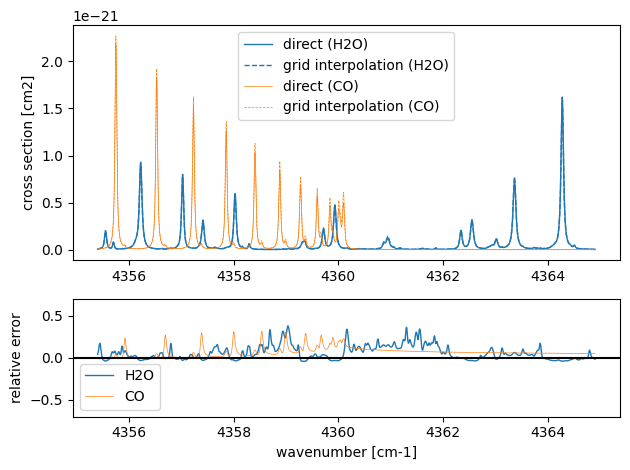}
    \includegraphics[width=0.49\linewidth]{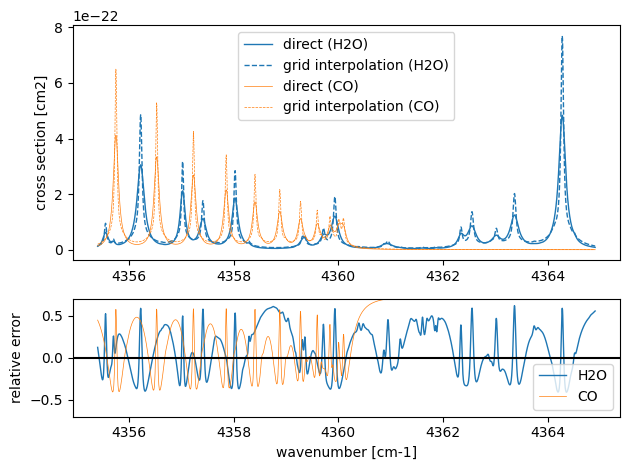}
    \caption{Comparison between the directly calculated cross-section (solid) and the grid interpolation (dashed; average value) (top panel) and the residual, Grid interpolation/Direct -1 (bottom panel). Left: interpolation of temperature with constant pressure (1 bar), right: interpolation of pressure with constant temperature (899.5K).}
    \label{fig:tablexs_error}
\end{figure*}

We investigate the error in the cross-section obtained by interpolating precomputed cross-section data on a two-dimensional grid of temperature and pressure. Here, we assume that cross-section data is provided for two adjacent points at temperatures $T_0 = 899.5K$ and $T_1 = 1215.1K$. Using the linear midpoint temperature ($T_c = 1057.3$ K) we create interpolated data by averaging the cross-sections at these two temperatures. These two points are part of the actual high-resolution cross-section grid provided by the petitRADTRANS website (June 10th, 2024). We compare this interpolated data with the cross-section directly calculated at the actual logarithmic midpoint temperature. The pressure is set to 1 bar. This comparison is shown on the left panel of Figure \ref{fig:tablexs_error}. The results remain similar even when the average is taken at the logarithmic temperature ($T_c=$1045.5K). Similarly, we conduct a comparison for two points on the pressure grid, \(P_0 = 1 \text{ bar}\) and \(P_1 = 10 \text{ bar}\) from petitRADTRANS. The evaluation point is at the logarithmic midpoint, $P_c = 10^{0.5} \text{ bar}$, with the temperature fixed at 899.5K. The results are shown in on the right panel of Figure \ref{fig:tablexs_error}

The error observed in the temperature-direction interpolation is mainly due to the error in line strength. This error persists even when the wavelength resolution is reduced. On the other hand, the error in the pressure-direction interpolation is primarily due to the error in the line profile caused by pressure broadening, which can be canceled out by lowering the wavelength resolution. However, this could cause systematic errors in analyses that utilize line profile information.

\subsection{Temperature Grid-based Line Density}\label{ap:tgrid}
Instead of $E$, we may use the temperature grid of $\mathfrak{S} (T)$ in Equation (\ref{eq:sslog}). We here ignore $\nu$ and $\bpar$ for simplicity.  In this case, we integrate the values over $\Elow$ in the host and send the discretized density $\mathfrak{S} (T_i)$ to the device. Let us estimate the error of the line density using the temperature grid method. We use inverse temperature $t = 1/T$ and use the line strength at $t_1$ and $t_2$ ($t_2 > t_1$), that is $S(t_1)$ and $S(t_2)$. We consider approximating the line strength at the midpoint of $t_1$ and $t_2$, that is, $t^\prime = (t_1+t_2)/2$. The recovered line strength by the linear interpolation is the mean of $S(t_1)$ and $S(t_2)$, $\tilde{S} (t^\prime) = (S(t_1) + S(t_2))/2$. The difference from the true value $S(t^\prime)$ is expressed as a function of the interval $\Delta t = t_2 -t_1$
\begin{eqnarray}
    \delta S(t^\prime) &=& \frac{  \tilde{S} (t^\prime) }{S(t^\prime)} - 1 
    \approx (c_2 \Elow \Delta t)^2/2
\end{eqnarray}
where we ignore $q_r(T)$ and $g(\hat{\nu},T)$ and only considered $f(\Elow, T)$ as a source of the difference. If we admit 1 \% as the difference, we need to satisfy $
    \Elow \Delta t \approx 0.1 \,\, \mathrm{cm^{-1} K^{-1}}
$.

As shown in Figure \ref{fig:elower_distribution}, we computed the $\Elow$ distribution from several molecular databases. The maximum values of $\Elow$, are $1 - 3 \times 10^4$ for methane, water, and carbon monoxide. Assuming $\Elow = 10^4$ and $t_l - t_h =10^{-3} \mathrm{K}^{-1}$ (for instance, $T \ge 1000$K), we obtain the number of the inverse temperature grid required for the 1 \% accuracy as $n_t \sim (t_l - t_h)/\Delta t = 10^2$ and  $\Elow = 3 \times 10^4$ and $t_l - t_h =3 \times 10^{-3} \mathrm{K}^{-1}$  (for instance $T \ge 300K$) as $ n_t \sim 10^3$, where $t_l = 1/T_l$ and $t_h=1/T_h$ are the inverse of the minimum and maximum values of the temperature range, $T_l$ and $T_h$. Thus, we need a hundred to a thousand temperature grids of the line density to keep 1 \% accuracy. 

The above discussion is also for interpolating the line strength at $T$ using a temperature grid-based line density. If the cross-section itself is tabulated as a function of discrete temperature and one interpolates the values from the tabulated cross-section given temperature and pressure, the accuracy will be worse than in the above case because the line broadening also depends on temperature. We investigated the error from the interpolation of the tabulated cross-section data in Appendix \ref{ap:taberror}.

\subsection{Error of The Line Strength Reconstruction from the LBD}\label{ap:err_line_intensity}

We assume $E_l = E_1 + \Delta E/2 = E_2 - \Delta E/2$. Then, 
\begin{eqnarray}
w_{1,l}(t) &=& 
\frac{1 - e^{-c_2 \frac{\Delta E}{2} (t - \tref)}}{e^{-c_2 \frac{\Delta E}{2} (t - \tref)} - e^{c_2 \frac{\Delta E}{2} (t - \tref)}} \\
w_{2,l}(t) &=& 
\frac{e^{c_2 \frac{\Delta E}{2} (t - \tref)}-1}{e^{-c_2 \frac{\Delta E}{2} (t - \tref)} - e^{c_2 \frac{\Delta E}{2} (t - \tref)}},
\end{eqnarray}
where $t = 1/T$, $\tref = 1/\Tref$.
These weights depend on $\Delta E$ but not on $E_1$ nor $E_2$.
The relative error to the exact line strength for the first-order approximation is analytically given by 
\begin{eqnarray}
\label{eq:error_line_s}
\Delta_S (T; \Delta E) &=& \frac{\tilde{S}(T)}{S(T)} -1 = \epsilon^{(0)} (t, \Delta E) + \epsilon^{(1)} (t, \Delta E) - 1\\
\epsilon^{(0)} (t, \Delta E) &=& \displaystyle{ \frac{e^{c_2 \frac{\Delta E}{2} (t - \tref)}\left[1-e^{-c_2 \frac{\Delta E}{2} (\ttyp - \tref)}\right] - e^{-c_2 \frac{\Delta E}{2} (t - \tref)} \left[1-e^{c_2 \frac{\Delta E}{2} (\ttyp - \tref)}\right]}{e^{-c_2 \frac{\Delta E}{2} (\ttyp - \tref)} - e^{c_2 \frac{\Delta E}{2} (\ttyp - \tref)}} }\\
\epsilon^{(1)} (t, \Delta E) &=& \frac{\left[ e^{c_2 \frac{\Delta E}{2} (t - \tref)} - e^{-c_2 \frac{\Delta E}{2} (t - \tref)} \right] \left[ 2 - e^{-c_2 \frac{\Delta E}{2} (\ttyp - \tref)} - e^{c_2 \frac{\Delta E}{2} (\ttyp - \tref)}\right]}{\left[e^{-c_2 \frac{\Delta E}{2} (\ttyp - \tref)} - e^{c_2 \frac{\Delta E}{2} (\ttyp - \tref)}\right]^2} \frac{c_2 \Delta E}{2} (t - \ttyp),
\end{eqnarray}
where $\ttyp = 1/\Ttyp$.

\subsection{Taylor Expansion of the weight}
In Section \ref{ss:discretization}, the LBD was approximated at the value of $T = \Ttyp$ (Equation \ref{eq:lbd_s}). Furthermore, a more accurate approximation can be achieved by performing a Taylor expansion around $\ttyp = 1/\Ttyp$ as a function of $t = 1/T$. 
\begin{eqnarray}
\label{eq:lbd_ttyp}
{\mathfrak{L}}_T (E_h) =  \sum_l {w}_{1,l} (t) S_l \delta(E_h, E_{1,l}) + {w}_{2,l} (t) S_l \delta(E_h, E_{2,l}) 
\approx \sum_{n=0}^N \frac{1}{n!} \frac{\partial^n}{\partial t^n} {\mathfrak{L}}_{\Ttyp} (E_h) (t - \ttyp)^n
\end{eqnarray}
where we used the Taylor approximation
\begin{eqnarray}
w_{i,l}(t) \approx \sum_{n=0}^N \frac{1}{n!} w_{i,l}^{(n)}(\ttyp)  (t-\ttyp)^n 
\end{eqnarray}
for $i = 1 \mbox{\,or\,} 2$ and ${w}_{i,l}^{(n)}(\ttyp) \equiv \partial^n w_{i,l}(t)/\partial t^n |_{t=\ttyp}$ and
\begin{eqnarray}
 \frac{\partial^n}{\partial t^n}{\mathfrak{L}}_{\Ttyp} (E_h) \equiv \sum_l w_{1,l}^{(n)} (\ttyp) S_l \delta(E_h, E_{1,l}) + w_{2,l}^{(n)} (\ttyp) S_l \delta(E_h, E_{2,l})  \nonumber \\ 
\end{eqnarray}
is the $n$-th derivative of the LBD at $T=\Ttyp$ by $t$.
It should be noted that we replaced the argument of the weight functions from $T$ to $t$. 

This implementation is available in {\sf ExoJAX}2 as the ``diffmode'' option, i.e., diffmode $= N$. Setting $N = 1$ or $N = 2$ allows for additional memory compression. However, since computational cost increases with $N$, the default is set to $N = 0$.

\section{Overlap and Add Convolution}\label{ap:ola}

The overlap and add (OLA) has an advantage in the case where the length of the signal is significantly greater than that of the finite impulse response (FIR) filter $M \equiv |\fv|$. The OLA algorithm divides the signal $\xv$ into $n$ segments of length $L$. 
This manipulation can be regarded as reshaping of the signal $\xv$ to the signal matrix $X \in \mathcal{R}^{n \times L}$, where we assume that the signal is divisible by $n$. The OLA algorithm extends the FIR filter $\xv \in \mathcal{R}^M$ to $\hat{\xv} \in \mathcal{R}^{L+M-1}$ and the signal matrix $X$ to $\hat{X} \in \mathcal{R}^{n \times (M + L -1)}$ by zero-padding. The convolution is applied to the column of $\hat{X}$ and $\hat{f}$ as 
\begin{eqnarray}
g_{i} = {\bf \hat{X}}_{i} \ast \hat{f} = \mathrm{IRFFT} [ \mathrm{RFFT}({\bf \hat{X}}_{i}) \, \mathrm{RFFT}(\hat{f}) ],
\end{eqnarray}
where $ \mathrm{RFFT}$ and $ \mathrm{IRFFT}$ are the real fast Fourier transform and its inverse. The convolution of $\xv$ and $\fv$ is computed by the overlap--and--add manipulation. The last manipulation generates the convolved vector of the length $n L + M - 1$ as the final output. The indexing part of the last manipulation has difficulties in implementation by {\sf JAX} because the index slicing is not as straightforward as it is done in pure Python. We used {\sf dynamic\_update\_slice} and {\sf scan} in {\sf jax.lax} to implement the overlap--and--add manipulation.

\section{Atmospheric layer model}\label{ap:layer_height}

\subsection{constant interval layer in the log-pressure}\label{ap:s:logp}

The default pressure layer model of {\sf ExoJAX}2 is composed of layers that are evenly divided along the logarithmic pressure axis. If we define $ q_n \equiv \log_{10} P_n$ (bar) as the logarithmic pressure, the logarithmic pressure difference of the layers $\Delta q$ is constant, $q_{n+1} = q_n + \Delta q$, and the difference in pressure between the upper and lower ends of each layer can be expressed as follows. 
\begin{align}
    \Delta P_n &= \underline{P}_n - \overline{P}_n = 10^{q_n + \Delta q/2} - 10^{q_n - \Delta q/2} = ( k^{-1/2} - k^{1/2}) P_n,
\end{align}
where $k \equiv 10^{- \Delta q}$ is the pressure decrease rate, i.e. $P_n = k P_{n+1}$.
This corresponds to the case where $ P_n $ is defined as the midpoint of the logarithmic pressure in the $n$-th layer. In general, by defining $ P_n $ as the point that corresponds to a proportion $ r $ (reference point, $p=0.5$ for the midpoint) in the logarithmic pressure coordinates, we obtain Equations (\ref{eq:pressure_overline}, \ref{eq:pressure_underline},\ref{eq:pressure_delta}).
The bottom pressure of all of the layers is given by $P_B \equiv \underline{P}_{N-1} = k^{r-1} P_{N-1}$.

\subsection{non-constant gravity}
We suppose all quantities with the unit of physical length are normalized by $R_0$. We start from the lowest layer $n=N-1$, with the boundary conditions of the radius $\underline{r}_{N-1}$ and the gravity $\underline{g}_{N-1} = G M_p/R_0^2$ at $\underline{r}_{N-1} = 1$. We ignore the atmospheric mass. 
Given the radius at the lower boundary of the $n$-th layer, $\underline{r}_n$, we obtain the analytic expression of the $n$-th layer height assuming a uniform temperature in the $n$-th layer of $T_n$ as 
\begin{align}
    \Delta h_n &=\underline{r}_{n-1} - \underline{r}_{n} = \underline{r}_{n} \left[ \left( 1 + \frac{\underline{H}_n}{\underline{r}_{n}} \log{k} \right)^{-1} - 1 \right]\\
    \underline{H}_n &\equiv \frac{k_B T_n}{\mu m_H \underline{g}_n} \\
    \underline{g}_n &= \underline{g}_{N-1}/\underline{r}_n^2
\end{align}
by integrating the hydrostatic equilibrium
\begin{align}
    dr = - \frac{k_B T_n}{\mu m_H g(r)} \frac{dP}{P} = \frac{k_B T_n}{\mu m_H \underline{g}_n} \frac{r^2}{\underline{r}_n^2 }\frac{dP}{P} = \underline{H}_n \frac{r^2}{\underline{r}_n^2 }\frac{dP}{P} 
\end{align}
from $r=\underline{r}_{n-1}$ to $r=\underline{r}_n$ for $dr$ and $P=\underline{P}_{n-1}$ to $P=\underline{P}_n$ for $dP$. Note that the isothermal temperature assumption simply yields 
\begin{align}
    \underline{r}_n = \underline{r}_{N-1} \left( 1 + \frac{\underline{H}_{N-1}}{\underline{r}_{N-1}} n \log{k} \right)^{-1} 
\end{align}

\subsection{Effective Transmission and Source (ETS) Treatment}\label{ap:ett_lart}
In Section \ref{ss:flux-adding}, using the equations for \( F^+ \) and \( F^- \), we derived the flux-adding treatment, which considers the transfer between layers as the sum of effective {\it reflection} and {\it sources}. In this section, we will first derive the equation for \( F^+ \) only and then describe a method that considers the transfer of \( F^+ \) between layers as the sum of effective {\it transmission} and {\it sources}. This form is a natural extension of the flux-based transfer for the pure absorption case, as used in {\sf ExoJAX}1.

\cite{1989JGR....9416287T} developed the tridiagonal form of radiative transfer. They transformed the two-stream equations (\ref{eq:2stream_1}, \ref{eq:2stream_2}) into a tridiagonal system, focusing on the coefficients of the equations post-conversion, in order to circumvent numerical instability. In contrast to their approach, which emphasizes the coefficients, we directly derive the tridiagonal system for $ F^+_n $. 
The primary advantage of formulating the tridiagonal system for $ F^+_n$ is twofold: first, it allows for the solution of $\Nlayer$ sets of the tridiagonal system as opposed to the $2 \Nlayer$ required by the coefficient-based approach; second, it simplifies the process of connecting the solution to the flux in the context of auto-differentiation.

Using Equations (\ref{eq:twosq1}, \ref{eq:twosq2}), we obtain a tridiagonal equation, 
\begin{eqnarray}
\label{eq:tridiag}
- a_n F^+_{n+1} + b_n F^+_n - c_{n-1} F^+_{n-1} &=& d_n  \mbox{ for $n=1,2,\cdots,N-2$}
\end{eqnarray}
or the matrix form of the tridiagonal system including the boundary conditions can be written as 
\begin{eqnarray}
\label{eq:completetri}
\begin{bmatrix}
b_0 & - a_0 & 0 & \cdots & 0 \\
- c_0 & b_1 & - a_1 & \cdots & \vdots \\
0 & \vdots & \ddots &  & \vdots \\
\vdots & \vdots & - c_{N-2} & b_{N-1} & - a_{N-1} \\
0 & \cdots & 0 & - c_{N-1} & b_{N} \\
\end{bmatrix} 
\begin{bmatrix}
F^+_0 \\
F^+_1 \\
\vdots \\
F^+_{N-1} \\
F_{B} \\
\end{bmatrix}
=
\begin{bmatrix}
d_0 \\
d_1 \\
\vdots \\
d_{N-1} \\
d_{N} \\
\end{bmatrix}
\nonumber \\
\end{eqnarray}
where
$a_n = \mathcal{S}_{n-1} \mbox{ ($n=1,2,\cdots,N-2$)}, b_n = {\mathcal{r}_n ( \mathcal{T}_{n-1}^2 - \mathcal{S}_{n-1}^2) +  \mathcal{r}^-_{n}}, c_n = \mathcal{r}_{n+1} \mathcal{T}_n \mbox{  ($n=0,1,\cdots,N-3$)}, d_n = - \mathcal{S}_{n-1} \pi G_n^+ + \mathcal{r}_n ( \mathcal{T}_{n-1}^2 - \mathcal{S}_{n-1}^2 ) \pi G_{n-1}^+ + {\mathcal{S}_{n-1} \mathcal{r}_n} \pi G_{n-1}^-,$
$\mathcal{r}_n \equiv \mathcal{S}_n/\mathcal{T}_n$, and $ \mathcal{r}^-_n \equiv \mathcal{S}_{n-1}/\mathcal{T}_n$. For the isothermal layer, $d_n$ reduces to ${\mathcal{r}^-_n} \pi \hat{\mathcal{B}}_n -  {\mathcal{r}_{n}}(\mathcal{T}_{n-1} - \mathcal{S}_{n-1}) \pi \hat{\mathcal{B}}_{n-1}$, 
where $\hat{\mathcal{B}}_n \equiv (1 - \mathcal{T}_{n} - \mathcal{S}_{n} ) \mathcal{B}_n$.

In solving the tridiagonal equation, one can utilize the Thomas algorithm, which offers a computational complexity of $O(N)$. To bridge the radiative transfer equations with the pure emission limit, we re-express the tridiagonal equation, as given in Eq. (\ref{eq:tridiag}), in the {\sf ExoJAX}1 form (Section 3 in Paper I). This transformation is accomplished using the inductive method. It's worth noting that the effective transmission form can be viewed as the Gauss elimination process specifically tailored for tridiagonal systems.

Let us assume the recurrence relation is expressed as
\begin{align}
\label{eq:induc}
    F_n^+ = \hat{T}_n F_{n+1}^+ + \hat{Q}_n. 
\end{align}
Computing $\hat{T}_n \times $ (\ref{eq:tridiag}) - $a_n \times$ (\ref{eq:induc}), we obtain
\begin{eqnarray}
\label{eq:recursive_f}
F_{n-1}^+ = \frac{b_n \hat{T}_n - a_n}{c_{n-1} \hat{T}_n} F_{n}^+ + \frac{a_n \hat{Q}_n - \hat{T}_n d_n}{c_{n-1} \hat{T}_n}
\end{eqnarray}

By assigning the first and second coefficients to $\hat{T}_{n-1}$ and $\hat{Q}_{n-1}$, respectively, we can formulate the recurrence relation as follows
\begin{align}
    \label{eq:gausselim_T}
    \hat{T}_{n} &= \frac{a_n}{b_n - c_{n-1} \hat{T}_{n-1}} \\
    \label{eq:gausselim_Q}
    \hat{Q}_{n} &= \frac{d_n + c_{n-1} \hat{Q}_{n-1}}{b_n - c_{n-1} \hat{T}_{n-1}} 
\end{align}
These quantities can be computed once the initial conditions $\hat{T}_0$ and $\hat{Q}_0$ are determined\footnote{In the lower atmosphere, the transmittance approaches zero. Therefore, computing the recursive formula from the bottom layer upward ($\hat{T}_{n-1} = (b_n \hat{T}_n - a_n)/(c_{n-1} \hat{T}_n)$) results in numerical instability. }.

For the emission spectrum with no reflection of the external stellar light, the boundary condition $F_-(0) = 0$ with (\ref{eq:twosq1}) yields
\begin{eqnarray}
F_0^+ = \mathcal{T}_0 F_1^+ - \mathcal{T}_0 \pi G_0^+.
\end{eqnarray}
We obtain the initial condition as
    $\hat{T}_0 = \mathcal{T}_0$ and
    $\hat{Q}_0 = - \mathcal{T}_0 \pi G_0^+$
or equivalently we can set 
$a_0 = \hat{T}_0 = \mathcal{T}_0$ and
$d_0 = \hat{Q}_0 = - \mathcal{T}_0 \pi G_0^+$
when we set $b_0 = 1$.

For the isothermal layer, we obtain
\begin{align}
\label{eq:boundary_isothermal}
\hat{Q}_0 &=  \frac{{{\zeta^+_0}} -{{\zeta^-_0}} }{{\zeta^+_0}^2  - (\zeta^-_0 \mathsf{T}_0)^2 } [ \zeta^+_0 ( 1 - \mathsf{T}_0) - \zeta^-_0 \mathsf{T}_0 ( 1- \mathsf{T}_0) ] \pi \mathcal{B}_0. 
\end{align}
If the top layer is taken to be sufficiently high in the atmosphere, then $\mathsf{T}_0$ can be set to 1, resulting in $\hat{Q}_0 = 0$.

Replacing $F^+_{N}$ to $F_B$, the bottom boundary is expressed as 
\begin{align}
    F^+_{N-1} = \hat{T}_{N-1} F_B + \hat{Q}_{N-1}.
\end{align}
For the surface flux is dominated by the thermal emission (thermal surface), we assume
    $F_B = B(T_B)$ (thermal surface),
while we assume 
    $F_B = 0$ (no surface)
for the gaseous planets but the bottom layer should be deep enough ($\hat{T}_0 \hat{T}_1 \cdots \hat{T}_{N-1} \approx 0$).

The upward flux at the ToA is computed by 
\begin{eqnarray}
\label{eq:recur}
F_0^+ &=&  \hat{T}_0 ( \hat{T}_1 ( \hat{T}_2 (\cdots\hat{T}_{N-2} (\hat{T}_{N-1} F_B + \hat{Q}_{N-1} ) + \hat{Q}_{N-2}) + \cdots + \hat{Q}_2) + \hat{Q}_1) + \hat{Q}_0.
\end{eqnarray}

Same as Equation (\ref{eq:rerr_tq}), defining cumulative transmission as
$\tcv = (1, \hat{T}_0,  \hat{T}_0 \hat{T}_1,\cdots,\hat{T}_0 \hat{T}_1 \cdots \hat{T}_{N-2}, \hat{T}_0 \hat{T}_1 \cdots \hat{T}_{N-1})^\top$ and the source vector $\qv \equiv (\hat{Q}_0, \hat{Q}_1, \cdots , \hat{Q}_{N} )^\top$, we obtain
\begin{eqnarray}
\label{eq:rerr_tq}
F_0^+ = \tcv \cdot \qv = \sum_{i=0}^{N} w_i, 
\end{eqnarray}
where $w_i = \phi_i q_i$ can also be interpreted as the contribution function from the $i$-th layer for $i \le N-1$ and the surface ($i=N$) to the planet spectrum for the scattering case.


For the two-stream case, Equation (\ref{eq:gausselim_Q}) can be approximated under the pure absorption limit as
\begin{align}
    \hat{Q}_n &\approx \frac{\pi \hat{B}_n - (\mathcal{T}_{n-1} - \mathcal{S}_{n-1} + \mathcal{T}_{n-1} \hat{Q}_{n-1})}{ \mathcal{T}_{n-1}^2 - \mathcal{S}_{n-1}^2 + 1 - \mathcal{T}_{n-1} \tilde{T}_{n-1}} \approx \pi \hat{B}_n - \mathcal{T}_{n-1} (\hat{Q}_{n-1} - \pi \hat{B}_{n-1})
\end{align}
We assume that $\mathcal{S}_{n} \approx \mathcal{S}_{n-1}$ for the first line, $\mathcal{S}_{n-1} \ll \mathcal{T}_{n-1}$ and $\tilde{T}_{n-1} \approx \mathcal{T}_{n-1}$, and $\mathcal{S}_{n-1} \ll 1$ for the second line. Then we find
\begin{align}
\label{eq:pureabs_Q}
    \hat{Q}_{n} \to \pi \hat{B}_n = (1 - \mathcal{T}_n - \mathcal{S}_n) \pi B_n \to (1 - \mathsf{T}_n) \pi B_n \mbox{\,\,\, for the pure absorption limit}.
\end{align}
The above is the same formulation of the radiative transfer adopted in {\sf ExoJAX}1. We also confirmed that the effective transmission form also numerically converges to Equation (\ref{eq:pureabs_Q}) when the single-scattering albedo is very small.

For the boundary condition of the pure absorption limit ($\zeta^+_0 \to 1, \zeta^-_0 \to 0$), Equation (\ref{eq:boundary_isothermal}) reduces to 
\begin{align}
\hat{Q}_0 \to ( 1- \mathsf{T}_0) \pi B_0.
\end{align}

\section{Cloud Opacity Model}\label{ap:clouds}

\subsection{Cloud Model}
As this section outlines, {\sf ExoJAX}2 implements a cloud model roughly based on \cite{ackerman2001precipitating}. Consider clouds formed by the phase transition of a single-component gas to solid or liquid. Clouds form when the gas undergoes a phase transition to solid or liquid above a certain altitude. Although the formation and dissipation of clouds involve complex physics, we will simply define the cloud base as the height where the vapor pressure curve matches the partial pressure. Specifically, if the volume mixing ratio of the gas component is \(\xi_v(P)\), then the cloud base is defined by the \((P, T)\) that satisfies $ P(T) = P_\mathrm{sat}(T)/\xi_v(P)$.
where $P_\mathrm{sat}(T)$ is the vapor pressure and $\xi_v$ is the vapor volume mixing ratio.
The vapor pressure \(P_\mathrm{sat}(T)\) can be expressed using the Clausius-Clapeyron equation with the latent heat \(l\) as
$P_\mathrm{sat}(T) = P_\mathrm{sat,0} e^{-l/RT}$.

For simplicity, we here consider a single molecule that condenses to form clouds. For example, clouds formed by water in the atmosphere exist in two states: the vapor state ($\vapor$) and the condensed state ($\cond$). The total amount of these two states is denoted by $\total$. To avoid confusion, we consider the mixing ratio as the mass mixing ratio instead of the volume mixing ratio\footnote{The volume mixing ratio, or number mixing ratio, requires a definition of how to count one unit. This could either be the number of molecules in the entire cloud or the sum of the number of cloud particles of different sizes (each containing different numbers of molecules).}. The total mass mixing ratio \(X_\total (z)\) is the sum of the mass mixing ratio of the condensed phase \(X_\cond (z)\) and the mass mixing ratio of the gaseous phase \(X_\vapor (z)\), $X_\total (z) = X_\cond (z) + X_\vapor (z)$.

Considering representative-sized cloud particles, the balance of vertical transport and the sedimentation of condensed particles due to precipitation can be described by the following equation:
\begin{align}
\label{eq:AM01}
- \Kzz (z) \frac{\partial}{\partial z} {X_\total (z)} - \overline{v}_f (z) X_\cond (z) = 0
\end{align}
where \(\Kzz (z) \) is the vertical eddy diffusion coefficient in units of \(\mathrm{cm^2/s}\), and \(\overline{v}_f(z)\) is the typical sedimentation velocity of the cloud particles. To solve equation (\ref{eq:AM01}), we need another constraint for the three states of \(X\). Let's assume $X_\cond (z)/X_\total (z) = \mathrm{constant} \equiv k_c$. With this assumption, the differential equation (\ref{eq:AM01}) becomes
\begin{align}
\label{eq:AM01c}
  \frac{\partial}{\partial z} {X_\cond (z)} = - \frac{k_c \overline{v}_f (z)}{K_{zz}(z)} X_\cond (z).
\end{align}
From equation (\ref{eq:AM01c}), we see that if we can assume \(\overline{v}_f (z) \propto \Kzz (z)\), the differential equation becomes easier to solve. In particular, if we assume that the sedimentation velocity \(\overline{v}_f(z)\) is proportional to the vertical eddy velocity scale \(v_\mathrm{eddy}(z) = \Kzz (z)/L\), where \(L\) is a typical convective scale and is assumed to be constant, we can write \(\overline{v}_f(z)\) as
\begin{align}
\overline{v}_f(z) = \fsed v_\mathrm{eddy}(z) = \fsed \frac{\Kzz(z)}{L} \quad \text{ (the \(\fsed\) assumption)},
\end{align}
where $\fsed$ is the proportional constant. With this assumption, the solution above the cloud base becomes
\begin{eqnarray}
\label{eq:AM01cx}
X_\cond (z) = X_\cond(0) \exp{\left( - \frac{ k_c \fsed}{L} z \right) }.
\end{eqnarray}
This equation describes the vertical mass distribution of the condensates above the cloud base.

Assuming \(L = H(z=0) = H_0\) (pressure scale height at $z=0$) and that the relation between \(z\) and \(P\) is determined by the hydrostatic equilbrium, Equation (\ref{eq:AM01cx}) can be rewritten as:
\begin{align}
X_\cond(P) = 
\begin{cases}
X_\cond(0) \left( \frac{P}{P_0} \right)^{f_{sed}/k_c} & P \le P_0 \\
0 & P > P_0
\end{cases}
\end{align}
where \(P = P_0\) represents the pressure at the cloud base. This form expresses the mass mixing ratio of the cloud particles as a function of pressure above the cloud base.

Let's introduce a size distribution model for cloud particles. The number density of cloud particles per unit radius is defined as
$q(r) \equiv \frac{d \mathcal{N}_\cond (r)}{d r}$, where \(\mathcal{N}_\cond(r)\) is the number density of particles with a radius less than or equal to \(r\). Thus,
\[
\int_0^\infty q(r) \, dr = \int_0^\infty \frac{d \mathcal{N}_\cond (r)}{d r} \, dr = \mathcal{N}_\cond (\infty) \equiv N
\]
gives the total number density of cloud particles. Note that we use a different symbol \(n\) for the number density of gas molecules because this does not count the number of molecules constituting the cloud particles. If \(n_\cond\) denotes the number density of molecules in cloud particles, it relates to the mass mixing ratio as
$X_\cond = \rho_\cond/\rho = (\mu_\cond/\mu) (n_\cond/n)$,
where \(\mu_\cond\) is the molecular weight of the molecules constituting the cloud particles, \(\rho_\cond\) is the mass density of the cloud particles in the atmosphere (in \(\mathrm{g/cm^3}\)), and \(\rho\) is the total mass density of the atmosphere. Note that \(\rho_\cond\) should not be confused with \(\delta_\cond\), which is the mass density of the cloud particles themselves (also in \(\mathrm{g/cm^3}\)).

If \(m_\cond\) is the mass of an individual cloud particle, then \(\displaystyle{\frac{dm_\cond}{dr}}\) is the mass distribution function of cloud particles with respect to their radius. Assuming the cloud particles are spherical, we have

\[
\frac{dm_\cond}{dr} = \frac{4}{3} \pi r^3 \delta_\cond \frac{d \mathcal{N}_\cond (r)}{d r} = \frac{4}{3} \pi r^3 \delta_\cond q(r)
\]

The eddy diffusion coefficient and the sedimentation velocity \(v_f(z; r)\) for each particle size \(r\) can be given separately. However, if given separately, the \(\fsed\) assumption generally does not hold. The strongest assumption to make the \(\fsed\) assumption hold is to assume for each size as $v_f(z; r) = \fsed v_\mathrm{eddy}(z) = \fsed {\Kzz(z)}/{L}$. However, this does not account for the size dependence of the sedimentation velocity. The original equation (\ref{eq:AM01}) does not necessarily apply to each particle size but holds for some representative value. Thus, applying the \(\fsed\) assumption involves imposing it on a representative operation. Now, let's consider the representative operation for the sedimentation velocity as the mass-weighted average. Thus,
\begin{eqnarray} 
\label{eq:fsedw}
\overline{v}_f(z) = \frac{\int v_f(m_\cond) dm_\cond}{M_c(z)} = \frac{\int_0^\infty v_f(z; r) (dm_\cond/dr) dr}{ \int_0^\infty (dm_\cond/dr) dr } = \frac{\int_0^\infty dr \, v_f(z;r) r^3 q(r)}{ \int_0^\infty dr \, r^3 q(r)}.
\end{eqnarray}
In the atmospheric layer at height \(z\), the total mass of cloud particles per unit volume is \(M_c(z) = X_\cond(z) \rho\), where \(\rho\) is the mass density of the atmosphere. To satisfy the \(\fsed\) assumption, the following condition must hold:
\begin{eqnarray} 
\label{eq:fsedw2}
\frac{\fsed \Kzz(z)}{L} = \frac{\int_0^\infty dr \, v_f(z;r) r^3 q(r)}{ \int_0^\infty dr \, r^3 q(r)}
\end{eqnarray}
This ensures that the mass-weighted average sedimentation velocity \(\overline{v}_f(z)\) aligns with the \(\fsed\) assumption.

In the cloud particle size distribution layer at height \(z\), we assume a log-normal distribution:
\begin{eqnarray} 
\label{eq:lognormal_am01}
q(r) dr &=& N(z) p_z(r) dr \\
\label{eq:lognormal_am01_pz}
p_z(r) &=& \frac{1}{r \sqrt{2 \pi} \log{\sigma_g}} \exp{\left\{ - \frac{1}{2 (\log{\sigma_g})^2} \left[\log{\left(\frac{r}{r_g(z)}\right)}\right]^2 \right\}} 
\end{eqnarray}
and \(\log{\sigma_g} > 0\) (\(\sigma_g > 1\)). Note that \(\sigma_g\) represents the spread of the distribution, and we use a common value for each layer. Using the moment formula for the log-normal distribution:
\[
E_k \equiv \int_0^\infty r^k p_z(r) \, dr = r_g^k(z) e^{k^2 \log^2{\sigma_g}/2},
\]
we can express the total mass of cloud particles per unit volume at height \(z\) as:
\[
M_c(z) = X_\cond \rho = \int_0^\infty \frac{4 \pi \delta_\cond r^3}{3} q(r) \, dr = N(z) \frac{4 \pi \delta_\cond r_g^3(z)}{3} \exp{\left(\frac{9}{2} \log^2 \sigma_g\right)}.
\]
From this, we find the number density of particles \(N(z)\) as 
\begin{align}
\label{eq:lognormal_am01N}
N(z) &= \frac{X_c \rho}{\tilde{m}},
\end{align}
where
\begin{align}
\label{eq:lognormal_mean_mass}
\tilde{m} &\equiv \frac{4 \pi r_g^3}{3} \delta_c \exp{\left(\frac{9}{2} \log^2{\sigma_g} \right)}  
\end{align}
can be interpreted as the average mass of a single cloud particle, linking the cloud mass density to the particle number density. From the requirement derived from the \(\fsed\) assumption (Equation \ref{eq:fsedw2}):
\[
\int_0^\infty v_f(z; r) r^3 q(r) \, dr = \fsed \frac{\Kzz(z)}{L} E_3
\]
we need to assume a \(v_f(r; z)\). Based on previous observations, despite minor variations due to Reynolds number, the terminal velocity generally follows a power-law distribution. Therefore, we assume a power-law distribution for sedimentation velocity as $v_f(r, z) = A r^\alpha$.
Substituting this into equation \ref{eq:fsedw2} gives:
\[
A = \frac{\Kzz(z)}{r_g^\alpha(z)} \frac{\fsed}{L} \frac{E_3}{E_{\alpha+3}} = \frac{\Kzz(z)}{r_g^\alpha(z)} \frac{\fsed}{L} e^{ -(\alpha^2 + 6 \alpha) \log^2{\sigma_g}/2 }
\]
Thus,
\[
\label{eq:vf_rw}
v_f(r; z) = \frac{\Kzz(z)}{r_g^\alpha(z)} \frac{\fsed}{L} e^{ -(\alpha^2 + 6 \alpha) \log^2{\sigma_g}/2 } r^\alpha = \frac{\Kzz(z)}{L} \left( \frac{r}{r_w(z)} \right)^\alpha,
\]
where
\begin{eqnarray} 
\label{eq:lognormal_am01rg}
r_w(z) = r_g(z) \fsed^{-1/\alpha} \exp{\left[ \left(\frac{\alpha+6}{2} \right) \log^2{\sigma_g}\right]}.
\end{eqnarray}
This ensures that the \(\fsed\) assumption holds with such a distribution function. Note that \(r_w(z)\) provides a typical scale, indicating the particle size whose sedimentation velocity balances with the vertical transport at a given \(z\).

To close the equations, we need to know \(r_g(z)\) and \(\alpha\) using the terminal velocity model. Depending on the Davis number $N_D = C_d N_{re}^2$ (or the Reynolds number $N_{re}$), the terminal velocity of the droplet is computed as follows, 
\begin{eqnarray}
\label{eq:vterm}
v_f =
\begin{cases}
\displaystyle{\frac{2}{9 \eta} g r^2 (\rho_c - \rho) (1+1.26 N_{Kn}) } & \mbox{\, for $N_D < 42$ ($N_{re} < 2$)} \\
\displaystyle{ \frac{\eta}{2 \rho r} \exp{(-0.0088 \log^2{N_D}+0.85 \log{N_D} -2.49)}} & \mbox{\, for $42 \le N_D < 10^5$ ($2 \le N_{re} < 500$)} \\
\displaystyle{\frac{\eta}{2 \rho r} \sqrt{\frac{N_D}{C_d}}}& \mbox{\, for $10^5 \ge N_D$ ($500 \ge N_{re}$)} 
\end{cases}
\end{eqnarray}
In the middle range of $N_D$ in Equation (\ref{eq:vterm}), we fit the polynomial equation to the $\log{N_D}$ -- $\log{N_{re}}$ relation given in Table 10.1 of \cite{pruppacher2010microstructure}. Following \cite{ackerman2001precipitating}, we adopt 0.45 to $C_d$ for $10^5 \ge N_D$. The dynamic viscosity $\eta$ can be estimated by 
\begin{eqnarray}
\label{eq:dyvis}
\eta = \frac{5}{16} \frac{\sqrt{\pi m k_B T}}{\pi d^2} \frac{(k_B T/\varepsilon)^{0.16}}{1.22}
\end{eqnarray}
where $d$ and $\varepsilon$ are Lennard-Jones potential parameters \citep{rosner2012transport}.
By considering \( r_w \) as the \( r \) at which \( v_f(r) \) equals \(\Kzz(z)/L\), \( r_g \) can be calculated from equation (\ref{eq:vterm}).

In addition, the effective droplet radius is expressed as
\begin{align}
r_\mathrm{eff} \equiv \frac{\int r^3 p_z(r) dr}{\int r^2 p_z(r) dr} =  e^{5 \log^2 \sigma_g/2} \, r_g = \fsed^{1/\alpha} \exp{\left(-\frac{1 + \alpha}{2} \log^2 \sigma_g\right)}  \, r_w,
\end{align}

In short, the features of this cloud model (constant \(k_c\) and \(\fsed\)) are summarized as follows: Given \(\fsed\), \(\Kzz(z)\), \(\sigma_g\), and \(k_c\), the vertical distribution of cloud mixing ratios in each layer (\ref{eq:AM01cx}) and the cloud particle distribution in each layer assuming a log-normal distribution (\ref{eq:lognormal_am01}, \ref{eq:lognormal_am01N}, \ref{eq:lognormal_am01rg}) are provided.

\vspace{\baselineskip}

\bibliography{sample63}{}
\bibliographystyle{aasjournal}
\end{document}